%% file: paper_arxiv.tex
\documentclass[aps,prl,reprint, superscriptaddress, ,longbibliography]{revtex4-1}
\usepackage[section]{placeins}

\include{header}

\include{acronyms}

\usepackage[normalem]{ulem}

\begin{document}
\author{Mattia Moroder}
\affiliation{Department of Physics, Arnold Sommerfeld Center for Theoretical Physics (ASC), Munich Center for Quantum Science and Technology (MCQST), Ludwig-Maximilians-Universit\"{a}t M\"{u}nchen, 80333 M\"{u}nchen, Germany}
\author{Matteo Mitrano}
\affiliation{Department of Physics, Harvard University, Cambridge, Massachusetts 02138, USA}
\author{Ulrich Schollw\"ock}
\affiliation{Department of Physics, Arnold Sommerfeld Center for Theoretical Physics (ASC), Munich Center for Quantum Science and Technology (MCQST), Ludwig-Maximilians-Universit\"{a}t M\"{u}nchen, 80333 M\"{u}nchen, Germany}

\author{Sebastian Paeckel}\thanks{Author to whom correspondence should be address; sebastian.paeckel@physik.uni-muenchen.de. These authors contributed equally and are listed alphabetically.}
\affiliation{Department of Physics, Arnold Sommerfeld Center for Theoretical Physics (ASC), Munich Center for Quantum Science and Technology (MCQST), Ludwig-Maximilians-Universit\"{a}t M\"{u}nchen, 80333 M\"{u}nchen, Germany}

\author{John Sous}\thanks{Author to whom correspondence should be address; john.sous@yale.edu.  These authors contributed equally and are listed alphabetically.}
\affiliation{Department of Chemistry and Biochemistry, University of California San Diego, La Jolla, California 92093, USA}
\affiliation{Department of Applied Physics and the Energy Sciences Institute, Yale University, New Haven, Connecticut 06511, USA}

\def\thetitle{Phonon state tomography of electron correlation dynamics in optically excited solids} 
\title{\thetitle}
\begin{abstract}
We introduce phonon state tomography (PST) as a diagnostic probe of electron dynamics in solids whose phonons are optically excited by a laser pulse at initial time. Using a projected-purified matrix-product states algorithm, PST decomposes the exact correlated electron-phonon wavefunction into contributions from purely electronic states corresponding to statistically typical configurations of the optically accessible phononic response, enabling a ‘tomographic’ reconstruction of the electronic dynamics generated by the phonons. Thus, PST may be used to diagnose electronic behavior in experiments that access only the phonon response, such as thermal diffuse x-ray and electron scattering. We study the dynamics of a metal whose infrared phonons are excited by an optical pulse at initial time and use it to simulate the sample-averaged momentum-resolved phonon occupancy and accurately reconstruct the electronic correlations. We also use PST to analyze the influence of different pulse shapes on the light-induced enhancement and suppression of electronic correlations.
\end{abstract}
\maketitle
\paragraph{Introduction.}
Optical driving of quantum materials has opened a door for engineering and controlling novel electronic states of matter~\cite{Basov2017-gu,RevModPhys.93.041002},  with a prime example being the generation of non-equilibrium superconducting states ~\cite{Fausti2011,Mitrano2016,Cantaluppi2018,C8CC01745J,Schwarz2020,PhysRevLett.124.153602,Budden2021,Buzzi2021,campi_prediction_2021} and emergent Floquet phases~\cite{Gedik2013, Mahmood2016-co, Shan2021-og, Zhou2023-ws}.  
One important class of experiments involves the optical excitation of specific phonons that couple strongly to the electronic states~\cite{Rini2007-eo, Forst2011-nr, Subedi2014, Nova2017-vd, Disa2021-jt,PhysRevX.11.021067,Henstridge2022-yl}.
These infrared-active phonons inherit the light's dipole character, meaning that dipole selection rules govern the coupling between the electrons and photoexcited phonons.
As a result, in centrosymmetric crystals, lattice nonlinearities govern the light-induced dynamics, giving rise to novel behavior inaccessible in equilibrium where nonlinear couplings are subdominant fluctuations~\cite{Sous2021}.  
\begin{figure}[!t]
    \centering
    \includegraphics[width=0.45\textwidth]{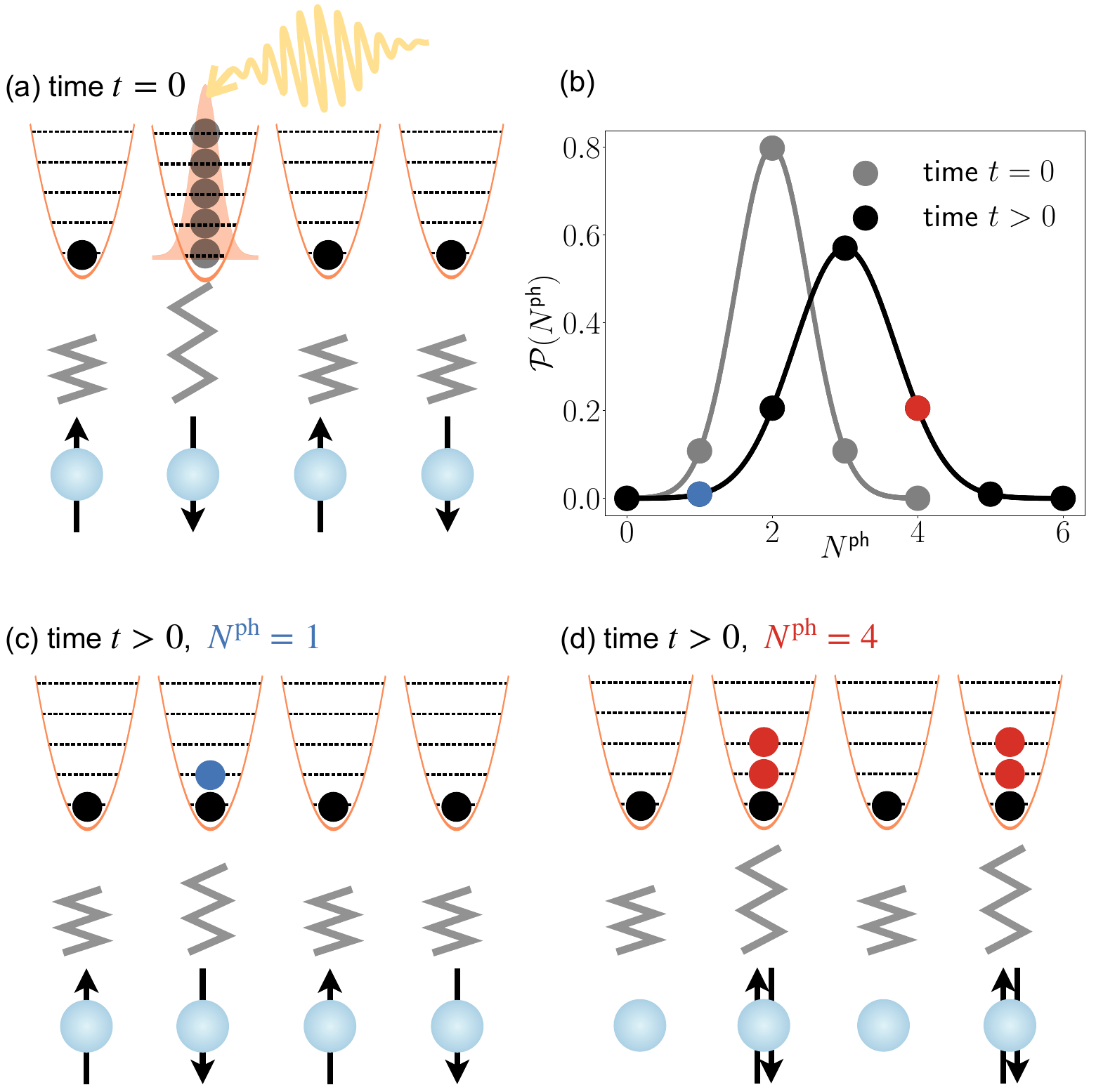}
    \caption{ 
        \Acrfull{PST}. (a) A metal whose electrons (blue spheres) couple locally to phonons (black dots), which are excited at $t=0$ by a pump pulse.
        (b) The probability $\mathcal{P}(N^{\text{ph}})$ to find $N^{\text{ph}}$ phonons evolves over time.
        (c) and (d) Cartoon of  the~\gls{PST} sampling approach used here to extract the phononic configurations of most relevance to the electronic state, thus enabling a tomographic decomposition of electronic observables and correlations.
 }
    \label{fig:PST:cartoon}
\vspace{-6mm}
\end{figure}

Probing and interpreting this rich behavior in the post-pump dynamics is a rapidly growing area of research whose goal is to extract a reliable understanding of light-induced energy transport and electronic correlations.  Optical probes provide direct access to charge transport~\cite{Mitrano2016,Budden2021}.
However, despite vigorous activity~\cite{PhysRevX.3.041033, PhysRevB.92.224517, PhysRevB.93.144506,
PhysRevB.94.214504_2016, PhysRevB.96.014512_2017, Kennes2017, PhysRevB.96.054506, PhysRevB.96.045125,PhysRevLett.120.246402,Hubener2018-cd, Paeckel2020,PhysRevX.11.041028,eckhardt2023theory}, accurate simulation of transport far away from equilibrium in strongly coupled electron-phonon systems remains a daunting challenge~\cite{Sous2021}, complicating interpretation of experiments.
Direct phonon probes such as x-ray diffuse scattering~\cite{Trigo2010imaging,Trigo2013fourier} and electron scattering~\cite{Konstantinova2018nonequilbrium} provide access to the phononic response induced by coupling to the electrons.
Here we present a~\gls{MPS}\hyp based method, dubbed~\acrfull{PST}, to simulate the sample-averaged dynamics of the phonon distribution function as would be obtained in x-ray diffuse scattering and electron scattering experiments.
\gls{PST} enables us to accurately represent the full electron-phonon wavefunction in terms of electronic components associated with the experimentally measured phononic quantity, e.g., the sample-averaged momentum-resolved phonon occupation.
Using this decomposition, we can accurately reconstruct electronic correlation functions by sampling electronic contributions associated with the statistically typical phononic configurations inferred from the spatially resolved phonon density matrix.
This means that \gls{PST} can be used as a diagnostic tool to relate the experimentally obtained sample-averaged phonon response following various time delays to the dynamics of electronic correlation functions, providing a \textit{one\hyp to\hyp one relation} between  the statistically measured phononic response and the underlying electronic behavior.
Once matched to experimental measurements, \gls{PST} can be used to tomographically reconstruct the electronic correlations without needing the full exact electron-phonon wavefunction.
This is pictorially summarized in Fig.~\ref{fig:PST:cartoon}, where an optical pulse excites a coherent phonon population (panel a). The probability distribution of having $N^\mathrm{ph}$ phonons excited globally evolves over time (panel b). At a given post-pump time delay $t>0$, the electronic state can be reconstructed statistically in terms of contributions from distinct $N^\mathrm{ph}$ configurations.
The tomographic decomposition of the electron\hyp phonon wavefunction and sampling over statistically weighted electronic states is made possible thanks to a new class of \acrfull{PP-MPS} methods~\cite{Koehler2021,Stolpp2021} which enables efficient computation of the spatially resolved global and local phonon reduced density matrices of the coupled electron-phonon system.
Given this, \gls{PST} can provide insights into the microscopic physics of materials probed in experiments that access the frequency, damping, occupation number, and couplings to other degrees of freedom of the optically excited phonon subsystem interrogated with multiple time\hyp resolved probes, such as infrared and Raman spectroscopy, x-ray, and electron scattering.
This may shed light into the mechanism behind dynamically\hyp enhanced superconductivity in the copper oxides~\cite{Hu2014optically,Kaiser2014optically} and organic materials~\cite{Mitrano2016,Buzzi2020photomolecular}, as well as of light-induced ferroelectricity~\cite{Fechner2024-gk} and high-temperature ferromagnetism~\cite{Disa2023-wq}.
In these materials, specific phonon resonances in the dynamical optical response~\cite{Liu2020pump,Rowe2023resonant} as well as signatures of structural distortions due to nonlinearities~\cite{Mankowsky2014nonlinear,Mankowsky2015coherent} highlight the significant role phonons play in the dynamics.
A particularly intriguing class of experiments are those based on diffuse x-ray and electron scattering, in which the scattering intensity is directly proportional to a weighted sum of phonon populations for different branches and at different momenta.
In the first-order kinematical approximation \cite{Xu2005determination}, the diffuse scattering intensity $I_S$ is given by 
\begin{equation}
 I_{\mathrm{S}}\sim\sum_i\frac{1}{\omega_i(\mathbf{k})}\bigg(n_i(\mathbf{k})+\frac{1}{2}\bigg)|F_i(\mathbf{k})|^2,
    \label{eq:n_q_from_TDS}
\end{equation}
where $\omega_i(\mathbf{k})$, $F_i(\mathbf{k})$, and $n_i(\mathbf{k})$ are frequency, structure factor and occupation number of the $i^{\mathrm{th}}$ phonon branch at momentum $\mathbf{k}$.
Given the Hamiltonian governing the electronic and phononic dynamics and their couplings, \gls{PST} allows us to simulate $I_{\mathrm{S}}$ and reconstruct the associated electronic correlations, thus potentially enabling interpretation of the ultrafast electron diffraction and x-ray diffuse scattering on cuprate superconductors (e.g. Bi$_2$Sr$_2$CaCu$_2$O$_{8+\delta}$ and YBa$_2$Cu$_3$O$_{6+\delta}$) under resonant excitation of the $c$-axis apical oxygen distortions \cite{Liu2020pump}.
\paragraph{\label{sec:method}\Acrlong{PST}.}
\begin{figure}[t!]
    \centering
    \includegraphics[width=0.48\textwidth]{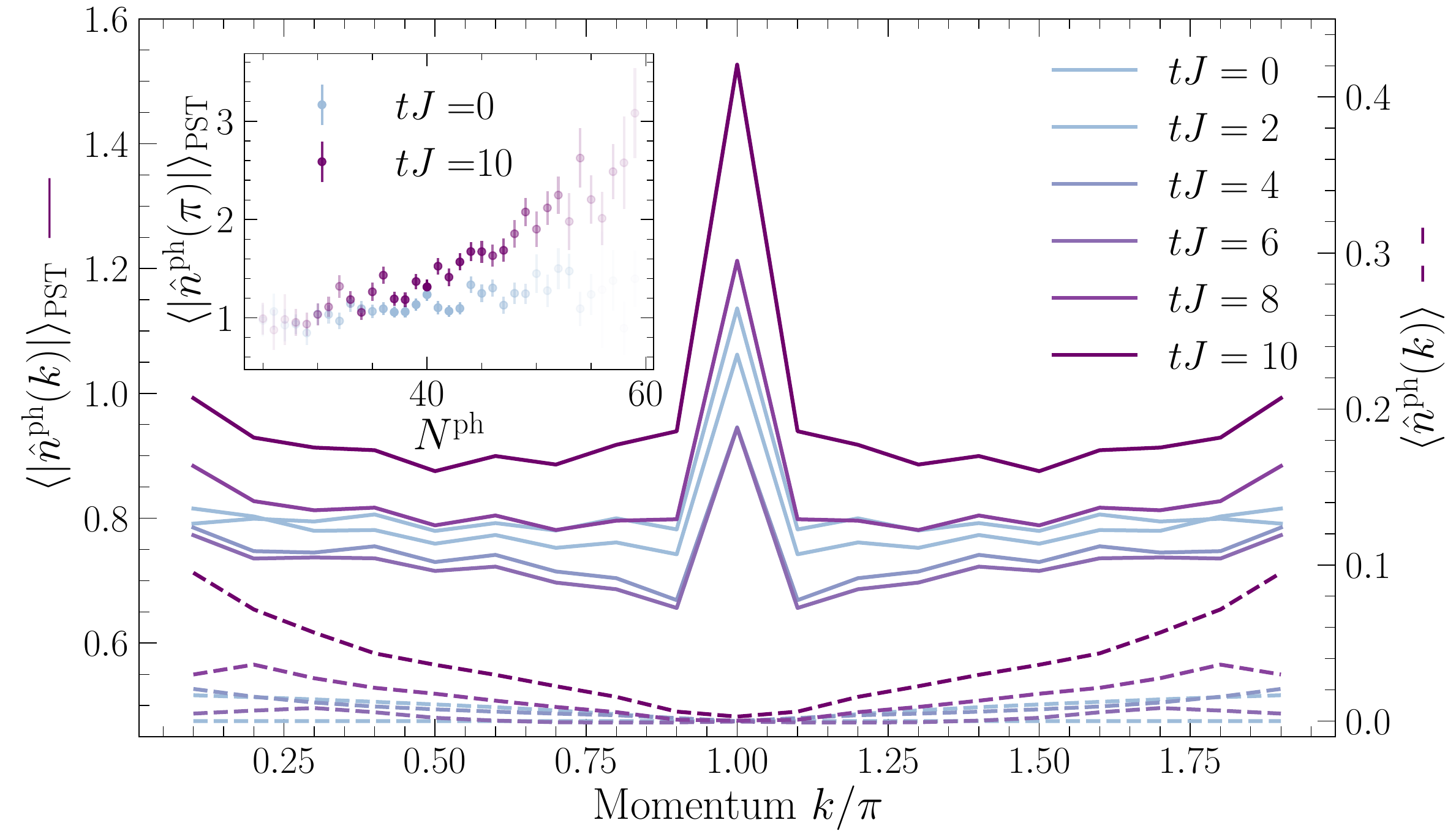}
    \caption{
    \label{fig:nk:sampled}
 Sample-averaged  momentum-resolved amplitude of the phonon distribution function $\langle \mathbf |{\hat{n}^\mathrm{ph}(k)}|\rangle_\mathrm{PST}$ from \gls{PST} and the corresponding expectation value of the momentum-resolved phonon distribution function $\langle {\hat{n}^\mathrm{ph}(k)}\rangle$ following different time delays. 
 We decompose $\langle \mathbf |{\hat{n}^\mathrm{ph}(\pi)}|\rangle_\mathrm{PST}$ into contributions from different total phonon number sectors $N^\mathrm{ph}$ in the inset. The  color intensity of the symbols indicates the fraction of all drawn samples found to have the corresponding value of $N^\mathrm{ph}$.
 We constructed this data using $3000$ samples. Error bars represent statistical errors corresponding to 1 standard deviation scaled by the inverse of the square root of the number of samples.
 }
\vspace{-6mm}    
\end{figure}
Given the exact electron\hyp phonon wavefunction, the idea behind~\gls{PST} is to draw samples of electronic states $\ket{\psi^\mathrm{el}(\mathbf n^\mathrm{ph})}$ associated with statistically typical phononic configurations labeled by their global spatially-resolved phonon occupation over the lattice $\mathbf n^\mathrm{ph} = (n^\mathrm{ph}_1, n^\mathrm{ph}_1, \ldots, \ n^\mathrm{ph}_L)$, where $L$ is the system size.
We thus decompose the electron\hyp phonon wavefunction as
\begin{equation}
    \ket{\psi}
 =
    \sum_{\{ \mathbf n^\mathrm{ph} \}} \ket{\psi^\mathrm{el}(\mathbf n^\mathrm{ph})} \otimes \left( \sqrt{ p(\mathbf n^\mathrm{ph})} \ket{\mathbf n^\mathrm{ph}} \right) \;,
\label{Eq:eq2}
\end{equation}
where $\ket{\psi^\mathrm{el}(\mathbf n^\mathrm{ph})}$ are normalized electronic states in the electronic Hilbert space $\mathcal H^\mathrm{el}$, $\ket{\mathbf n^\mathrm{ph}} \in \mathcal H^\mathrm{ph}$ are phonon number states which span the phononic Hilbert space $\mathcal H^\mathrm{ph}$, and $p(\mathbf n^\mathrm{ph})$ is the probability of finding the phonon configuration $\mathbf{n}^\mathrm{ph}$ in the state $\ket{\psi}$.
We represent $\ket{\psi}$ as a \gls{MPS}~\cite{white_1992,white_1993,schollwoeck_2005,schollwoeck_2011}.
We generalize the \gls{PSS}~\cite{Ferris2012} employed in snapshot techniques used in the study of ultracold atoms~\cite{Bohrdt2019,Bohrdt2021,Bohrdt2021b,Buser2022,Palm2022,pauw2023,Hirthe2023} to draw phonon configurations $\mathbf n^\mathrm{ph}$ distributed according to $p(\mathbf n^\mathrm{ph})$ and the associated electronic states $\ket{\psi^\mathrm{el}(\mathbf n^\mathrm{ph})}$.
The sampling is performed by successively decomposing $p(\mathbf n^\mathrm{ph})$ into the probability over a given lattice site conditioned on the probability over the remaining sites: $p(\mathbf n^\mathrm{ph}) = p(n^\mathrm{ph}_j) p(\mathbf n^\mathrm{ph}\setminus n^\mathrm{ph}_j \vert n^\mathrm{ph}_j)$ (where $n^\mathrm{ph}_j$ is the phonon occupation on site $j$ and $\mathbf n^\mathrm{ph}\setminus n^\mathrm{ph}_j$ is the set of occupations associated with all sites except site $j$), see Supporting Information for details.

\paragraph{\gls{PST} of an optically pumped metal.}
%
To demonstrate the power of the \gls{PST}, we simulate the dynamics of the sample\hyp averaged momentum\hyp resolved phonon distribution function and reconstruct the associated electronic correlations of a one\hyp dimensional metal whose phonons are excited by a light pulse at initial time.
This system is governed by the Hamiltonian~\cite{Kennes2017,Sous2021}
\begin{equation}
\begin{aligned}
    \hat{H}
    &=
    \underbrace{
         -J \sum_{j = 1}^L \sum_{\sigma = \uparrow, \downarrow}
         \left(
            \hat{c}_{j,\sigma}^\dagger \hat{c}_{j+1,\sigma}^\nodagger + \text{h.c.}
        \right )
    }_{
        \hat{H}_{\text{el}}
    } \\
    &+
    \underbrace{
        \omega \sum_{j=1}^L
        \left(
            \hat{b}_j^\dagger \hat{b}_j^\nodagger + \frac{1}{2}
        \right )
    }_{
        \hat{H}_{\text{ph}}
    }
    +
    \underbrace{
 g \sum_{j=1}^L
        \left(
            \hat{n}_j - 1
        \right)
       \left (  \hat{b}_j^\dagger + \hat{b}_j^\nodagger \right )^2
    }_{
        \hat{H}_{\text{el-ph}}
    }\; . 
\end{aligned}
\label{eq:hamiltonian}
\end{equation}
Here, electrons with spin $\sigma$ and hopping amplitude $J$ are quadratically coupled with coupling coefficient $g$ to Einstein oscillators with frequency $\omega$ on a chain with size $L$.
This nonlinear coupling follows from the fact that optically accessible phonons are long\hyp wavelength dipole modes that cannot couple linearly to electrons in the presence of inversion symmetry~\cite{Forst2011-nr,Subedi2014,Mankowsky2016,Kennes2017,PhysRevB.95.205111}.
Note that PST is not limited to any specific form of the electron-phonon coupling and can be applied to study model Hamiltonians with both linear and nonlinear couplings.
We choose a very small value of $g$ so that the initial equilibrium state is essentially indistinguishable from a metal of free fermions in a product state with the vacuum of the phonons.
We simulate the dynamics following the application of a spatially homogeneous pump pulse $\hat{A}^{\mathrm{c}} \equiv \hat{\mathcal{D}}(\alpha) = \prod_j \hat{d}_j(\alpha)\equiv \prod_j e^{\alpha \hat{b}^\dagger_j - \alpha^* \hat{b}_j}$ ($\alpha$ is the pump fluence) which  coherently and uniformly displaces the oscillator on every site of the crystal.
To simulate the dynamics we use a hybrid time\hyp evolution scheme for the~\gls{MPS} combining global and local expansion techniques~\cite{Haegeman2011,Haegeman2016,mps_time_ev_methods_paeckel,Yang_2020prb} and represent the phononic degrees of freedom with ~\gls{PP}~\cite{Koehler2021,Stolpp2021}, see Supporting Information for details.
In the following we consider a system size with $L = 20$ with $g/J = 0.07$, $\omega/J = \pi/10$ and $\alpha=\sqrt{2}$.
We employ a local phonon Hilbert space dimension of up to $\mathrm{d}^{\mathrm{ph}}=40$, which we verify to be sufficient to achieve convergence to the limit of the infinite phonon Hilbert space. 
\begin{figure}[!h]
    \subfloat[\label{fig:coherent:dynamics:both}]{
        \centering
        \includegraphics[width=0.4\textwidth]{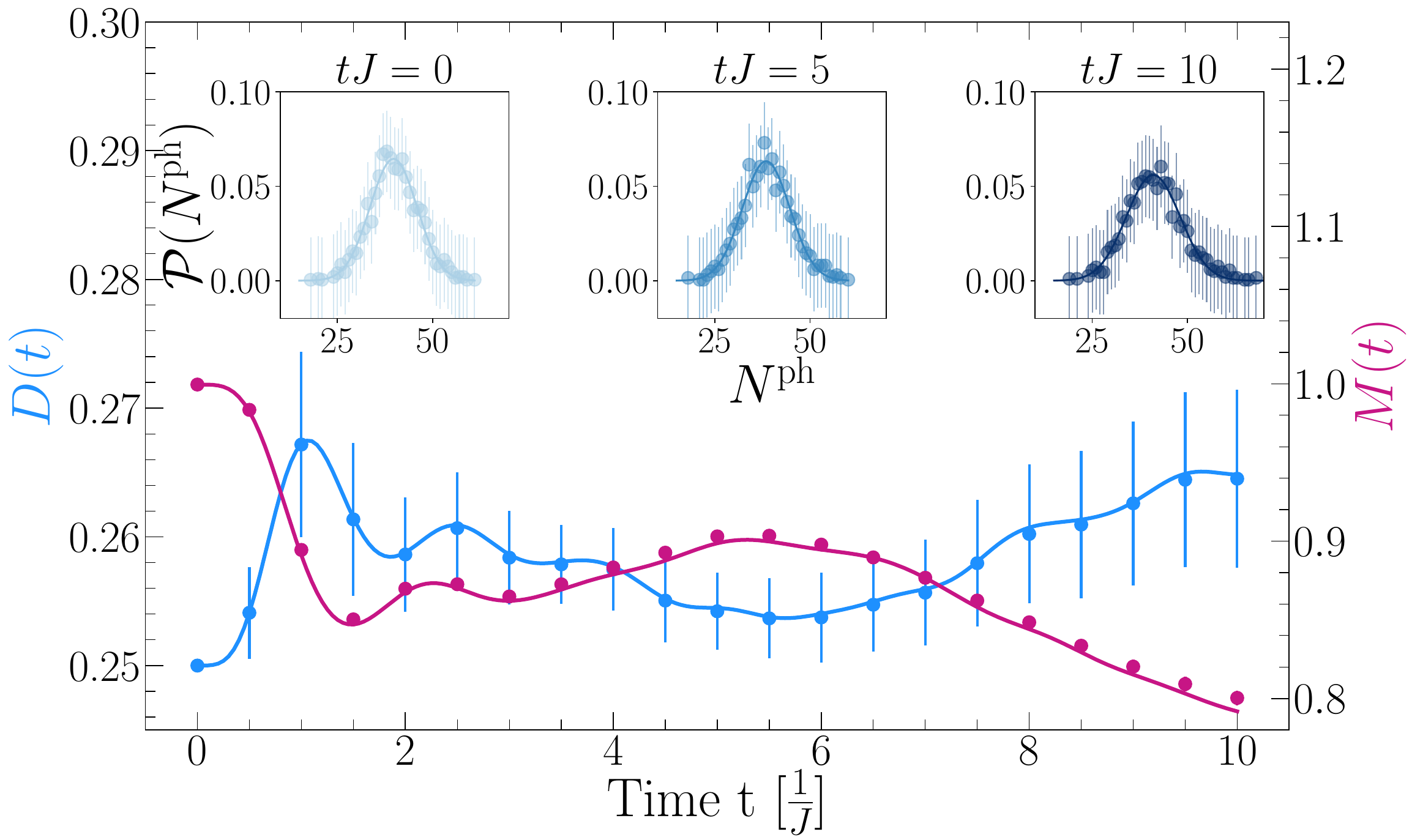}
 }\\
    \vspace*{-1em}
    \subfloat[\label{fig:coherent:dynamics:time-resolved:D}]{
        \centering
        \includegraphics[width=0.40\textwidth]{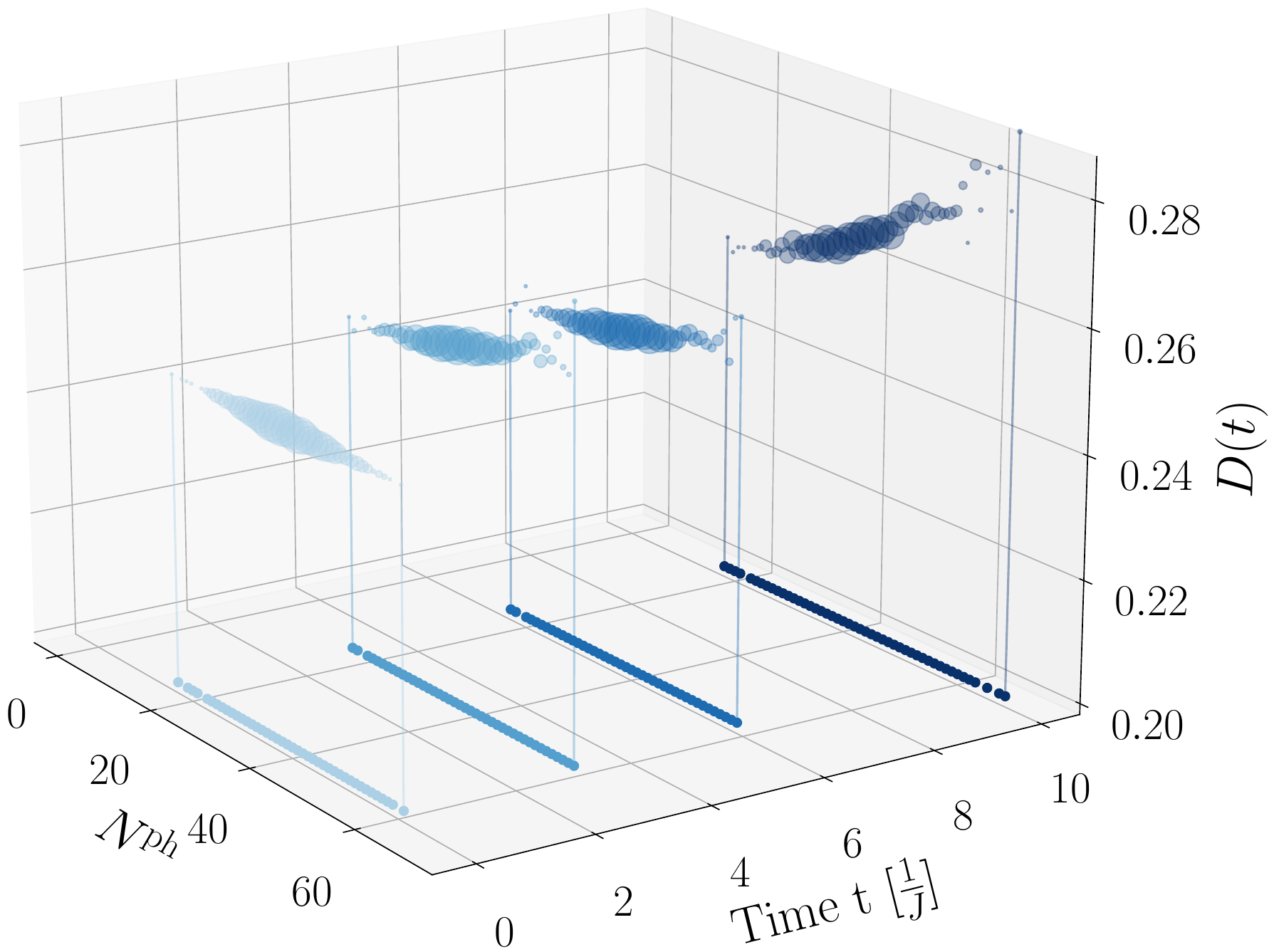}
 }\\
    \vspace*{-2em}
    \subfloat[\label{fig:coherent:dynamics:time-resolved:S}]{
        \centering
        \includegraphics[width=0.39\textwidth]{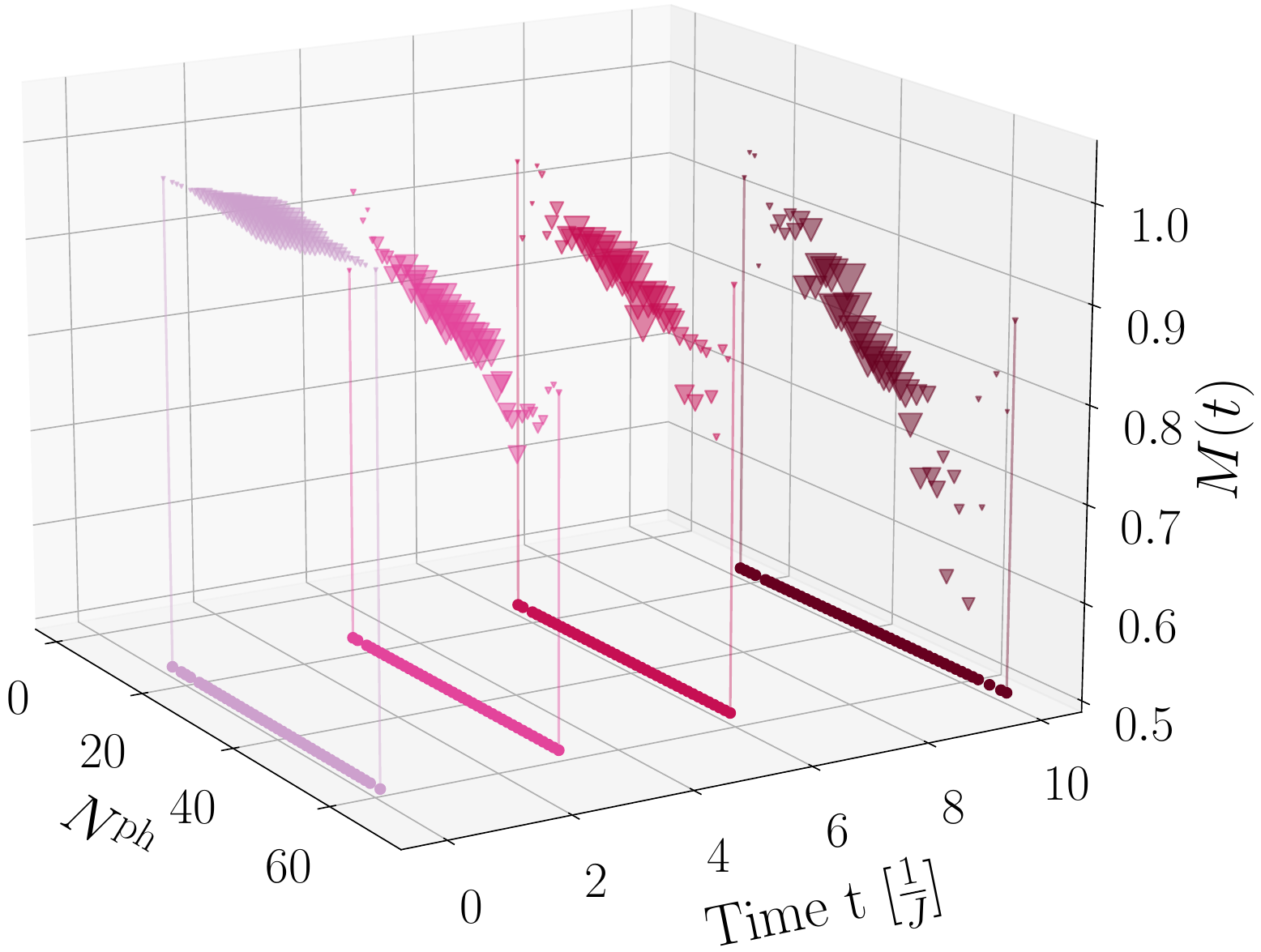}
 }
    \caption
 {
        \label{fig:coherent:dynamics}
 Dynamics of electronic correlations reconstructed using PST.
        \subfigref{fig:coherent:dynamics:both} The reconstructed charge double occupancy $D(t)$ (left axis) and staggered magnetic correlation $M(t)$ (right axis) (symbols with error bars) are compared with the exact  correlations (solid lines).
 In the inset, we show the sampled global phonon distribution function $\mathcal{P}(N^{\mathrm{ph}})$ at different times.
 In panels~\subfigref{fig:coherent:dynamics:time-resolved:D} and~\subfigref{fig:coherent:dynamics:time-resolved:S} we obtain a decomposition of $D(t)$ and $M(t)$, respectively, into contributions from different total phonon number sectors $N^\mathrm{ph}$.  Here, the size of the symbols indicates the fraction of all drawn samples found to have the corresponding value of $N^\mathrm{ph}$.
 The small dots at the bottom of panels \subfigref{fig:coherent:dynamics:time-resolved:D} and \subfigref{fig:coherent:dynamics:time-resolved:S} are projections onto the plane which serve as a guide to the eye.
 We have used $2000$ samples in this figure. Error bars represent statistical errors corresponding to 1 standard deviation scaled by the inverse of the square root of the number of samples.
 }
\vspace{-6mm}
\end{figure}

\paragraph{Sample-averaged momentum-resolved phonon distribution function.}
We use~\gls{PST} to simulate the experimental protocol for measuring the sample-averaged momentum-resolved phonon distribution function.
Phonon probes such as x-ray diffuse and electron scattering measure the momentum-resolved phonon distribution function for a given sample, and statistically significant results are constructed by averaging over a sufficiently large number of samples.
Experiments do not access the phase associated with the momentum-resolved phonon distribution function for a given sample. Thus, the sample-averaged momentum-resolved phonon occupation function corresponds to the sample-averaged amplitude of the Fourier-transformed spatially-resolved phonon distribution function measured for every given sample.
We thus use \gls{PST} to simulate the sample-averaged absolute value of the Fourier-transformed phonon occupation $\langle \mathbf |{\hat{n}^\mathrm{ph}(k)}|\rangle$ (which we henceforth refer to as $\langle \mathbf |{\hat{n}^\mathrm{ph}(k)}|\rangle_\mathrm{PST}$).
In~\cref{fig:nk:sampled}, we show $\langle \mathbf |{\hat{n}^\mathrm{ph}(k)}|\rangle_\mathrm{PST}$ and the expectation value of the phonon occupation in the exact electron\hyp phonon wavefunction $\langle {\hat{n}^\mathrm{ph}(k)}\rangle$.
Averaging the amplitude of the occupation function over sample realizations  after applying the Fourier transform provides qualitatively different information than that obtained from the exact expectation value of the phonon occupation containing the phase relation between different components of the full wavefunction (which corresponds to averaging over samples without taking the absolute value of each sample).
This is an important result because it means that access to the full electron\hyp phonon wavefunction is {\em not} sufficient to directly interpret sample\hyp averaged momentum-resolved phononic observables obtained in experiments and that appropriately decomposing the full wavefunction into a suitable basis corresponding to the experimentally measured quantity, the momentum-resolved phonon occupation in this case, is essential.
That is precisely what~\gls{PST} achieves.
An important feature in the simulated $\langle \mathbf |{\hat{n}^\mathrm{ph}(k)}|\rangle_\mathrm{PST}$ is the appearance of a peak at $k=\pi$ absent in the exact expectation value of the phonon occupation.
We understand this behavior qualitatively as follows.
To construct a sample with a given  total phonon number $N^\mathrm{ph}$ one samples over different phonon occupations whose sum adds up to $N^\mathrm{ph}$; these occur with probability given by the onsite local phonon distribution function (which is Poisson at initial time).
This means that exciting different phonon numbers at different sites occurs with unequal probabilities (because the local distribution function is not uniform), leading to an overall inhomogeneity in the distribution of phonon states over the lattice.
This manifests as a nearest-neighbor structure, which gives rise to a peak at $k=\pi$ in the sample-averaged phonon occupation.
Importantly, at the initial time, just after the application of the pump, the Poisson distribution over phonon number states on each site means that low\hyp lying phonon number states are more likely to be occupied with an average of $|\alpha|^2 = 2$ phonons excited on each site. At later times, more phonons are excited during the course of the dynamics.
Indeed this is what we observe numerically in the inset of~\cref{fig:nk:sampled} where we decompose $\langle \mathbf |{\hat{n}^\mathrm{ph}(\pi)}|\rangle_\mathrm{PST}$ into contributions from sectors with total phonon number $N^\mathrm{ph}\equiv \sum_j n^\mathrm{ph}_j$ both at initial and late times.
In the following, we will connect this observation to the analysis of electronic correlations.
\begin{figure}[!t]
    \subfloat[\label{fig:squeezed:dynamics}]
 {
        \centering
        \includegraphics[width=0.46\textwidth]{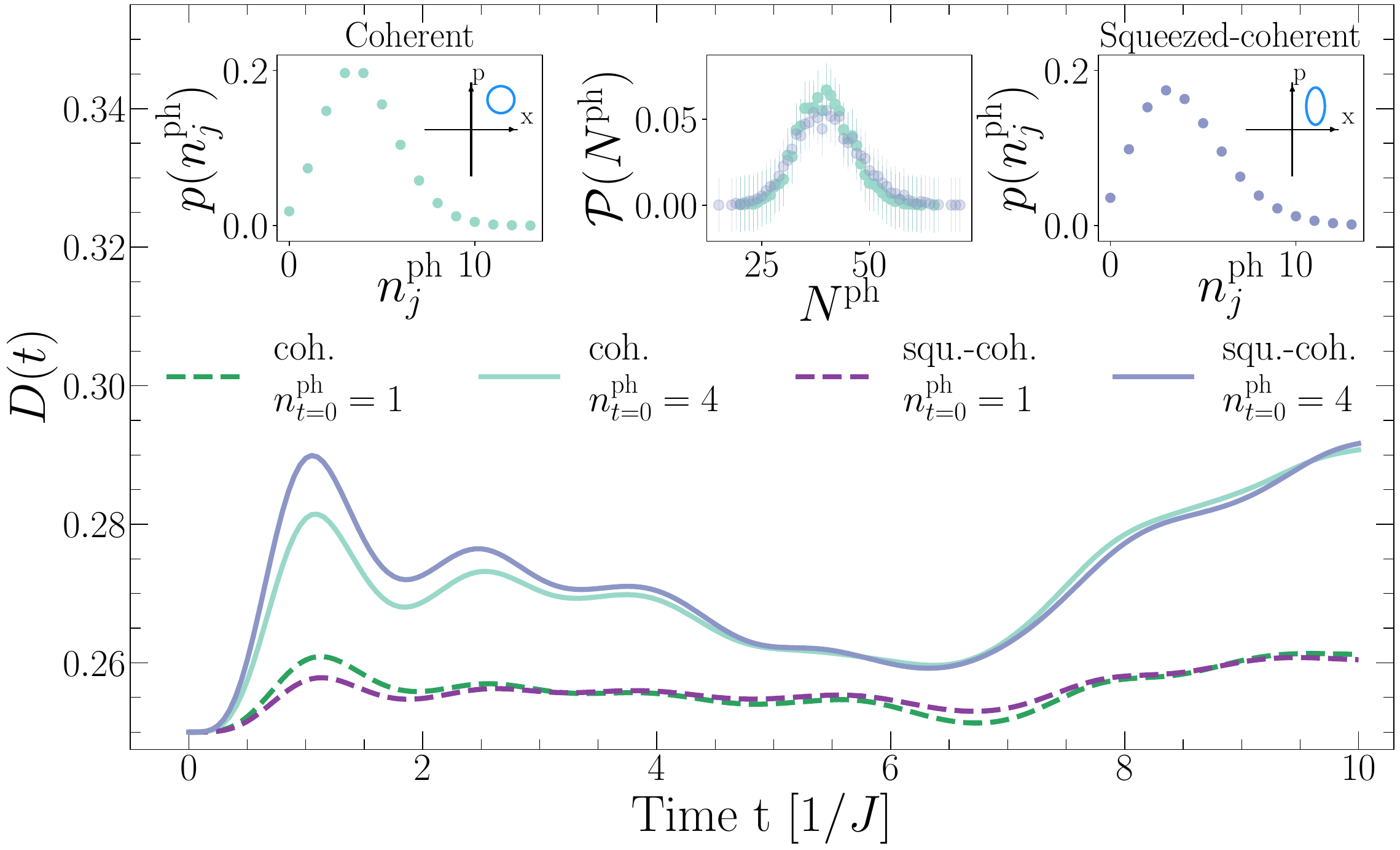}
 }\\
    \vspace*{-2em}
    \subfloat[\label{fig:beta:dynamics}]
 {
        \centering 
        \includegraphics[width=0.46\textwidth]{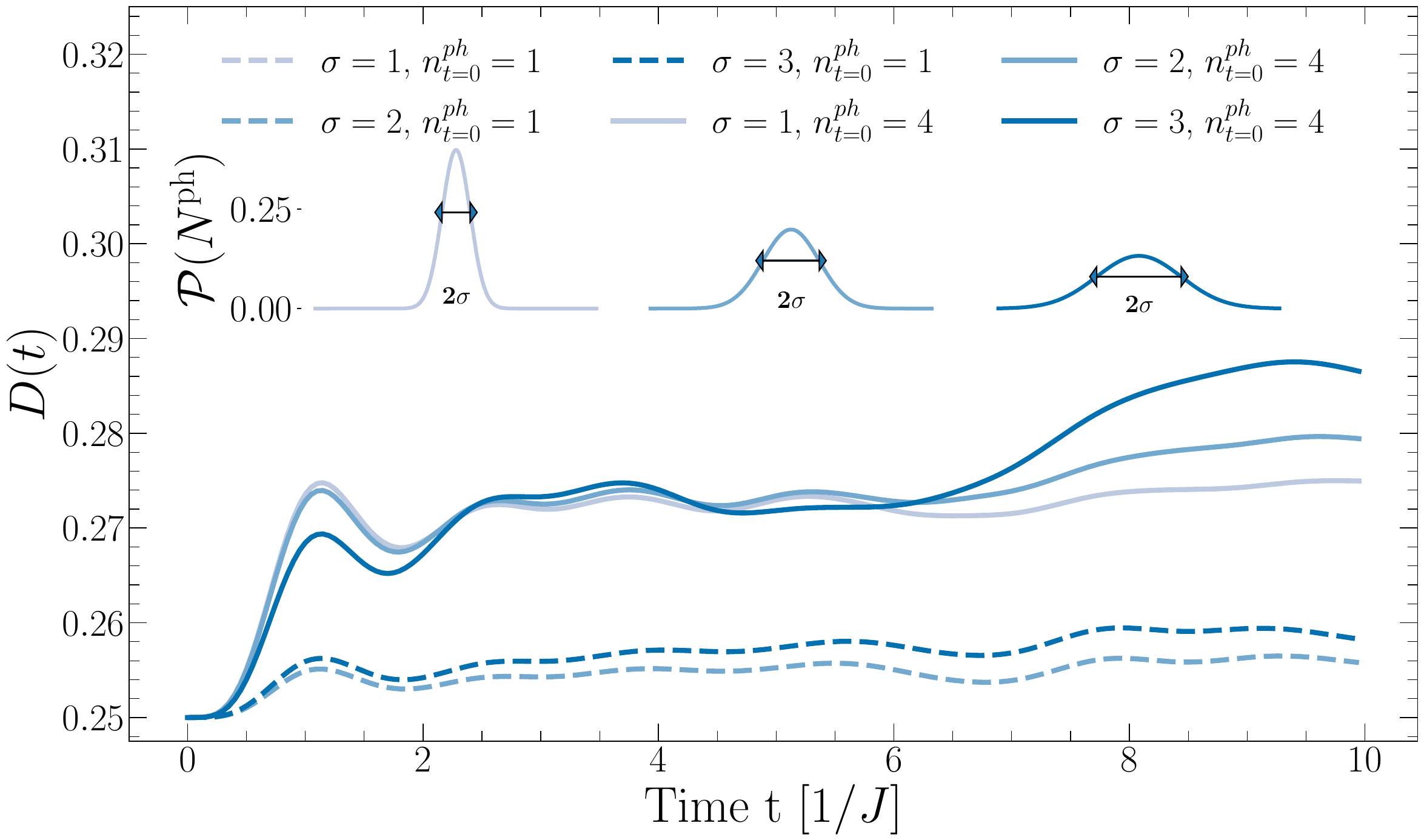}
 }
    \caption
 {
        \label{fig:dynamics}
 Analysis of the effect of the pulse shape on the induced dynamics of charge double occupancy $D(t)$ obtained by constructing the phonon probability distribution $P(N^\mathrm{ph})$ using \gls{PST}. In~\subfigref{fig:squeezed:dynamics} we compare the dynamics following the application of a coherent-exciting pulse $\hat A^\mathrm{c}$ with that following a squeezed-coherent-exciting pulse $\hat A^\mathrm{sc}$, while in~\subfigref{fig:beta:dynamics} we analyze the dynamics for different values of $\beta$ following the application of a $\beta$-pulse $\hat A^\beta$.
 The left and right insets in~\subfigref{fig:squeezed:dynamics} show the onsite local phonon probability distribution function over phonon number states for a coherent excitation (left) and a squeezed-coherent excitation (right), while the central inset shows the total phonon number distribution function for both pulses. 
 (Note that the pump fluence corresponds to  the expected value of  $n^\mathrm{ph}$ given its probability distribution function $P(n^\mathrm{ph})$).
 The dynamics following both pulses appear to be very similar, suggesting that it is controlled by $P(N^\mathrm{ph})$.
 This is further corroborated in~\subfigref{fig:beta:dynamics} where we find that the larger the width of $\mathcal{P}(N^\mathrm{ph})$, $\sigma$, the larger the induced double occupancy.
 We use $L=10$ sites and $4000$ samples in this figure.
 }
\vspace{-6mm}
\end{figure}
\paragraph{Tomographically reconstructed electronic correlations.}
Given the tomographic representation of the wavefunction \cref{Eq:eq2}, we reconstruct the electronic double occupancy $D(t) = \frac{1}{L} \sum_{j=0}^{L-1} \langle \hat n^\nodagger_{j,\uparrow}\hat n^\nodagger_{j,\downarrow} \rangle (t)$ and the staggered magnetic correlation $M(t) = \frac{1}{L} \sum_{r=0}^{L-1} (-1)^{r} \langle\hat s^z_j \hat s^z_{j+r}\rangle (t)$ and show their dynamics in~\cref{fig:coherent:dynamics:both}.
We observe evolution towards a state with a slight increase of the electronic double occupancy accompanied by suppression of the staggered magnetic correlation consistent with the observations of Ref.~\cite{Sous2021}.
Importantly, the tomographically reconstructed data (symbols with error bars) and the exact results (solid lines) are in perfect agreement, indicating that \gls{PST} can accurately reconstruct electronic correlations.
We find convergence to accurate results is possible for less than $2000$ samples of phonon configurations, implying that while the exact electron-phonon wavefunction is exponentially large in the phonon Hilbert space ($40^L$), a reasonably small part of the phononic Hilbert space is all that is needed in order to fully quantify and faithfully reconstruct the electronic correlation dynamics.
\gls{PST} also allows us to decompose $D(t)$ and $M(t)$ into contributions associated with sectors with different total phonon number $N^\mathrm{ph}$, which we study in~\cref{fig:coherent:dynamics:time-resolved:D} and \cref{fig:coherent:dynamics:time-resolved:S}, finding that the dynamics of correlations is associated with a global phonon distribution function $\mathcal{P}(N^\mathrm{ph})$ (insets of~\cref{fig:coherent:dynamics:both}) which has a roughly constant-in-time value of the mean (around 40) and a slightly increasing value of the width. 
Note that the fraction of drawn samples represented as the size of the symbols in ~\cref{fig:coherent:dynamics:time-resolved:D} and \cref{fig:coherent:dynamics:time-resolved:S} are significantly smaller for values of $N^\mathrm{ph}$ smaller than the mean of $\sim 40$, indicating the importance of intermediate- and large\hyp $N^\mathrm{ph}$ configurations.
Notably, we also find that at late times the electronic double occupancy correlates with $N^\mathrm{ph}$, while the staggered magnetic correlation anticorrelates with $N^\mathrm{ph}$.
Combining all these observations, \gls{PST} enables relating the formation of a $k=\pi$ peak in the sample-averaged phonon occupation driven by contributions from large\hyp $N^\mathrm{ph}$ configurations to the evolution of electronic correlations.
\paragraph{Tomographic analysis of pulse shapes.}
The wavefunction decomposition obtained from~\gls{PST} suggests dependence of the dynamics of electronic correlations on the population of intermediate- and high\hyp $N^\mathrm{ph}$ configurations.
This hints at the possibility for controllable enhancement of electronic correlations using tailored optical pulses which can tune the initial state global phonon distribution $\mathcal{P}(N^\mathrm{ph})$, as we demonstrate in the following.
Besides $\hat A^\mathrm{c}$, we consider $\hat A^\mathrm{sc} = \prod_j\hat{d}_j(\alpha) \mathrm e^{ \frac{1}{2} \left( z^* \hat b^2_j - z \hat b_j^{\dagger2} \right ) }$ which generates squeezed-coherent phonon states with squeezing parameter $z$. 
This pulse shape enables control over the ratio between the standard deviation of the position and momentum operators $\Delta x / \Delta p$ at each site by varying $z$, therefore acts as a knob of control for tuning the width of the onsite local phonon distribution created by the pump (c.f. left and right insets of~\cref{fig:squeezed:dynamics}). 
We also design an artificial pulse that depends on a single real parameter $\beta$ which enables explicit control over the width of the global phonon distribution function $\mathcal P(N^\mathrm{ph})$, we dub the $\beta$\hyp pulse $\hat A^\beta = \prod_j \left[\hat b^\dagger_j\right]^{n^\mathrm{ph}}  \left(\frac{\sqrt{1-2\beta^2}}{\sqrt{n^\mathrm{ph}!}}  + \frac{\beta \hat b^\dagger_j}{\sqrt{n^\mathrm{ph}+1!}} + \frac{\beta \hat b^\nodagger_j}{\sqrt{n^\mathrm{ph}-1!}} \right)$. Here, the generated initial global phonon distribution $\mathcal{P}(N^{\mathrm{ph}}_m) = \sum_{i=0}^{L-m-2i \geq 0} \zeta(\beta; i, m, L) / \xi(i, m, L)$ (where $\zeta(\beta; i, m, L) = { \left( (1-2\beta^2)^{\frac{L-m-2i}{2}} \beta^{m+2i} \right )^2 L!}$ and $\xi(i, m, L) = {(L-m-2i)!(m+2i)!} $) is Gaussian with mean $L \cdot n^\mathrm{ph}$ and encompasses the phonon sectors $N^{\mathrm{ph}}_m = Ln+m$, with $m = -L, -L+1, \dots,+L-1,+L$ (see Supporting Information).
In~\cref{fig:squeezed:dynamics} we compare the post\hyp pump dynamics following the application of $\hat A^\mathrm{c}$ with that obtained following the application of $\hat A^\mathrm{sc}$ for pump fluence corresponding to exciting $n^\mathrm{ph}=1,4$ phonons locally on each site.
Notably, the ensuing dynamics for the two pulse shapes are nearly identical.
Examining the global phonon distribution $\mathcal P(N^\mathrm{ph})$ for both types of pulses (middle inset of~\cref{fig:squeezed:dynamics}), we find good agreement within the error bars.
This suggests that manipulating the local onsite phonon distribution created by the pump bears no effect on the electronic dynamics while tuning the global phonon distribution $\mathcal P(N^\mathrm{ph})$ may allow better control at least over short and intermediate timescales.
This is what we find in~\cref{fig:beta:dynamics} where we show the dynamics following the application of various $\beta$ pulses corresponding to global phonon distribution with varied width $\sigma(\beta)$.
Here, we observe a strong correlation between increasing the value of $\sigma$ and generating a larger light-induced double occupancy.
\paragraph{\gls{PST}\hyp based experimental protocol.}

The implementation of the \gls{PST} algorithm in pump-probe experiments proceeds as follows. After exciting specific phonon modes of the material with infrared pump pulses~\cite{Liu2020pump}, one measures the time-dependent evolution of the diffuse scattering intensity $I_s$ as a function of momentum and time delay using time-resolved x-ray or electron diffraction. Knowing the phonon energy and structure factor of the unperturbed material, the initial phonon occupation $n^\mathrm{ph}(\mathbf{k},t_0)$ after photoexcitation is determined via~\cref{eq:n_q_from_TDS}. With $n^\mathrm{ph}(\mathbf{k},t_0)$ and an {\em ab-initio} description of the electron-phonon system, time-dependent electronic observables (energy density, etc.) and correlation functions (density-density, pairing, etc.) that are not directly accessible are computed and decomposed into contributions from different total phonon number excitations using \gls{PST} (Fig. 3). Measuring $n^\mathrm{ph}(\mathbf{k},t>t_0)$ serves as a test for the accuracy of the numerical modeling. Information about electronic observables can then guide the tailoring of optical excitations to enhance desired electronic properties (Fig. 4b), which can be indirectly detected with other experimental probes. For instance, enhanced pairing can be confirmed through optical conductivity measurements. An intriguing application of PST could be the study of electronic pairing after photoexcitation of the B$_{1u}$ infrared-active modes in YBa$_2$Cu$_3$O$_{6.5}$ ($\sim$15 $\mu$m, 83 meV), which is known to induce transient distortions of the crystal lattice via nonlinear phonon coupling \cite{Mankowsky2014nonlinear} and transient superconducting correlations at temperatures far above the equilibrium $T_c$ \cite{Hu2014optically,Liu2020pump}. The phonon occupation could be measured with transient diffuse scattering measurements on bulk samples or freestanding membranes and used to address changes in the Fermi surface~\cite{Mankowsky2014nonlinear} or the pairing interaction.
\paragraph{Conclusion.}\label{sec:conclusion}
We introduced \gls{PST} as a tool to decompose the fully coupled electron-phonon wavefunction into electronic contributions weighted by phonon occupations, which can be sampled efficiently to reconstruct electronic observables.
We demonstrate the usefulness of \gls{PST} by simulating the experimental protocol for probing the phononic response of a metal whose phonons are excited by a pump pulse at an initial time, obtaining sample-averaged phonon occupations that can be matched to experiments and accurately reconstructing the dynamics of electronic correlations.
\gls{PST} thus enables building a connection between phononic observables measured using direct phonon probes and the underlying electronic correlations, which is relevant to experiments based on diffuse x\hyp ray and electron scattering.

We also show how to use \gls{PST} to analyze and design different pulse shapes tailored to enhance or suppress certain electronic correlations in the post-pump dynamics.
\gls{PST} may thus aid in the experimental control of light\hyp matter interaction and the analysis of experimental data.
\FloatBarrier
\paragraph{Acknowledgements}
M.~Mor., U.~S., and S.~P. acknowledge support by the Deutsche Forschungsgemeinschaft (DFG, German Research Foundation) under Germany’s Excellence Strategy-426 EXC-2111-390814868 and by Grant No. INST 86/1885-1 FUGG of the German Research Foundation (DFG).
M. Mit. acknowledges support by the U.S. Department of Energy, Office of Basic Energy Sciences, Early Career Award Program, under Award No. DE-SC0022883.

\bibliography{Literature}
\end{document}


\author{Mattia Moroder}
\affiliation{Department of Physics, Arnold Sommerfeld Center for Theoretical Physics (ASC), Munich Center for Quantum Science and Technology (MCQST), Ludwig-Maximilians-Universit\"{a}t M\"{u}nchen, 80333 M\"{u}nchen, Germany}
\author{Matteo Mitrano}
\affiliation{Department of Physics, Harvard University, Cambridge, Massachusetts 02138, USA}
\author{Ulrich Schollw\"ock}
\affiliation{Department of Physics, Arnold Sommerfeld Center for Theoretical Physics (ASC), Munich Center for Quantum Science and Technology (MCQST), Ludwig-Maximilians-Universit\"{a}t M\"{u}nchen, 80333 M\"{u}nchen, Germany}

\author{Sebastian Paeckel}\thanks{Author to whom correspondence should be address; sebastian.paeckel@physik.uni-muenchen.de. These authors contributed equally and are listed alphabetically.}
\affiliation{Department of Physics, Arnold Sommerfeld Center for Theoretical Physics (ASC), Munich Center for Quantum Science and Technology (MCQST), Ludwig-Maximilians-Universit\"{a}t M\"{u}nchen, 80333 M\"{u}nchen, Germany}

\author{John Sous}\thanks{Author to whom correspondence should be address; john.sous@yale.edu.  These authors contributed equally and are listed alphabetically.}
\affiliation{Department of Chemistry and Biochemistry, University of California San Diego, La Jolla, California 92093, USA}
\affiliation{Department of Applied Physics and the Energy Sciences Institute, Yale University, New Haven, Connecticut 06511, USA}
\def\thetitle{Supporting Information \\ for \\ ``Phonon state tomography of electron correlation dynamics in optically excited solids''} 

\title{\thetitle}
\maketitle
%
%
%
\section{\Gls{PST}\label{app:sec:pst:method}}
We consider a lattice model with electronic and phononic degrees of freedom spanning the Hilbert space $\mathcal H = \mathcal H^\mathrm{el} \otimes \mathcal H^\mathrm{ph}$.
$\mathcal H^\mathrm{ph} = \bigotimes_{j=0}^{L^\mathrm{ph}-1} \mathcal H^\mathrm{ph}_{j}$  has a tensor\hyp product structure, with $j$ labeling the different $L^\mathrm{ph}$  spatial phonon sites, which, for the model we consider, is equivalent to the $L$ lattice sites, henceforth we drop the $\mathrm{ph}$ superscript in $L^\mathrm{ph}$.
In general, there is no constraint on the type of phonons (whether Holstein~\cite{HOLSTEIN1959325} or Peierls~\cite{Marchand2010,Sous2018}; dispersive or not; etc.) we consider.
Introducing the vector $\mathbf n^\mathrm{ph} = \left(n^\mathrm{ph}_0, n^\mathrm{ph}_1, \ldots, n^\mathrm{ph}_{L-1} \right)$ as a label of any   possible configuration of the global phonon subsystem, any state $\ket{\psi} \in \mathcal H$ in the full Hilbert space $\mathcal H$ can be expanded as
\begin{equation}
    \ket{\psi}
 =
    \sum_{\{ \mathbf n^\mathrm{ph} \}} \varphi(\mathbf n^\mathrm{ph}) \ket{\psi^\mathrm{el}(\mathbf n^\mathrm{ph})} \otimes \ket{\mathbf n^\mathrm{ph}} \;,
\end{equation}
where $\ket{\psi^\mathrm{el}(\mathbf n^\mathrm{ph})} \in \mathcal H^\mathrm{el}$ is a normalized electronic wavefunction, and the sum is over the set of all possible  $\mathbf n^\mathrm{ph}$ configurations of the phononic subsystem. 
In this representation, $\lvert \varphi(\mathbf n^\mathrm{ph}) \rvert^2 \equiv p(\mathbf n^\mathrm{ph})$ denotes the probability to measure the phonon configuration $\mathbf n^\mathrm{ph}$, so we can obtain the global phonon distribution function $\mathcal P(N^\mathrm{ph})$ by summing over the  configurations $\mathbf n^\mathrm{ph} \in \Omega_{N^\mathrm{ph}}$ in the global phononic subspace $\Omega_{N^\mathrm{ph}}$ with a total number of phonons $N^\mathrm{ph}$.
To illustrate the practical relevance of $p(\mathbf n^\mathrm{ph})$ and $\mathcal P(N^\mathrm{ph})$, let us consider an electronic observable $\hat O$ acting only on the electronic subspace $\mathcal H^\mathrm{el}$.
We can write the expectation value as
\begin{align}
    \braket{\hat O}
    &=
    \sum_{\left\{\mathbf n^\mathrm{ph}\right\}} p(\mathbf n^\mathrm{ph}) O^\mathrm{el}(\mathbf n^\mathrm{ph}) \label{eq:pst:exact0}\\
    &=
    \sum_{N^\mathrm{ph}} \mathcal P(N^\mathrm{ph}) \sum_{\mathbf n^\mathrm{ph}\in \Omega_{N^\mathrm{ph}}} \frac{p(\mathbf n^\mathrm{ph})}{\mathcal P(N^\mathrm{ph})} O^\mathrm{el}(\mathbf n^\mathrm{ph}) \; , \label{eq:pst:exact}
\end{align}
with $O^\mathrm{el}(\mathbf n^\mathrm{ph}) = \braket{\psi^\mathrm{el}(\mathbf n^\mathrm{ph}) | \hat O | \psi^\mathrm{el}(\mathbf n^\mathrm{ph})}$.
In the last line, we recognize the conditional probability $\frac{p(\mathbf n^\mathrm{ph})}{\mathcal P(N^\mathrm{ph})}$ of measuring a phonon configuration $\mathbf n^\mathrm{ph}$ conditioned upon having a total of  $N^\mathrm{ph}$ phonons in the system.
While an exact evaluation of $\mathcal P(N^\mathrm{ph})$ is not possible in practice due to the exponential scaling associated with computing its value, here we propose to approximate~\cref{eq:pst:exact0} by an average over a set of samples $\Omega$ (with sample size $|\Omega|$) which are drawn according to the distribution $p(\mathbf n^\mathrm{ph})$
\begin{equation}
    \braket{\hat O} \approx \frac{1}{\lvert \Omega \rvert } \sum_{\mathbf n^{ph}\in\Omega} p(\mathbf n^\mathrm{ph}) O^\mathrm{el}(\mathbf n^\mathrm{ph}) \; . \label{eq:pst:estimation}
\end{equation}
We extend the \acrfull{PSS} introduced for~\glspl{MPS} in the context of ultracold-atom quantum simulators~\cite{Ferris2012,Bohrdt2019} to efficiently draw samples from $p(\mathbf n^\mathrm{ph})$.
The idea is to decompose the joint probability distribution function of a phononic configuration into a product of conditional probabilities
\begin{equation}
 p(\mathbf n^\mathrm{ph})
 =
 p(n^\mathrm{ph}_0)
 p(n^\mathrm{ph}_1 \vert n^\mathrm{ph}_0)
    \cdots
 p(n^\mathrm{ph}_{L-1} \vert n^\mathrm{ph}_{L-2},\ldots,n^\mathrm{ph}_0) \; .
\end{equation}
The local probabilities $p(n^\mathrm{ph}_j \lvert n^\mathrm{ph}_{j-1}, \ldots, n^\mathrm{ph}_0)$ can be computed efficiently in a recursive manner.
For a given site $j$, a measurement at a given time $t$ is simulated by drawing a sample from the local distribution (for example, at initial time $t=0$ this is the distribution imprinted by the pump pulse, which is Poisson in case of the coherent state) and choosing a measurement outcome $n^\mathrm{ph}_j$ according to $p(n^\mathrm{ph}_j \lvert n^\mathrm{ph}_{j-1}, \ldots, n^\mathrm{ph}_0)$ obtained from the diagonal elements of the \gls{1phRDM} $\hat{\rho}^\mathrm{ph}_j$.
Then, the state is projected onto the corresponding local phonon subspace by applying the projection operator $\ket{n^\mathrm{ph}_j}\bra{n^\mathrm{ph}_j}$.  This automatically satisfies the condition associated with the conditional probability $p(n^\mathrm{ph}_{j+1} \lvert n^\mathrm{ph}_{j}, \ldots, n^\mathrm{ph}_0)$ for the next site $j+1$.  This is iteratively applied from site $0$ to site $L-1$.

We represent the combined electron\hyp phonon system as an~\gls{MPS}~\cite{white_1992,white_1993,schollwoeck_2005,schollwoeck_2011} and exploit the~\gls{PP}~\cite{Koehler2021,Stolpp2021,PhysRevB.107.214310}, which associates a fictitious auxiliary boson with the physical bosons, to efficiently encode the large local Hilbert space of the phononic degrees of freedom.
For a naive computation of the conditional probabilities $p(n^\mathrm{ph}_j \lvert n^\mathrm{ph}_{j-1}, \ldots, n^\mathrm{ph}_0)$, the electronic subsystem has to be traced out to obtain the \gls{1phRDM} $\hat \rho^\mathrm{ph}_{j}$, which renders the computations inefficient and impractical.
However, using~\gls{PP} we can exploit the gauge\hyp fixing property of its mapping such that the explicit computation of each $\hat \rho^\mathrm{ph}_{j}$ can be avoided and, instead, the desirable diagonal elements $\braket{n^\mathrm{ph}_j | \hat \rho^\mathrm{ph}_{j} | n^\mathrm{ph}_j }$ can be evaluated directly with negligible computational costs.
This is achieved via a \gls{SVD} of the block associated with the physical phononic site $j$, exploiting the following relation
\begin{equation}
    \rho_{n_j^{\text{ph}}, n_j^{\text{ph}}} = \sum_{\lambda^{\text{ph}}_j} \abs*{ S_{\lambda^{\text{ph}}_j} }^2,
\label{eq:pp:rdm:svd}    
\end{equation}
which relates the $n_j^{\text{ph}}$th diagonal entry of the \gls{1phRDM} to the singular values $S_{\lambda^{\text{ph}}_j}$ of the block corresponding to $n_j^{\text{ph}}$~\cite{Koehler2021}.
Note that the cost of this operation is independent of the system size $L$.

\begin{figure}[tb] 
    \centering
    \includegraphics[width=\textwidth]{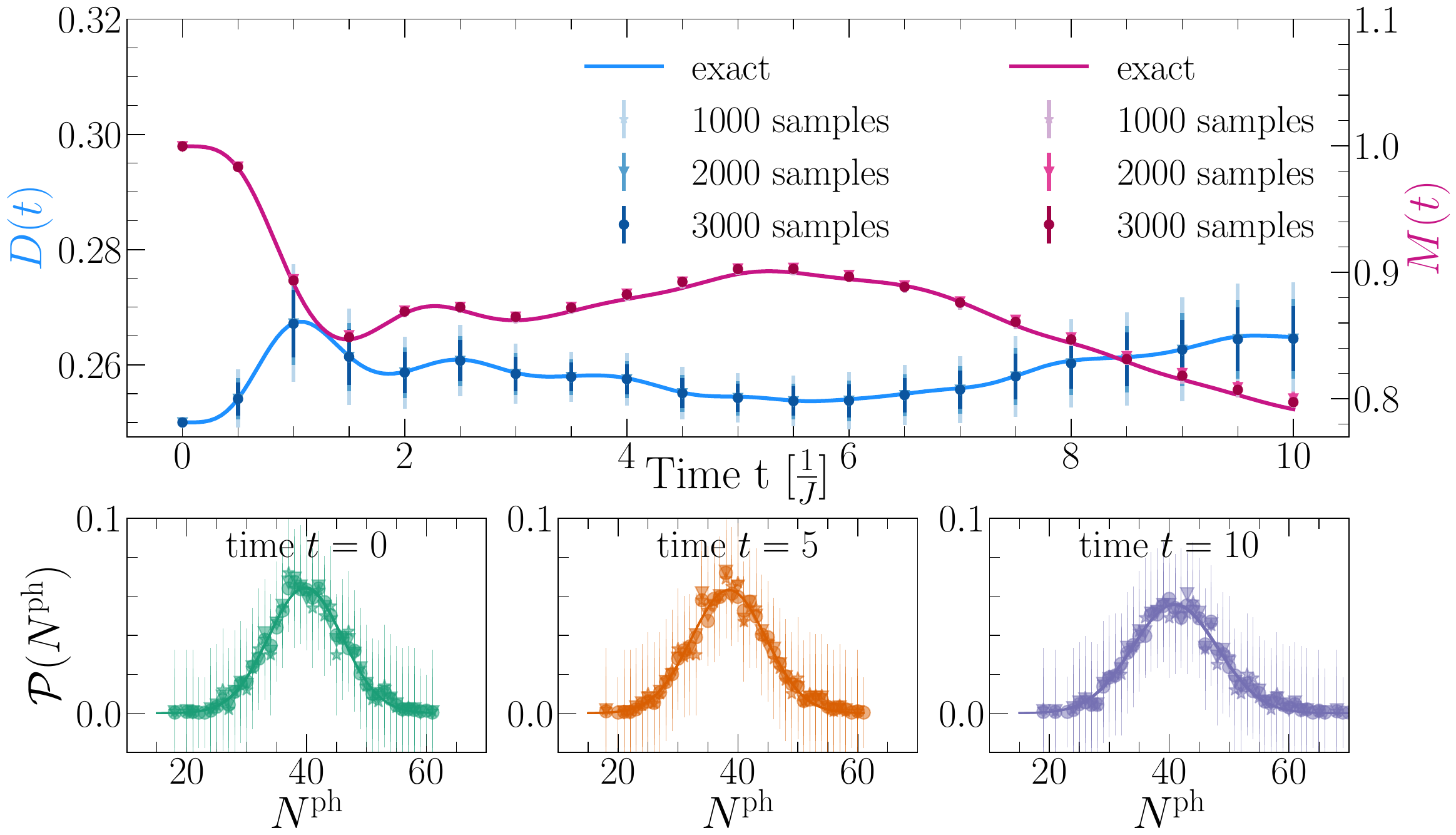}
    \caption
 {
    \label{fig:trajectories_convergence}
 Convergence of \gls{PST} with respect to the number of computed samples for the model of a metal whose phonons are photoexcited at initial time,  studied in the main text, see also~\cref{app:sec:numerical-conv}.
 Upper panel: electronic double occupancy $D(t)$ defined in~\cref{eq:double:occ} (left axis) and of the staggered magnetic correlation $M(t)$ defined in~\cref{eq:staggered:mag} (right axis).
 The comparison between the~\gls{PST} results (symbols with error bars) and the exact observables (full lines) shows that already $\lvert \Omega \rvert=1000$ samples are sufficient to obtain converged results.
 Lower panel: global phonon distribution function $\mathcal{P}(N^\mathrm{ph})$ at times $t = 0, \, 5, \, 10$. 
 }
    \end{figure}
Given an arbitrary pure state $\ket{\psi}$ in the coupled electron-phonon system, the \gls{PST} algorithm for the computation of a single sample (starting from the leftmost site $j=0$) can be summarized as follows:
\begin{enumerate}[topsep=0pt,nosep]
    \item On site $j$, compute the \gls{1phRDM} $\hat \rho^\mathrm{ph}_{j} \equiv \tr_{\text{el}}{ \tr_{\text{ph}_{\setminus j}}  \ket{\psi}\bra{\psi} }$ and identify its diagonal entries with $p(n_{j}^\text{ph})$. When adopting the~\gls{PP} representation, the diagonal elements of $\hat \rho^\mathrm{ph}_{j}$ can be efficiently computed according to the relation~\cref{eq:pp:rdm:svd}.
    \item Draw the projector $\ket{n_{j}^\text{ph}} \bra{n_{j}^\text{ph}}$ according to $p(n_j^\text{ph})$ and apply it to $\ket{\psi}$.
    \item Increment $j$ by $1$ and repeat steps $1)$ and $2)$ until the last site has been reached.
\end{enumerate}
Samples constructed in this way are grouped according to the total phonon number in the global phonon subspace $N^\text{ph} = \sum_{j=0}^{L-1} n_j^\text{ph}$. 
We use a sufficient number of samples is used to ensure convergence to the true distribution.
Then, an arbitrary electronic observable $\braket{\hat O}$ can be reconstructed using \cref{eq:pst:estimation}, and, by analyzing the contributions from the various total phonon sectors $N^\text{ph}$, we obtain the decomposition shown in panels b and c of Fig.~3 in the main text.
Analogously, we estimate $\mathcal P(N^\mathrm{ph})$ as the ratio of the number of drawn configurations  corresponding to a given $N^\mathrm{ph}$ to the total number of samples drawn $\lvert \Omega \rvert$.

Note that PST enables a  decomposition of electronic observables in terms of contributions associated with any given phonon property.  For example, we can use PST to decompose electronic observables in terms of contributions associated with  different phonon momenta $\mathbf{k}^\mathrm{ph}$ by transforming the sampled phonon configurations to momentum space and grouping the corresponding electronic wavefunction according to  $\mathbf{k}^\mathrm{ph}$.

In the main text, we apply~\gls{PST} to study a model of a metal whose phonons are photoexcited at initial time (see~\cref{app:sec:numerical-conv} for details).
In \cref{fig:trajectories_convergence}, we analyze the convergence of \gls{PST} with respect to the number of drawn samples for the electronic double occupancy in the model
\begin{equation}
    D(t) = \frac{1}{L} \sum_{j=0}^{L-1} \langle \hat n^\nodagger_{j,\uparrow}\hat n^\nodagger_{j,\downarrow} \rangle (t) \;,
    \label{eq:double:occ}
\end{equation}
and the staggered magnetic correlation
\begin{equation}
    M(t) = \frac{1}{L} \sum_{r=0}^{L-1} (-1)^{r} \langle\hat s^z_j \hat s^z_{j+r}\rangle (t) \;.
    \label{eq:staggered:mag}
\end{equation}
We find that for both observables (see also Fig. 3 in the main text) $1000$ samples suffice to yield converged results despite that the phonon Hilbert space dimension is $\mathrm{dim} \left(\mathcal H^\mathrm{ph} \right ) = 40^{20}$, highlighting the utility of the method in extracting the most significant configurations in the Hilbert space.

\begin{figure}[tb] 
    \centering
    \includegraphics[width=0.65\textwidth]{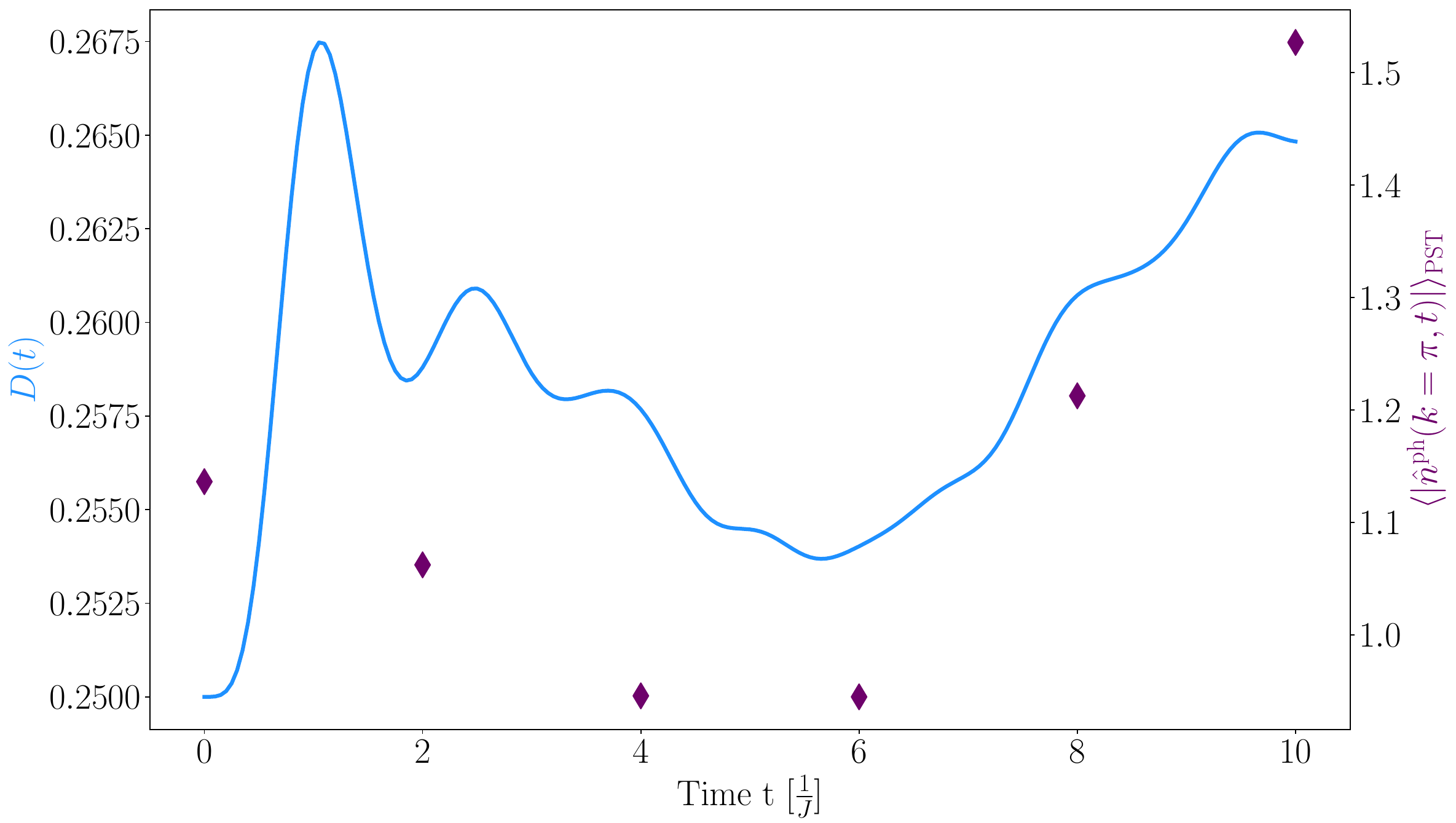}
    \caption
 {
    \label{fig:double_occ_vs_n_ph_pi} 
Time evolution of the double occupancy and sample-averaged phonon occupations. After a short time delay, electrons and phonons become ``time-locked" and the electronic double occupancy $D(t)$ (left axis) tracks over time the sample-averaged momentum-dependent phonon occupation obtained with~{PST} (right axis).
 }
    \end{figure}

Using \gls{PST} one can also simulate the momentum-dependent occupation of a phonon mode, which can be experimentally obtained using x-ray or electron scattering, as shown in Fig. 2 in the main text.
This is achieved by computing the Fourier transform of each sampled phonon configuration $\mathbf{n}^\mathrm{ph}$, taking the absolute value and then averaging over all samples.
We label the sample-averaged, momentum-dependent phonon occupation $\langle \mathbf |{\hat n^\mathrm{ph}(k)}|\rangle_\mathrm{PST}$.
This experimental quantity can serve as an input for the~\gls{PST} algorithm, as it specifies the initial state of the phonon subsystem.
Moreover, as electrons and phonons become correlated during the dynamics, $\langle \mathbf |{\hat n^\mathrm{ph}(k=\pi,t)}|\rangle_\mathrm{PST}$ encodes important information about the electronic subsystem. For instance,in~\cref{fig:double_occ_vs_n_ph_pi}, we find that the phonon occupation at momentum $k=\pi$ tracks the evolution with time of the electron double occupancy, indicating that after a very short time delay electrons and phonons become ``time locked''.

\section{Numerical simulation of time evolution}\label{app:sec:numerical-conv}

To simulate the dynamics of the optically excited metal, we combine the efficient \gls{MPS} representation for the phononic degrees of freedom in \gls{PP} \cite{Koehler2021} with state-of-the-art time-evolution algorithms.
~\gls{PP} implements enlarged bosonic sites (composed of the physical phonons supplemented with fictitious bosons) with $\mathrm{U(1)}$ symmetry (which is otherwise broken by the interaction term of the physical bosons (c.f. Eq. (3) in the main text)).
The block structure of the $\mathrm{U(1)}$\hyp symmetric site tensors greatly simplifies~\gls{MPS} calculations~\cite{Singh2011} and enables parallelization over different $\mathrm{U(1)}$ sectors.
We further truncate the local dimension of the physical phonons, $d^{\mathrm{ph}}$ during the time evolution in a manner similar to how the bond dimension $m$ is truncated in standard~\gls{DMRG} calculations~\cite{Koehler2021}. 
The growth of $d^{\mathrm{ph}}$ and $m$ in the time evolution is controlled by the discarded weight $\delta$, which is  the sum of the squared discarded singular values for each site tensor~\cite{schollwoeck_2011}.
For all the results presented here and in the main text, we use $\delta = 10^{-10}$.  We use a sufficiently large bond dimension $\sim 3000$ (see below) to ensure that our simulations do not saturate $\delta$ and thus achieve convergence.

A powerful method for time\hyp evolving \gls{MPS} with large local Hilbert space dimensions is the \gls{LSE-TDVP} \cite{GSE_with_LSE_arxiv,grundner2023cooperpaired}.
It is based on a~\gls{1TDVP} combined with a local expansion of the basis which results in growth of the bond dimension in the time evolution~\cite{PhysRevB.91.155115}, enabling a more efficient encoding of the spread of correlations in the course of the dynamics.
A drawback of this method is that it is prone to errors when applied to states with a very small bond dimension.
Thus, in practice, we find it helpful to use a slower but more robust variant of the algorithm, the \gls{GSE-TDVP}~\cite{Yang_2020prb}, which is based on \textit{global} basis updates, for the first few timesteps and then switch to \gls{LSE-TDVP}. 

As a first test for our implementation, we reproduce the dynamics of the electronic, phononic, and interaction energy densities simulated in Ref.~\cite{Sous2021}.
\begin{figure}[tb] 
    \centering
    \includegraphics[width=0.75\textwidth]{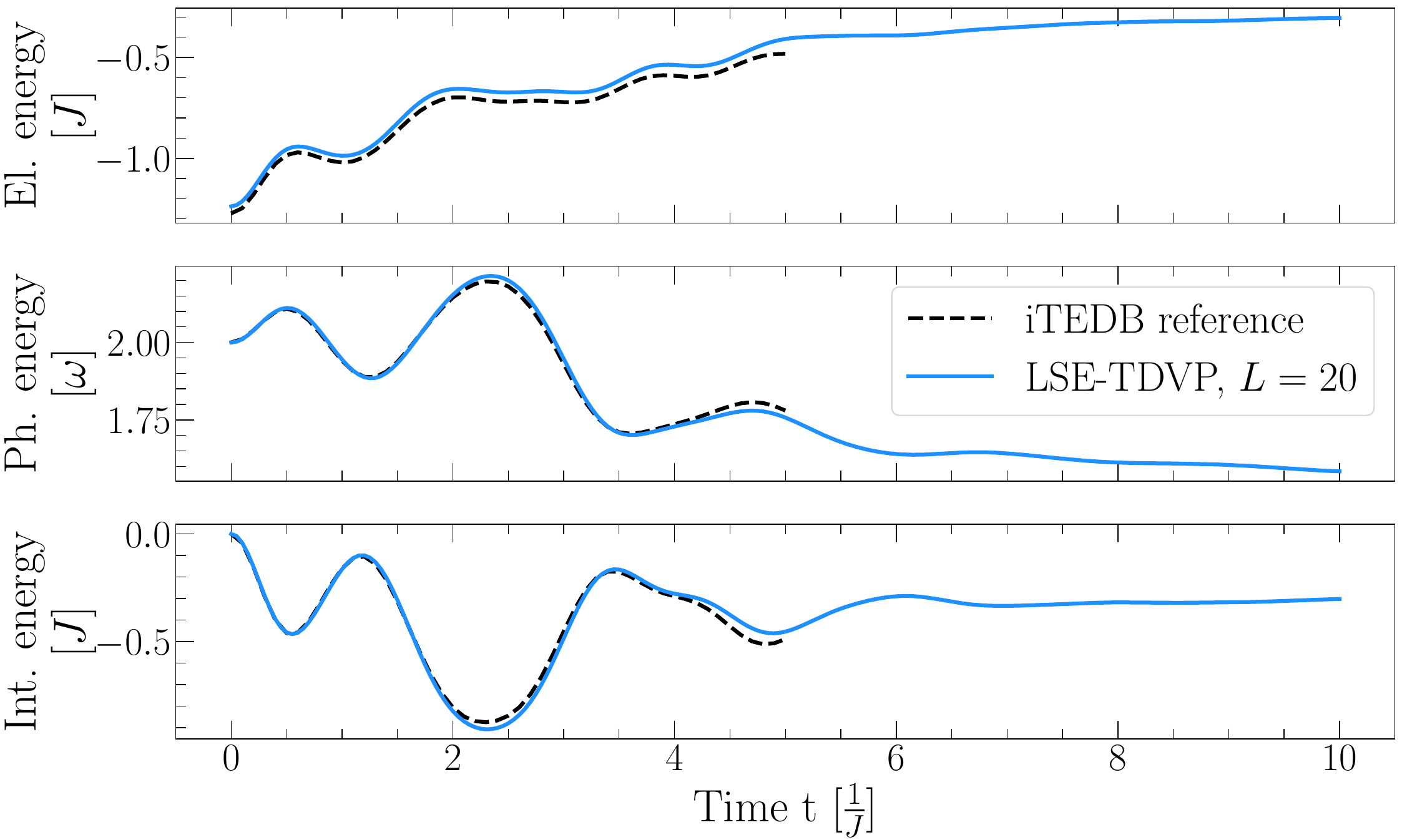}
    \caption
 {
    \label{fig:iTEDB_energy_benchmark} 
 Comparison with the results from Ref.~\cite{Sous2021}.
 Our \gls{LSE-TDVP} results for the electron (upper panel), phonon (middle panel), and interaction (lower panel) energy densities are shown in the solid blue lines, while the \gls{iTEBD} results from \cite{Sous2021}  are shown as dashed black lines. 
 The model parameters used here are $g=0.25\,J$, $\omega = \pi/2\,J$, $\alpha=\sqrt{2}$.
 For our \gls{LSE-TDVP} calculation we use $L=20$ sites, a local Hilbert space dimension for the phonons of $d^{\mathrm{ph}}=12$, a timestep of $\delta t = 0.05\, 1/J$, a bond dimension of $m=2000$ and a discarded weight $\delta = 10^{-10}$.
 Overall, our finite-system calculations with $L=20$ sites reproduce the \gls{iTEBD} results very well.
 }
    \end{figure}
\cref{fig:iTEDB_energy_benchmark} shows that the reference \acrfull{iTEBD} results of ~Ref.~\cite{Sous2021} are in excellent agreement with our results obtained with \gls{LSE-TDVP} for a finite system with $L=20$ sites. 
(A small offset between the electronic energies (upper panel) at time $t=0$ can be noted. 
This is due to the fact that in our \gls{LSE-TDVP} calculation we consider \gls{OBC}, while in \cite{Sous2021} \gls{PBC} were used. (We have checked that this discrepancy disappears when we use \gls{PBC}.)) 
We remark that the combination of \gls{LSE-TDVP} with \gls{PP} allows us to double the simulation times and to consider more intense optical pulses (see Fig. 4 in the main text).

\begin{figure}[tb] 
    \centering
    \includegraphics[width=0.75\textwidth]{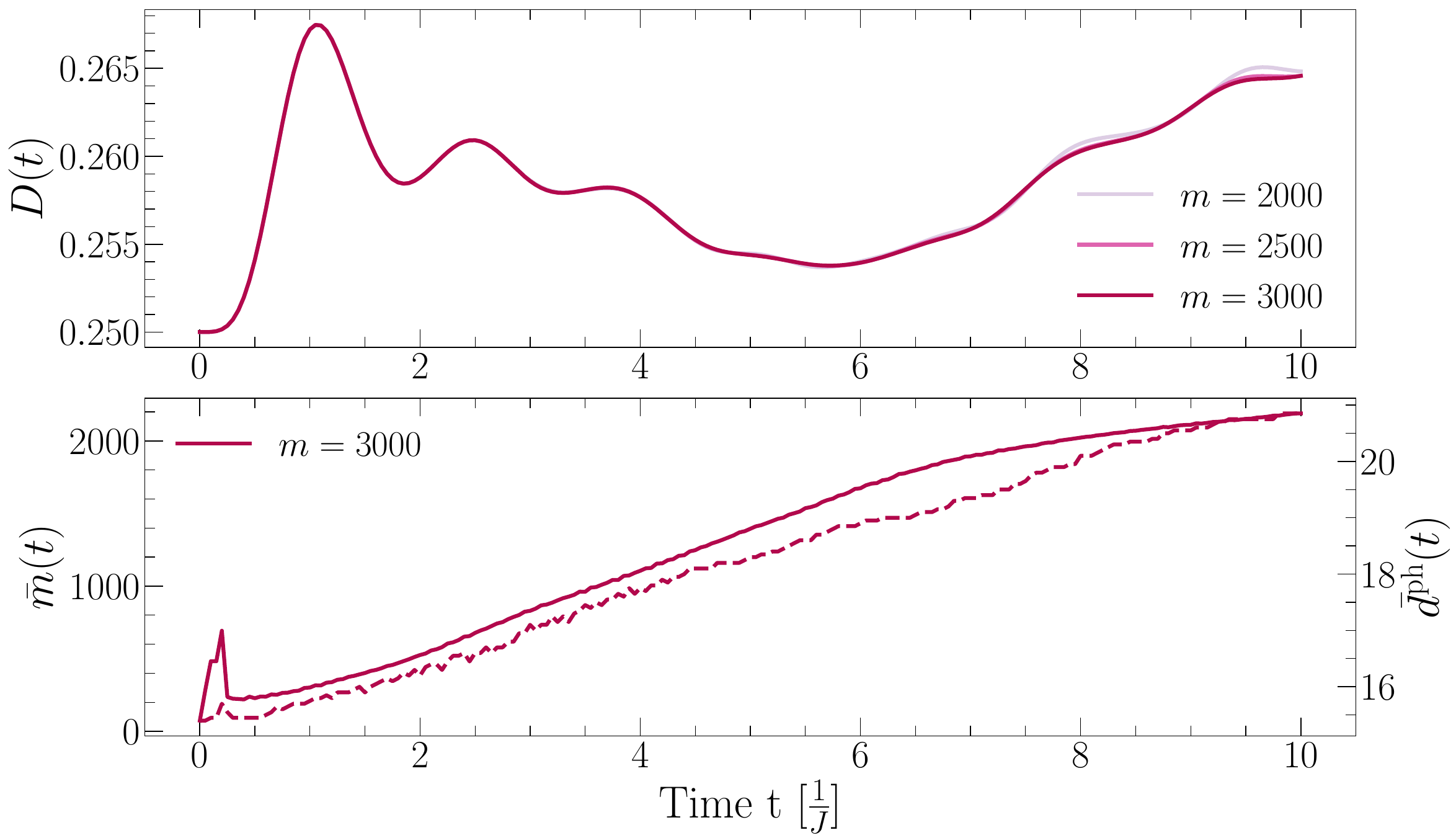}
    \caption
 {
    \label{fig:bdim_physdim} 
 Convergence of the electronic double occupancy $D(t)$ with respect to the bond dimension. Upper panel: $D(t)$ for the same parameters as in Fig. 3 in the main text.
 Increasing the maximally allowed bond dimension from $m=2000$ to $m=3000$ does not change the results.
 Lower panel: growth of the site-averaged bond dimension $\bar{m}$ (left axis, full line) and site-averaged phonon physical dimension $\bar{d}^{\mathrm{ph}}$ (right axis, dashed line) during the time evolution.
 The small spikes in $\bar{m}$, typical for the \gls{GSE-TDVP} method \cite{Yang_2020prb}, are absent after the first 4 timesteps when we switch to \gls{LSE-TDVP}~\cite{GSE_with_LSE_arxiv}. 
 }
    \end{figure}

Next, we study the electronic double occupancy for the same parameters as those of Fig. 3 of the main text as a function of the bond dimension $m$.
\cref{fig:bdim_physdim} shows that our calculations are well converged (upper panel), and the maximal average bond dimension and phonon physical dimension do not saturate within the simulation time (lower panel).

\begin{figure}[tb] 
    \centering
    \includegraphics[width=0.75\textwidth]{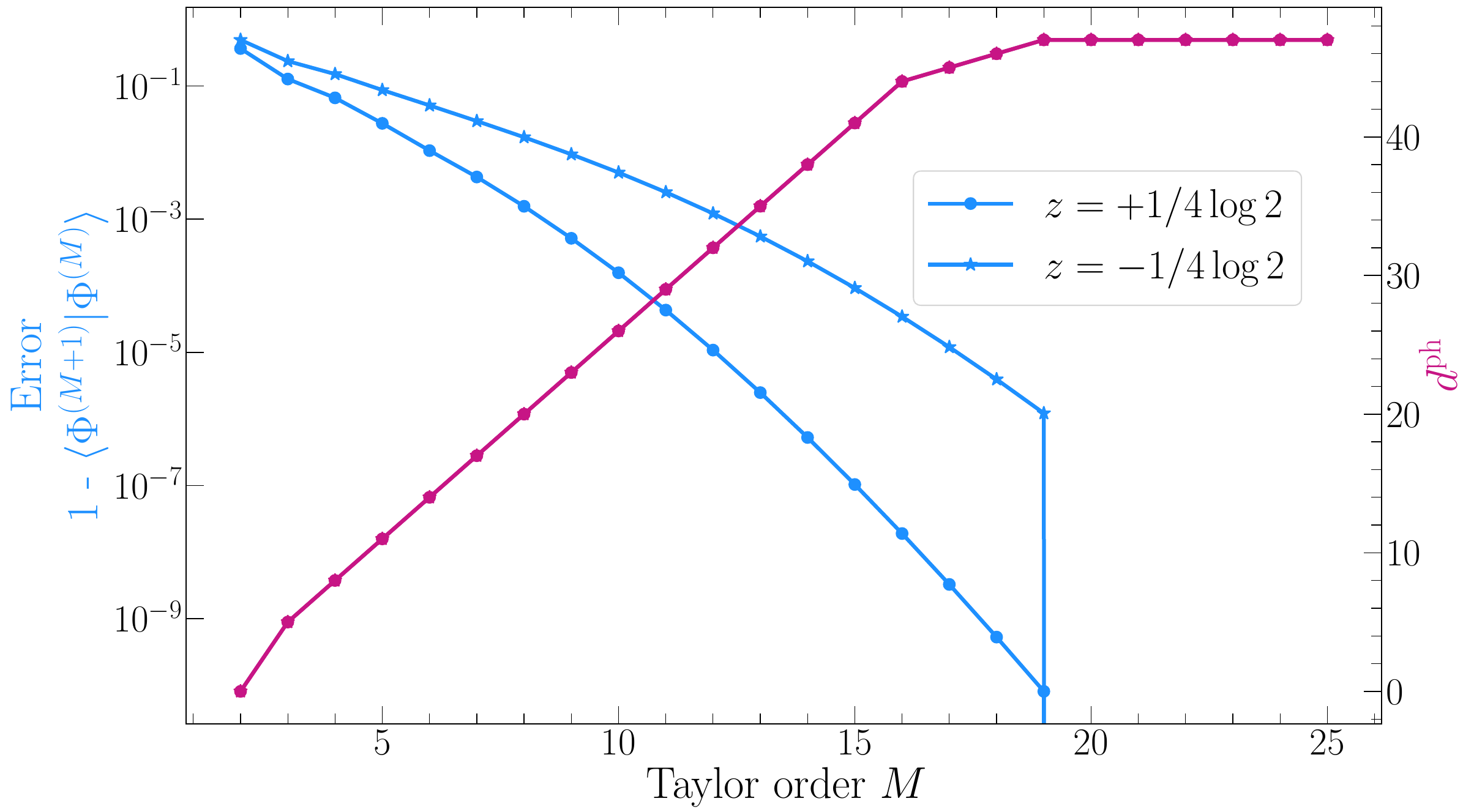}
    \caption
 {
    \label{fig:squeezed_coherent_convergence} 
 Convergence of the Taylor-expanded squeezed-coherent state. 
 Left axis: error as a function of the Taylor order $M$ for $z= + 1/4 \log{2}$ (corresponding to an initial-time value of $\bar{n}^{\mathrm{ph}}=1$) and $z= - 1/4 \log{2}$ (corresponding to an initial-time value of $\bar{n}^{\mathrm{ph}}=4$).
 Right axis: local phonon dimension $d^{\mathrm{ph}}$ as a function of $M$.
 }
    \end{figure}

To generate the initial squeezed-coherent phonon states $\ket{z, \alpha}_j \equiv \hat{D}_j(\alpha) \hat{S}_j(z) \ket{\mathrm{vac}}$, where $\ket{\mathrm{vac}}$ is the vacuum of the physical phonons (c.f. Fig. 4 in the main text), we use a  Taylor-series expansion to order $M$ for the displacement and squeezing operators
\begin{equation}
    \begin{split}
        \hat{D}^{(M)}_{\mathrm{taylor}} (\alpha) & \equiv \sum_{m=0}^M \frac{1}{m!} \left( \alpha \hat{b}^\dagger - \alpha^* \hat{b}^\nodagger  \right)^m  \\ 
        \hat{S}^{(M)}_{\mathrm{taylor}}(z) &\equiv \sum_{m=0}^M \frac{1}{m!} \left( \frac{1}{2} \left( z^* \hat{b}^2 - z \hat{b}^{\dagger 2} \right ) \right)^m . 
    \end{split}
    \label{app:eq:displacement:squeeze:op:taylor}
\end{equation}
We drop the site index $j$ since we apply these operators coherently on every site at initial time.
To check the convergence of the Taylor expansion up to  order $M$, we compute the overlap between $\ket{{\tilde\Phi}^{(M)}} \equiv\hat{D}^{(M)}_{\mathrm{taylor}} (\alpha) \hat{S}^{(M)}_{\mathrm{taylor}}(z)\ket{\mathrm{vac}} $ and $\ket{{\tilde\Phi}^{(M+1)}} \equiv\hat{D}^{(M+1)}_{\mathrm{taylor}} (\alpha) \hat{S}^{(M+1)}_{\mathrm{taylor}}(z)\ket{\mathrm{vac}} $.
\cref{fig:squeezed_coherent_convergence} shows that, for both $z=\pm 1/4 \log{2}$, the error decreases steadily when increasing $M$ and becomes zero for $M=19$. Moreover, the local phonon Hilbert space dimension (right axis) is found to saturate at $d^{\mathrm{ph}} = 47$.

\begin{figure}[tb] 
    \centering
    \includegraphics[width=0.75\textwidth]{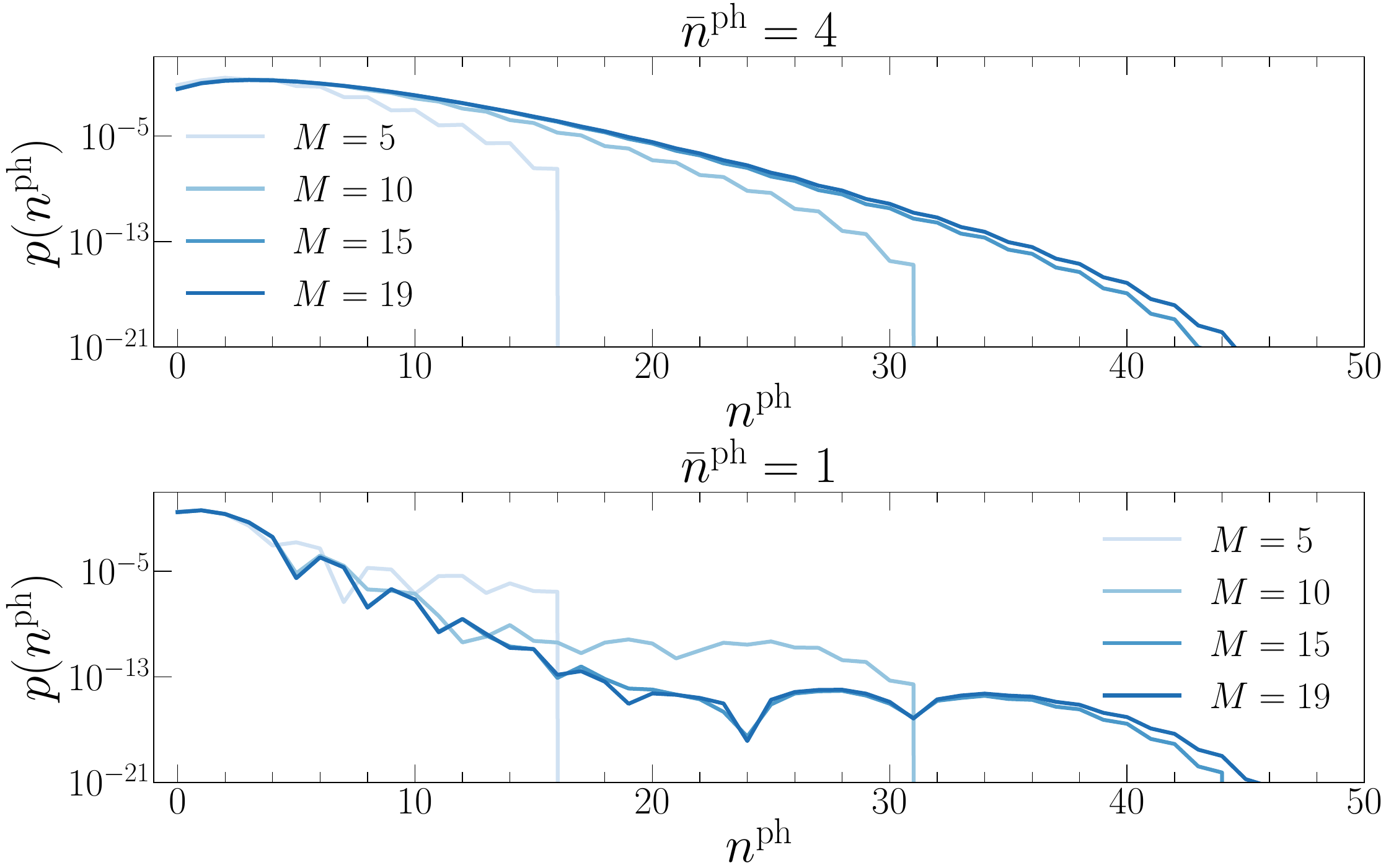}
    \caption
 {
    \label{fig:rdm_squeezed_coh_taylor} 
 Convergence of the local phonon probability distribution for the Taylor-expanded squeezed-coherent state. 
 The upper panel shows $p\left( n^{\mathrm{ph}} \right)$ for the case where $\bar{n}^{\mathrm{ph}}=4$ are excited at initial time and the lower panel for $\bar{n}^{\mathrm{ph}}=1$ as a function of the Taylor order $M$.
 }
    \end{figure}

Finally, we compute the local phonon probability distribution $p\left( n^{\mathrm{ph}} \right)$ (which corresponds to the diagonal part of the  \gls{1phRDM}) for the two squeezed\hyp coherent states considered in  \cref{fig:squeezed_coherent_convergence} as a function of the Taylor order $M$. 
\cref{fig:rdm_squeezed_coh_taylor} shows that when an average of $4$ phonons per site are excited at initial time by the pump (upper panel), $p\left(n^{\mathrm{ph}} \right)$ is a smooth function and converges very well in the Taylor order.
For $\bar{n}^{\mathrm{ph}}=1$ (lower panel) we instead see that $p\left(n^{\mathrm{ph}} \right)$ decays much faster and displays a jagged pattern.
Since these calculations are converged with respect to the Hilbert space dimension, these artifacts are probably caused by numerical instabilities which implied that $p\left(n^{\mathrm{ph}} \right)$ is accurate only up to $\sim 10^{-6}$.
\section{$\beta$-pulse}\label{app:sec:beta-pulse}
To systematically investigate the impact of the width $\sigma$ of the global phonon distribution function $\mathcal{P}(N^\mathrm{ph})$ on the electron dynamics, we seek to find a state simple enough to allow an exact analytical treatment.
We make an ansatz for the many-body phonon state
\begin{equation}
    \ket{\psi} = \bigotimes_{j=0}^{L-1} \left (\frac{ \alpha \ket{n}_j + \beta \ket{n+1}_j + \beta \ket{n-1}_j }{\sqrt{\alpha^2 + \beta^2}} \right ) \; ,
 \label{eq:beta_state}
\end{equation}
where $\alpha$ and $\beta$ are real constants satisfying $\alpha^2 + 2\beta^2 = 1$ and $\ket{n}_j$ is the $n$th phononic Fock state of the $j$th phonon site.
This state is generated by acting with the pump pulse
\begin{equation}
    \hat A^\beta
 =
    \prod_{j=0}^{L-1}
    \left[\hat b^\dagger_j\right]^{n^\mathrm{ph}}
    \left(\frac{\alpha}{\sqrt{n^\mathrm{ph}!}} + \frac{\beta \hat b^\dagger_j}{\sqrt{n^\mathrm{ph}+1!}} + \frac{\beta \hat b^\nodagger_j}{\sqrt{n^\mathrm{ph}-1!}} \right)
    \label{eq:app:beta_pulse_operator}
\end{equation}
on the phononic vacuum.
The simplicity of \cref{eq:beta_state} lies in the fact that the single parameter $\beta$ controls the global phonon probability distribution $\mathcal{P}(N^\mathrm{ph})$ (since normalization implies $\alpha = \pm \sqrt{1-2\beta^2}$).
The state (\cref{eq:beta_state}) has contributions from the phonon sectors $N^{\mathrm{ph}}_m = Ln+m$, with $m = -L, -L+1, \dots,+L-1,+L$. 
To determine $\mathcal{P}(N^{\mathrm{ph}}_m)$, we perform three steps. 
First, we introduce an index $i$ that labels all non-permutationally equivalent states corresponding to the same $N^{\mathrm{ph}}_m$.
For example, for $L=2$ sites and $m=0$, these are $\ket{L=2,m=0,i=0} \equiv \{\ket{n,n}\}$ and $\ket{L=2,m=0,i=1} \equiv \{ \ket{n+1,n-1}, \ket{n-1,n+1} \}$.
Second, we note that all permutationally equivalent states occur with the same probability density which a simple algebraic calculation shows to be $\alpha^{L-m-2i} \beta^{m+2i}$.
The square of this  probability density multiplied by the degeneracy factor accounting for the number of permutations $L!/(L-m-2i)!(m+1)!i!$ gives the $i$th contribution to $\mathcal{P}(N^{\mathrm{ph}}_m)$.
Finally, we observe that for a sector fixed by $m$, the sum over the non-permutationally equivalent contributions runs from $0$ to $L-m-2i \geq 0$.
Putting all of this together yields
\begin{equation}
    \mathcal{P}(N^{\mathrm{ph}}_m) = \sum_{i=0}^{L-m-2i \geq 0} \frac{ \left( \alpha^{L-m-2i} \beta^{m+2i} \right )^2 L!}{(L-m-2i)!(m+2i)!} \;.
 \label{eq:b_state_exact_probability}
 \end{equation} 
Via a Gaussian fit, we can then extrapolate the standard deviation $\sigma(\beta)$ of $\mathcal{P}(N^{\mathrm{ph}}_m)$ associated with every $\beta$.
We checked \cref{eq:b_state_exact_probability} against brute-force numerical \gls{PST} calculations and found excellent agreement.
\begin{figure}[tb]
    \subfloat[\label{fig:beta:state:L10}]{
        \centering
        \includegraphics[width=0.75\textwidth]{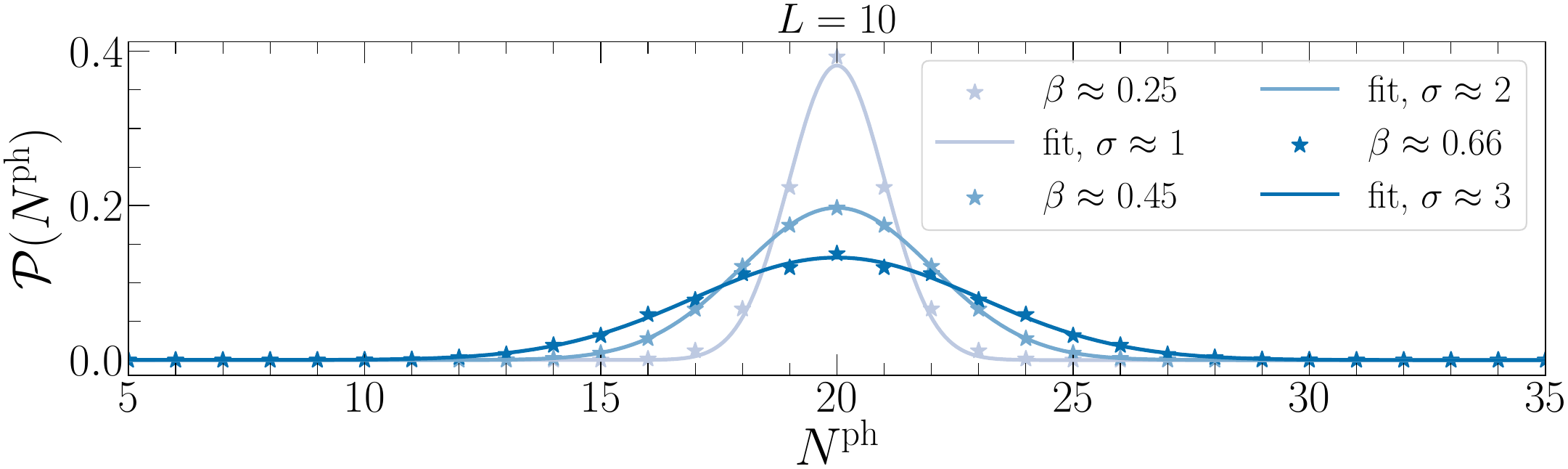}
 }\\
    \vspace*{-1em}
    \subfloat[\label{fig:beta:state:L40}]{
        \centering
        \includegraphics[width=0.75\textwidth]{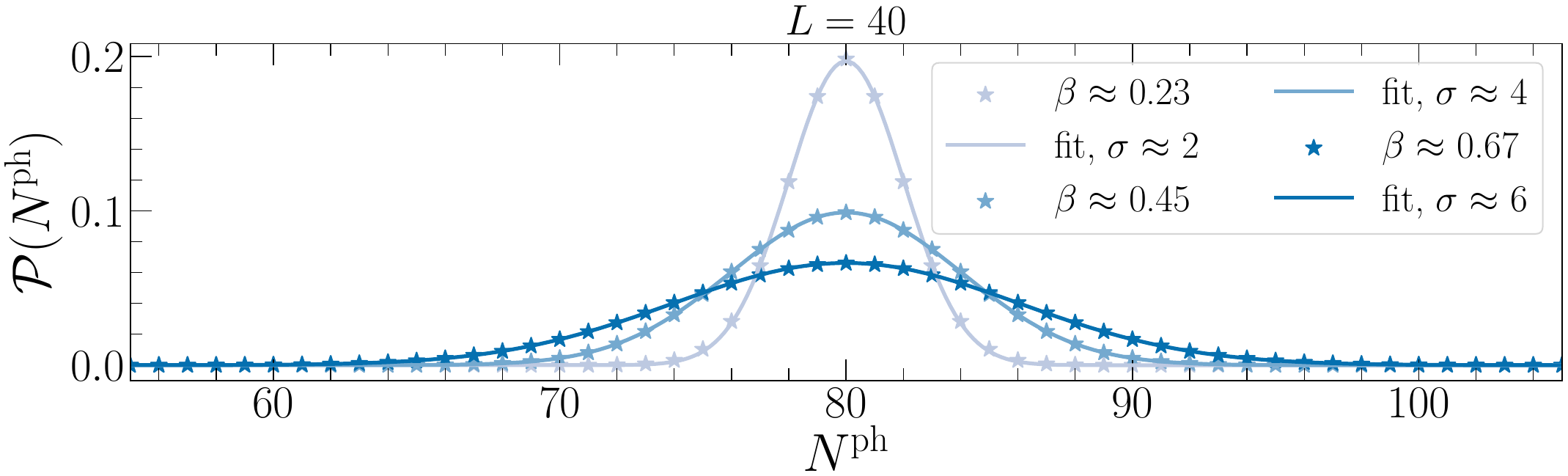}
 }
    \caption
 {
        \label{fig:beta:state:distribution}
 Global phonon distribution function $\mathcal P(N^\mathrm{ph})$ induced by the $\beta$-pulse \cref{eq:app:beta_pulse_operator}.
 The data points were obtained by evaluating \cref{eq:b_state_exact_probability} for $L=10$ sites (upper panel) and $L=40$ sites (lower panels), while the solid lines are Gaussian fits. 
 }
\end{figure}
In \cref{fig:beta:state:distribution}, we show the global phonon probability distribution $\mathcal P(N^\mathrm{ph})$ (we drop the index $m$) for $L=10$ and $L=40$ sites, obtained by evaluating \cref{eq:b_state_exact_probability} for different values of $\beta$.
Note that while in principle any value of $\beta$ between $0$ and $\sqrt{1/2}$ is allowed, in practice, when the number of sites $L$ is moderate, very small and very large values of $\beta$ lead to non-Gaussian distributions.
\cref{fig:beta:state:L10} shows that for $\sigma=1, 3$ some deviations from the Gaussian shape exist, indicating that $\beta$ should be chosen approximately in the range $[0.25,0.66]$.
At the same time, the excellent agreement between the $\beta$-pulse distributions and the Gaussian functions in \cref{fig:beta:state:L40} indicates that for larger system sizes $L$ the range of allowed $\beta$ may increase.

\bibliography{Literature}

%% file: header.tex
\usepackage{lineno}
\usepackage{graphicx}  
\usepackage{dcolumn}   
\usepackage{bm}        
\usepackage{amssymb}   
\usepackage{amsmath,stmaryrd}
\usepackage{blkarray, multirow, graphicx, diagbox, color, colortbl}
\usepackage[dvipsnames]{xcolor}
\usepackage{bbm, bbold}
\usepackage{glossaries}
\usepackage{hyphenat}
\usepackage{ifthen}
\usepackage[colorlinks, linkcolor = blue, citecolor = blue, filecolor = black, urlcolor = blue]{hyperref}
\usepackage{xkeyval}
\usepackage{moreverb}
\usepackage{rotating}
\usepackage{wrapfig}
\usepackage{slashbox}
\usepackage{xspace}
\usepackage{nicefrac}
\usepackage[]{units}
\usepackage{physics}
\usepackage{booktabs}
\usepackage{braket}
\usepackage[inline]{enumitem}
\usepackage{tabto}
\usepackage{listings}
\usepackage{xstring}
\def\ReplaceStr#1{%
	\IfSubStr{#1}{p}{%
		\StrSubstitute{#1}{p}{.}}{#1}}

\usepackage[caption=false]{subfig}
\newcommand\subfigref[1]{\protect\subref{#1}}

\usepackage[capitalise]{cleveref} %
\usepackage{chemformula}
\usepackage{algorithm}

\newcommand{\nodagger}[0]{{\vphantom{\dagger}}}

%% file: acronyms.tex
\newacronym{iTEBD}{iTEBD}{infinite time-evolving block decimation}
\newacronym{TDVP}{TDVP}{time-dependent variational principle}
\newacronym{1TDVP}{1TDVP}{single-site time-dependent variational principle}
\newacronym{2TDVP}{2TDVP}{two-site time-dependent variational principle}
\newacronym{GSE-TDVP}{GSE-TDVP}{global subspace expansion time-dependent variational principle}
\newacronym{PP-TDVP}{PP-TDVP}{projected purified time-dependent variational principle}
\newacronym{CDW}{CDW}{charge\hyp density wave}
\newacronym{PP}{PP}{projected purification}
\newacronym{PP-MPS}{PP-MPS}{projected\hyp purified matrix\hyp product states}
\newacronym{GSE}{GSE}{global subspace expansion}
\newacronym{MPS}{MPS}{matrix\hyp product state}
\newacronym{1RDM}{1RDM}{single\hyp site reduced density\hyp matrix}
\newacronym{1phRDM}{1phRDM}{single\hyp site phonon reduced density\hyp matrix}
\newacronym{DMRG}{DMRG}{density\hyp matrix renormalization group}
\newacronym{PST}{PST}{phonon state tomography}
\newacronym{SVD}{SVD}{singular value decomposition}
\newacronym{OBC}{OBC}{open boundary conditions}
\newacronym{PBC}{PBC}{periodic boundary conditions}
\newacronym{LSE-TDVP}{LSE-TDVP}{local subspace expansion time-dependent variational principle}
\newacronym{TDS}{TDS}{thermal diffuse scattering}
\newacronym{PSS}{PSS}{perfect sampling scheme}
\newacronym{SA-MR-PDF}{SA-MR-PDF}{sample\hyp averaged, momentum\hyp resolved phonon distribution function}

%% file: paper_arxiv.bbl
\begin{thebibliography}{69}%
\makeatletter
\providecommand \@ifxundefined [1]{%
 \@ifx{#1\undefined}
}%
\providecommand \@ifnum [1]{%
 \ifnum #1\expandafter \@firstoftwo
 \else \expandafter \@secondoftwo
 \fi
}%
\providecommand \@ifx [1]{%
 \ifx #1\expandafter \@firstoftwo
 \else \expandafter \@secondoftwo
 \fi
}%
\providecommand \natexlab [1]{#1}%
\providecommand \enquote  [1]{``#1''}%
\providecommand \bibnamefont  [1]{#1}%
\providecommand \bibfnamefont [1]{#1}%
\providecommand \citenamefont [1]{#1}%
\providecommand \href@noop [0]{\@secondoftwo}%
\providecommand \href [0]{\begingroup \@sanitize@url \@href}%
\providecommand \@href[1]{\@@startlink{#1}\@@href}%
\providecommand \@@href[1]{\endgroup#1\@@endlink}%
\providecommand \@sanitize@url [0]{\catcode `\\12\catcode `\$12\catcode `\&12\catcode `\#12\catcode `\^12\catcode `\_12\catcode `\%12\relax}%
\providecommand \@@startlink[1]{}%
\providecommand \@@endlink[0]{}%
\providecommand \url  [0]{\begingroup\@sanitize@url \@url }%
\providecommand \@url [1]{\endgroup\@href {#1}{\urlprefix }}%
\providecommand \urlprefix  [0]{URL }%
\providecommand \Eprint [0]{\href }%
\providecommand \doibase [0]{http://dx.doi.org/}%
\providecommand \selectlanguage [0]{\@gobble}%
\providecommand \bibinfo  [0]{\@secondoftwo}%
\providecommand \bibfield  [0]{\@secondoftwo}%
\providecommand \translation [1]{[#1]}%
\providecommand \BibitemOpen [0]{}%
\providecommand \bibitemStop [0]{}%
\providecommand \bibitemNoStop [0]{.\EOS\space}%
\providecommand \EOS [0]{\spacefactor3000\relax}%
\providecommand \BibitemShut  [1]{\csname bibitem#1\endcsname}%
\let\auto@bib@innerbib\@empty
\bibitem [{\citenamefont {Basov}\ \emph {et~al.}(2017)\citenamefont {Basov}, \citenamefont {Averitt},\ and\ \citenamefont {Hsieh}}]{Basov2017-gu}%
  \BibitemOpen
  \bibfield  {author} {\bibinfo {author} {\bibfnamefont {D~N}\ \bibnamefont {Basov}}, \bibinfo {author} {\bibfnamefont {R~D}\ \bibnamefont {Averitt}}, \ and\ \bibinfo {author} {\bibfnamefont {D}~\bibnamefont {Hsieh}},\ }\bibfield  {title} {\enquote {\bibinfo {title} {Towards properties on demand in quantum materials},}\ }\href {https://www.nature.com/articles/nmat5017} {\bibfield  {journal} {\bibinfo  {journal} {Nat. Mater.}\ }\textbf {\bibinfo {volume} {16}},\ \bibinfo {pages} {1077--1088} (\bibinfo {year} {2017})}\BibitemShut {NoStop}%
\bibitem [{\citenamefont {de~la Torre}\ \emph {et~al.}(2021)\citenamefont {de~la Torre}, \citenamefont {Kennes}, \citenamefont {Claassen}, \citenamefont {Gerber}, \citenamefont {McIver},\ and\ \citenamefont {Sentef}}]{RevModPhys.93.041002}%
  \BibitemOpen
  \bibfield  {author} {\bibinfo {author} {\bibfnamefont {Alberto}\ \bibnamefont {de~la Torre}}, \bibinfo {author} {\bibfnamefont {Dante~M.}\ \bibnamefont {Kennes}}, \bibinfo {author} {\bibfnamefont {Martin}\ \bibnamefont {Claassen}}, \bibinfo {author} {\bibfnamefont {Simon}\ \bibnamefont {Gerber}}, \bibinfo {author} {\bibfnamefont {James~W.}\ \bibnamefont {McIver}}, \ and\ \bibinfo {author} {\bibfnamefont {Michael~A.}\ \bibnamefont {Sentef}},\ }\bibfield  {title} {\enquote {\bibinfo {title} {Colloquium: Nonthermal pathways to ultrafast control in quantum materials},}\ }\href {https://link.aps.org/doi/10.1103/RevModPhys.93.041002} {\bibfield  {journal} {\bibinfo  {journal} {Rev. Mod. Phys.}\ }\textbf {\bibinfo {volume} {93}},\ \bibinfo {pages} {041002} (\bibinfo {year} {2021})}\BibitemShut {NoStop}%
\bibitem [{\citenamefont {Fausti}\ \emph {et~al.}(2011)\citenamefont {Fausti}, \citenamefont {Tobey}, \citenamefont {Dean}, \citenamefont {Kaiser}, \citenamefont {Dienst}, \citenamefont {Hoffmann}, \citenamefont {Pyon}, \citenamefont {Takayama}, \citenamefont {Takagi},\ and\ \citenamefont {Cavalleri}}]{Fausti2011}%
  \BibitemOpen
  \bibfield  {author} {\bibinfo {author} {\bibfnamefont {D.}~\bibnamefont {Fausti}}, \bibinfo {author} {\bibfnamefont {R.~I.}\ \bibnamefont {Tobey}}, \bibinfo {author} {\bibfnamefont {N.}~\bibnamefont {Dean}}, \bibinfo {author} {\bibfnamefont {S.}~\bibnamefont {Kaiser}}, \bibinfo {author} {\bibfnamefont {A.}~\bibnamefont {Dienst}}, \bibinfo {author} {\bibfnamefont {M.~C.}\ \bibnamefont {Hoffmann}}, \bibinfo {author} {\bibfnamefont {S.}~\bibnamefont {Pyon}}, \bibinfo {author} {\bibfnamefont {T.}~\bibnamefont {Takayama}}, \bibinfo {author} {\bibfnamefont {H.}~\bibnamefont {Takagi}}, \ and\ \bibinfo {author} {\bibfnamefont {A.}~\bibnamefont {Cavalleri}},\ }\bibfield  {title} {\enquote {\bibinfo {title} {Light-induced superconductivity in a stripe-ordered cuprate},}\ }\href {https://www.science.org/doi/abs/10.1126/science.1197294} {\bibfield  {journal} {\bibinfo  {journal} {Science}\ }\textbf {\bibinfo {volume} {331}},\ \bibinfo {pages} {189--191} (\bibinfo {year} {2011})}\BibitemShut {NoStop}%
\bibitem [{\citenamefont {Mitrano}\ \emph {et~al.}(2016)\citenamefont {Mitrano}, \citenamefont {Cantaluppi}, \citenamefont {Nicoletti}, \citenamefont {Kaiser}, \citenamefont {Perucchi}, \citenamefont {Lupi}, \citenamefont {Di~Pietro}, \citenamefont {Pontiroli}, \citenamefont {Ricc{\`o}}, \citenamefont {Clark}, \citenamefont {Jaksch},\ and\ \citenamefont {Cavalleri}}]{Mitrano2016}%
  \BibitemOpen
  \bibfield  {author} {\bibinfo {author} {\bibfnamefont {M.}~\bibnamefont {Mitrano}}, \bibinfo {author} {\bibfnamefont {A.}~\bibnamefont {Cantaluppi}}, \bibinfo {author} {\bibfnamefont {D.}~\bibnamefont {Nicoletti}}, \bibinfo {author} {\bibfnamefont {S.}~\bibnamefont {Kaiser}}, \bibinfo {author} {\bibfnamefont {A.}~\bibnamefont {Perucchi}}, \bibinfo {author} {\bibfnamefont {S.}~\bibnamefont {Lupi}}, \bibinfo {author} {\bibfnamefont {P.}~\bibnamefont {Di~Pietro}}, \bibinfo {author} {\bibfnamefont {D.}~\bibnamefont {Pontiroli}}, \bibinfo {author} {\bibfnamefont {M.}~\bibnamefont {Ricc{\`o}}}, \bibinfo {author} {\bibfnamefont {S.~R.}\ \bibnamefont {Clark}}, \bibinfo {author} {\bibfnamefont {D.}~\bibnamefont {Jaksch}}, \ and\ \bibinfo {author} {\bibfnamefont {A.}~\bibnamefont {Cavalleri}},\ }\bibfield  {title} {\enquote {\bibinfo {title} {Possible light-induced superconductivity in k3c60 at high temperature},}\ }\href {https://doi.org/10.1038/nature16522} {\bibfield  {journal} {\bibinfo  {journal} {Nature}\
  }\textbf {\bibinfo {volume} {530}},\ \bibinfo {pages} {461--464} (\bibinfo {year} {2016})}\BibitemShut {NoStop}%
\bibitem [{\citenamefont {Cantaluppi}\ \emph {et~al.}(2018)\citenamefont {Cantaluppi}, \citenamefont {Buzzi}, \citenamefont {Jotzu}, \citenamefont {Nicoletti}, \citenamefont {Mitrano}, \citenamefont {Pontiroli}, \citenamefont {Ricc{\`o}}, \citenamefont {Perucchi}, \citenamefont {Di~Pietro},\ and\ \citenamefont {Cavalleri}}]{Cantaluppi2018}%
  \BibitemOpen
  \bibfield  {author} {\bibinfo {author} {\bibfnamefont {A.}~\bibnamefont {Cantaluppi}}, \bibinfo {author} {\bibfnamefont {M.}~\bibnamefont {Buzzi}}, \bibinfo {author} {\bibfnamefont {G.}~\bibnamefont {Jotzu}}, \bibinfo {author} {\bibfnamefont {D.}~\bibnamefont {Nicoletti}}, \bibinfo {author} {\bibfnamefont {M.}~\bibnamefont {Mitrano}}, \bibinfo {author} {\bibfnamefont {D.}~\bibnamefont {Pontiroli}}, \bibinfo {author} {\bibfnamefont {M.}~\bibnamefont {Ricc{\`o}}}, \bibinfo {author} {\bibfnamefont {A.}~\bibnamefont {Perucchi}}, \bibinfo {author} {\bibfnamefont {P.}~\bibnamefont {Di~Pietro}}, \ and\ \bibinfo {author} {\bibfnamefont {A.}~\bibnamefont {Cavalleri}},\ }\bibfield  {title} {\enquote {\bibinfo {title} {Pressure tuning of light-induced superconductivity in k3c60},}\ }\href {https://doi.org/10.1038/s41567-018-0134-8} {\bibfield  {journal} {\bibinfo  {journal} {Nat. Phys.}\ }\textbf {\bibinfo {volume} {14}},\ \bibinfo {pages} {837--841} (\bibinfo {year} {2018})}\BibitemShut {NoStop}%
\bibitem [{\citenamefont {Biswas}\ \emph {et~al.}(2018)\citenamefont {Biswas}, \citenamefont {Husek},\ and\ \citenamefont {Baker}}]{C8CC01745J}%
  \BibitemOpen
  \bibfield  {author} {\bibinfo {author} {\bibfnamefont {Somnath}\ \bibnamefont {Biswas}}, \bibinfo {author} {\bibfnamefont {Jakub}\ \bibnamefont {Husek}}, \ and\ \bibinfo {author} {\bibfnamefont {L.~Robert}\ \bibnamefont {Baker}},\ }\bibfield  {title} {\enquote {\bibinfo {title} {Elucidating ultrafast electron dynamics at surfaces using extreme ultraviolet (xuv) reflection–absorption spectroscopy},}\ }\href {http://dx.doi.org/10.1039/C8CC01745J} {\bibfield  {journal} {\bibinfo  {journal} {Chem. Commun.}\ }\textbf {\bibinfo {volume} {54}},\ \bibinfo {pages} {4216--4230} (\bibinfo {year} {2018})}\BibitemShut {NoStop}%
\bibitem [{\citenamefont {Schwarz}\ and\ \citenamefont {Manske}(2020)}]{Schwarz2020}%
  \BibitemOpen
  \bibfield  {author} {\bibinfo {author} {\bibfnamefont {Lukas}\ \bibnamefont {Schwarz}}\ and\ \bibinfo {author} {\bibfnamefont {Dirk}\ \bibnamefont {Manske}},\ }\bibfield  {title} {\enquote {\bibinfo {title} {Theory of driven higgs oscillations and third-harmonic generation in unconventional superconductors},}\ }\href {https://link.aps.org/doi/10.1103/PhysRevB.101.184519} {\bibfield  {journal} {\bibinfo  {journal} {Phys. Rev. B}\ }\textbf {\bibinfo {volume} {101}},\ \bibinfo {pages} {184519} (\bibinfo {year} {2020})}\BibitemShut {NoStop}%
\bibitem [{\citenamefont {Clear}\ \emph {et~al.}(2020)\citenamefont {Clear}, \citenamefont {Schofield}, \citenamefont {Major}, \citenamefont {Iles-Smith}, \citenamefont {Clark},\ and\ \citenamefont {McCutcheon}}]{PhysRevLett.124.153602}%
  \BibitemOpen
  \bibfield  {author} {\bibinfo {author} {\bibfnamefont {Chloe}\ \bibnamefont {Clear}}, \bibinfo {author} {\bibfnamefont {Ross~C.}\ \bibnamefont {Schofield}}, \bibinfo {author} {\bibfnamefont {Kyle~D.}\ \bibnamefont {Major}}, \bibinfo {author} {\bibfnamefont {Jake}\ \bibnamefont {Iles-Smith}}, \bibinfo {author} {\bibfnamefont {Alex~S.}\ \bibnamefont {Clark}}, \ and\ \bibinfo {author} {\bibfnamefont {Dara P.~S.}\ \bibnamefont {McCutcheon}},\ }\bibfield  {title} {\enquote {\bibinfo {title} {Phonon-induced optical dephasing in single organic molecules},}\ }\href {https://link.aps.org/doi/10.1103/PhysRevLett.124.153602} {\bibfield  {journal} {\bibinfo  {journal} {Phys. Rev. Lett.}\ }\textbf {\bibinfo {volume} {124}},\ \bibinfo {pages} {153602} (\bibinfo {year} {2020})}\BibitemShut {NoStop}%
\bibitem [{\citenamefont {Budden}\ \emph {et~al.}(2021)\citenamefont {Budden}, \citenamefont {Gebert}, \citenamefont {Buzzi}, \citenamefont {Jotzu}, \citenamefont {Wang}, \citenamefont {Matsuyama}, \citenamefont {Meier}, \citenamefont {Laplace}, \citenamefont {Pontiroli}, \citenamefont {Ricc{\`o}}, \citenamefont {Schlawin}, \citenamefont {Jaksch},\ and\ \citenamefont {Cavalleri}}]{Budden2021}%
  \BibitemOpen
  \bibfield  {author} {\bibinfo {author} {\bibfnamefont {M.}~\bibnamefont {Budden}}, \bibinfo {author} {\bibfnamefont {T.}~\bibnamefont {Gebert}}, \bibinfo {author} {\bibfnamefont {M.}~\bibnamefont {Buzzi}}, \bibinfo {author} {\bibfnamefont {G.}~\bibnamefont {Jotzu}}, \bibinfo {author} {\bibfnamefont {E.}~\bibnamefont {Wang}}, \bibinfo {author} {\bibfnamefont {T.}~\bibnamefont {Matsuyama}}, \bibinfo {author} {\bibfnamefont {G.}~\bibnamefont {Meier}}, \bibinfo {author} {\bibfnamefont {Y.}~\bibnamefont {Laplace}}, \bibinfo {author} {\bibfnamefont {D.}~\bibnamefont {Pontiroli}}, \bibinfo {author} {\bibfnamefont {M.}~\bibnamefont {Ricc{\`o}}}, \bibinfo {author} {\bibfnamefont {F.}~\bibnamefont {Schlawin}}, \bibinfo {author} {\bibfnamefont {D.}~\bibnamefont {Jaksch}}, \ and\ \bibinfo {author} {\bibfnamefont {A.}~\bibnamefont {Cavalleri}},\ }\bibfield  {title} {\enquote {\bibinfo {title} {Evidence for metastable photo-induced superconductivity in k3c60},}\ }\href {https://doi.org/10.1038/s41567-020-01148-1}
  {\bibfield  {journal} {\bibinfo  {journal} {Nat. Phys.}\ }\textbf {\bibinfo {volume} {17}},\ \bibinfo {pages} {611--618} (\bibinfo {year} {2021})}\BibitemShut {NoStop}%
\bibitem [{\citenamefont {Buzzi}\ \emph {et~al.}(2021)\citenamefont {Buzzi}, \citenamefont {Jotzu}, \citenamefont {Cavalleri}, \citenamefont {Cirac}, \citenamefont {Demler}, \citenamefont {Halperin}, \citenamefont {Lukin}, \citenamefont {Shi}, \citenamefont {Wang},\ and\ \citenamefont {Podolsky}}]{Buzzi2021}%
  \BibitemOpen
  \bibfield  {author} {\bibinfo {author} {\bibfnamefont {Michele}\ \bibnamefont {Buzzi}}, \bibinfo {author} {\bibfnamefont {Gregor}\ \bibnamefont {Jotzu}}, \bibinfo {author} {\bibfnamefont {Andrea}\ \bibnamefont {Cavalleri}}, \bibinfo {author} {\bibfnamefont {J.~Ignacio}\ \bibnamefont {Cirac}}, \bibinfo {author} {\bibfnamefont {Eugene~A.}\ \bibnamefont {Demler}}, \bibinfo {author} {\bibfnamefont {Bertrand~I.}\ \bibnamefont {Halperin}}, \bibinfo {author} {\bibfnamefont {Mikhail~D.}\ \bibnamefont {Lukin}}, \bibinfo {author} {\bibfnamefont {Tao}\ \bibnamefont {Shi}}, \bibinfo {author} {\bibfnamefont {Yao}\ \bibnamefont {Wang}}, \ and\ \bibinfo {author} {\bibfnamefont {Daniel}\ \bibnamefont {Podolsky}},\ }\bibfield  {title} {\enquote {\bibinfo {title} {Higgs-mediated optical amplification in a nonequilibrium superconductor},}\ }\href {https://link.aps.org/doi/10.1103/PhysRevX.11.011055} {\bibfield  {journal} {\bibinfo  {journal} {Phys. Rev. X}\ }\textbf {\bibinfo {volume} {11}},\ \bibinfo {pages} {011055}
  (\bibinfo {year} {2021})}\BibitemShut {NoStop}%
\bibitem [{\citenamefont {Campi}\ \emph {et~al.}(2021)\citenamefont {Campi}, \citenamefont {Kumari},\ and\ \citenamefont {Marzari}}]{campi_prediction_2021}%
  \BibitemOpen
  \bibfield  {author} {\bibinfo {author} {\bibfnamefont {Davide}\ \bibnamefont {Campi}}, \bibinfo {author} {\bibfnamefont {Simran}\ \bibnamefont {Kumari}}, \ and\ \bibinfo {author} {\bibfnamefont {Nicola}\ \bibnamefont {Marzari}},\ }\bibfield  {title} {\enquote {\bibinfo {title} {Prediction of {Phonon}-{Mediated} {Superconductivity} with {High} {Critical} {Temperature} in the {Two}-{Dimensional} {Topological} {Semimetal} {W2N3}},}\ }\href {https://doi.org/10.1021/acs.nanolett.0c05125} {\bibfield  {journal} {\bibinfo  {journal} {Nano Lett.}\ }\textbf {\bibinfo {volume} {21}},\ \bibinfo {pages} {3435--3442} (\bibinfo {year} {2021})}\BibitemShut {NoStop}%
\bibitem [{\citenamefont {Wang}\ \emph {et~al.}(2013)\citenamefont {Wang}, \citenamefont {Steinberg}, \citenamefont {Jarillo-Herrero},\ and\ \citenamefont {Gedik}}]{Gedik2013}%
  \BibitemOpen
  \bibfield  {author} {\bibinfo {author} {\bibfnamefont {Y.~H.}\ \bibnamefont {Wang}}, \bibinfo {author} {\bibfnamefont {H.}~\bibnamefont {Steinberg}}, \bibinfo {author} {\bibfnamefont {P.}~\bibnamefont {Jarillo-Herrero}}, \ and\ \bibinfo {author} {\bibfnamefont {N.}~\bibnamefont {Gedik}},\ }\bibfield  {title} {\enquote {\bibinfo {title} {Observation of floquet-bloch states on the surface of a topological insulator},}\ }\href {https://www.science.org/doi/abs/10.1126/science.1239834} {\bibfield  {journal} {\bibinfo  {journal} {Science}\ }\textbf {\bibinfo {volume} {342}},\ \bibinfo {pages} {453--457} (\bibinfo {year} {2013})}\BibitemShut {NoStop}%
\bibitem [{\citenamefont {Mahmood}\ \emph {et~al.}(2016)\citenamefont {Mahmood}, \citenamefont {Chan}, \citenamefont {Alpichshev}, \citenamefont {Gardner}, \citenamefont {Lee}, \citenamefont {Lee},\ and\ \citenamefont {Gedik}}]{Mahmood2016-co}%
  \BibitemOpen
  \bibfield  {author} {\bibinfo {author} {\bibfnamefont {Fahad}\ \bibnamefont {Mahmood}}, \bibinfo {author} {\bibfnamefont {Ching-Kit}\ \bibnamefont {Chan}}, \bibinfo {author} {\bibfnamefont {Zhanybek}\ \bibnamefont {Alpichshev}}, \bibinfo {author} {\bibfnamefont {Dillon}\ \bibnamefont {Gardner}}, \bibinfo {author} {\bibfnamefont {Young}\ \bibnamefont {Lee}}, \bibinfo {author} {\bibfnamefont {Patrick~A}\ \bibnamefont {Lee}}, \ and\ \bibinfo {author} {\bibfnamefont {Nuh}\ \bibnamefont {Gedik}},\ }\bibfield  {title} {\enquote {\bibinfo {title} {Selective scattering between {Floquet--Bloch} and volkov states in a topological insulator},}\ }\href {https://doi.org/10.1038/nphys3609} {\bibfield  {journal} {\bibinfo  {journal} {Nat. Phys.}\ }\textbf {\bibinfo {volume} {12}},\ \bibinfo {pages} {306--310} (\bibinfo {year} {2016})}\BibitemShut {NoStop}%
\bibitem [{\citenamefont {Shan}\ \emph {et~al.}(2021)\citenamefont {Shan}, \citenamefont {Ye}, \citenamefont {Chu}, \citenamefont {Lee}, \citenamefont {Park}, \citenamefont {Balents},\ and\ \citenamefont {Hsieh}}]{Shan2021-og}%
  \BibitemOpen
  \bibfield  {author} {\bibinfo {author} {\bibfnamefont {Jun-Yi}\ \bibnamefont {Shan}}, \bibinfo {author} {\bibfnamefont {M}~\bibnamefont {Ye}}, \bibinfo {author} {\bibfnamefont {H}~\bibnamefont {Chu}}, \bibinfo {author} {\bibfnamefont {Sungmin}\ \bibnamefont {Lee}}, \bibinfo {author} {\bibfnamefont {Je-Geun}\ \bibnamefont {Park}}, \bibinfo {author} {\bibfnamefont {L}~\bibnamefont {Balents}}, \ and\ \bibinfo {author} {\bibfnamefont {D}~\bibnamefont {Hsieh}},\ }\bibfield  {title} {\enquote {\bibinfo {title} {Giant modulation of optical nonlinearity by floquet engineering},}\ }\href {https://www.nature.com/articles/s41586-021-04051-8} {\bibfield  {journal} {\bibinfo  {journal} {Nature}\ }\textbf {\bibinfo {volume} {600}},\ \bibinfo {pages} {235--239} (\bibinfo {year} {2021})}\BibitemShut {NoStop}%
\bibitem [{\citenamefont {Zhou}\ \emph {et~al.}(2023)\citenamefont {Zhou}, \citenamefont {Bao}, \citenamefont {Fan}, \citenamefont {Zhou}, \citenamefont {Gao}, \citenamefont {Zhong}, \citenamefont {Lin}, \citenamefont {Liu}, \citenamefont {Yu}, \citenamefont {Tang}, \citenamefont {Meng}, \citenamefont {Duan},\ and\ \citenamefont {Zhou}}]{Zhou2023-ws}%
  \BibitemOpen
  \bibfield  {author} {\bibinfo {author} {\bibfnamefont {Shaohua}\ \bibnamefont {Zhou}}, \bibinfo {author} {\bibfnamefont {Changhua}\ \bibnamefont {Bao}}, \bibinfo {author} {\bibfnamefont {Benshu}\ \bibnamefont {Fan}}, \bibinfo {author} {\bibfnamefont {Hui}\ \bibnamefont {Zhou}}, \bibinfo {author} {\bibfnamefont {Qixuan}\ \bibnamefont {Gao}}, \bibinfo {author} {\bibfnamefont {Haoyuan}\ \bibnamefont {Zhong}}, \bibinfo {author} {\bibfnamefont {Tianyun}\ \bibnamefont {Lin}}, \bibinfo {author} {\bibfnamefont {Hang}\ \bibnamefont {Liu}}, \bibinfo {author} {\bibfnamefont {Pu}~\bibnamefont {Yu}}, \bibinfo {author} {\bibfnamefont {Peizhe}\ \bibnamefont {Tang}}, \bibinfo {author} {\bibfnamefont {Sheng}\ \bibnamefont {Meng}}, \bibinfo {author} {\bibfnamefont {Wenhui}\ \bibnamefont {Duan}}, \ and\ \bibinfo {author} {\bibfnamefont {Shuyun}\ \bibnamefont {Zhou}},\ }\bibfield  {title} {\enquote {\bibinfo {title} {Pseudospin-selective floquet band engineering in black phosphorus},}\ }\href
  {https://www.nature.com/articles/s41586-022-05610-3} {\bibfield  {journal} {\bibinfo  {journal} {Nature}\ }\textbf {\bibinfo {volume} {614}},\ \bibinfo {pages} {75--80} (\bibinfo {year} {2023})}\BibitemShut {NoStop}%
\bibitem [{\citenamefont {Rini}\ \emph {et~al.}(2007)\citenamefont {Rini}, \citenamefont {Tobey}, \citenamefont {Dean}, \citenamefont {Itatani}, \citenamefont {Tomioka}, \citenamefont {Tokura}, \citenamefont {Schoenlein},\ and\ \citenamefont {Cavalleri}}]{Rini2007-eo}%
  \BibitemOpen
  \bibfield  {author} {\bibinfo {author} {\bibfnamefont {Matteo}\ \bibnamefont {Rini}}, \bibinfo {author} {\bibfnamefont {Ra'anan}\ \bibnamefont {Tobey}}, \bibinfo {author} {\bibfnamefont {Nicky}\ \bibnamefont {Dean}}, \bibinfo {author} {\bibfnamefont {Jiro}\ \bibnamefont {Itatani}}, \bibinfo {author} {\bibfnamefont {Yasuhide}\ \bibnamefont {Tomioka}}, \bibinfo {author} {\bibfnamefont {Yoshinori}\ \bibnamefont {Tokura}}, \bibinfo {author} {\bibfnamefont {Robert~W}\ \bibnamefont {Schoenlein}}, \ and\ \bibinfo {author} {\bibfnamefont {Andrea}\ \bibnamefont {Cavalleri}},\ }\bibfield  {title} {\enquote {\bibinfo {title} {Control of the electronic phase of a manganite by mode-selective vibrational excitation},}\ }\href {https://www.nature.com/articles/nature06119} {\bibfield  {journal} {\bibinfo  {journal} {Nature}\ }\textbf {\bibinfo {volume} {449}},\ \bibinfo {pages} {72--74} (\bibinfo {year} {2007})}\BibitemShut {NoStop}%
\bibitem [{\citenamefont {F{\"o}rst}\ \emph {et~al.}(2011)\citenamefont {F{\"o}rst}, \citenamefont {Manzoni}, \citenamefont {Kaiser}, \citenamefont {Tomioka}, \citenamefont {Tokura}, \citenamefont {Merlin},\ and\ \citenamefont {Cavalleri}}]{Forst2011-nr}%
  \BibitemOpen
  \bibfield  {author} {\bibinfo {author} {\bibfnamefont {M}~\bibnamefont {F{\"o}rst}}, \bibinfo {author} {\bibfnamefont {C}~\bibnamefont {Manzoni}}, \bibinfo {author} {\bibfnamefont {S}~\bibnamefont {Kaiser}}, \bibinfo {author} {\bibfnamefont {Y}~\bibnamefont {Tomioka}}, \bibinfo {author} {\bibfnamefont {Y}~\bibnamefont {Tokura}}, \bibinfo {author} {\bibfnamefont {R}~\bibnamefont {Merlin}}, \ and\ \bibinfo {author} {\bibfnamefont {A}~\bibnamefont {Cavalleri}},\ }\bibfield  {title} {\enquote {\bibinfo {title} {Nonlinear phononics as an ultrafast route to lattice control},}\ }\href {https://www.nature.com/articles/nphys2055} {\bibfield  {journal} {\bibinfo  {journal} {Nat. Phys.}\ }\textbf {\bibinfo {volume} {7}},\ \bibinfo {pages} {854--856} (\bibinfo {year} {2011})}\BibitemShut {NoStop}%
\bibitem [{\citenamefont {Subedi}\ \emph {et~al.}(2014)\citenamefont {Subedi}, \citenamefont {Cavalleri},\ and\ \citenamefont {Georges}}]{Subedi2014}%
  \BibitemOpen
  \bibfield  {author} {\bibinfo {author} {\bibfnamefont {Alaska}\ \bibnamefont {Subedi}}, \bibinfo {author} {\bibfnamefont {Andrea}\ \bibnamefont {Cavalleri}}, \ and\ \bibinfo {author} {\bibfnamefont {Antoine}\ \bibnamefont {Georges}},\ }\bibfield  {title} {\enquote {\bibinfo {title} {Theory of nonlinear phononics for coherent light control of solids},}\ }\href {https://link.aps.org/doi/10.1103/PhysRevB.89.220301} {\bibfield  {journal} {\bibinfo  {journal} {Phys. Rev. B}\ }\textbf {\bibinfo {volume} {89}},\ \bibinfo {pages} {220301} (\bibinfo {year} {2014})}\BibitemShut {NoStop}%
\bibitem [{\citenamefont {Nova}\ \emph {et~al.}(2017)\citenamefont {Nova}, \citenamefont {Cartella}, \citenamefont {Cantaluppi}, \citenamefont {F{\"o}rst}, \citenamefont {Bossini}, \citenamefont {Mikhaylovskiy}, \citenamefont {Kimel}, \citenamefont {Merlin},\ and\ \citenamefont {Cavalleri}}]{Nova2017-vd}%
  \BibitemOpen
  \bibfield  {author} {\bibinfo {author} {\bibfnamefont {T~F}\ \bibnamefont {Nova}}, \bibinfo {author} {\bibfnamefont {A}~\bibnamefont {Cartella}}, \bibinfo {author} {\bibfnamefont {A}~\bibnamefont {Cantaluppi}}, \bibinfo {author} {\bibfnamefont {M}~\bibnamefont {F{\"o}rst}}, \bibinfo {author} {\bibfnamefont {D}~\bibnamefont {Bossini}}, \bibinfo {author} {\bibfnamefont {R~V}\ \bibnamefont {Mikhaylovskiy}}, \bibinfo {author} {\bibfnamefont {A~V}\ \bibnamefont {Kimel}}, \bibinfo {author} {\bibfnamefont {R}~\bibnamefont {Merlin}}, \ and\ \bibinfo {author} {\bibfnamefont {A}~\bibnamefont {Cavalleri}},\ }\bibfield  {title} {\enquote {\bibinfo {title} {An effective magnetic field from optically driven phonons},}\ }\href {https://www.nature.com/articles/nphys3925} {\bibfield  {journal} {\bibinfo  {journal} {Nat. Phys.}\ }\textbf {\bibinfo {volume} {13}},\ \bibinfo {pages} {132--136} (\bibinfo {year} {2017})}\BibitemShut {NoStop}%
\bibitem [{\citenamefont {Disa}\ \emph {et~al.}(2021)\citenamefont {Disa}, \citenamefont {Nova},\ and\ \citenamefont {Cavalleri}}]{Disa2021-jt}%
  \BibitemOpen
  \bibfield  {author} {\bibinfo {author} {\bibfnamefont {Ankit~S}\ \bibnamefont {Disa}}, \bibinfo {author} {\bibfnamefont {Tobia~F}\ \bibnamefont {Nova}}, \ and\ \bibinfo {author} {\bibfnamefont {Andrea}\ \bibnamefont {Cavalleri}},\ }\bibfield  {title} {\enquote {\bibinfo {title} {Engineering crystal structures with light},}\ }\href {https://www.nature.com/articles/s41567-021-01366-1} {\bibfield  {journal} {\bibinfo  {journal} {Nat. Phys.}\ }\textbf {\bibinfo {volume} {17}},\ \bibinfo {pages} {1087--1092} (\bibinfo {year} {2021})}\BibitemShut {NoStop}%
\bibitem [{\citenamefont {Khalsa}\ \emph {et~al.}(2021)\citenamefont {Khalsa}, \citenamefont {Benedek},\ and\ \citenamefont {Moses}}]{PhysRevX.11.021067}%
  \BibitemOpen
  \bibfield  {author} {\bibinfo {author} {\bibfnamefont {Guru}\ \bibnamefont {Khalsa}}, \bibinfo {author} {\bibfnamefont {Nicole~A.}\ \bibnamefont {Benedek}}, \ and\ \bibinfo {author} {\bibfnamefont {Jeffrey}\ \bibnamefont {Moses}},\ }\bibfield  {title} {\enquote {\bibinfo {title} {Ultrafast control of material optical properties via the infrared resonant raman effect},}\ }\href {https://link.aps.org/doi/10.1103/PhysRevX.11.021067} {\bibfield  {journal} {\bibinfo  {journal} {Phys. Rev. X}\ }\textbf {\bibinfo {volume} {11}},\ \bibinfo {pages} {021067} (\bibinfo {year} {2021})}\BibitemShut {NoStop}%
\bibitem [{\citenamefont {Henstridge}\ \emph {et~al.}(2022)\citenamefont {Henstridge}, \citenamefont {F{\"o}rst}, \citenamefont {Rowe}, \citenamefont {Fechner},\ and\ \citenamefont {Cavalleri}}]{Henstridge2022-yl}%
  \BibitemOpen
  \bibfield  {author} {\bibinfo {author} {\bibfnamefont {M}~\bibnamefont {Henstridge}}, \bibinfo {author} {\bibfnamefont {M}~\bibnamefont {F{\"o}rst}}, \bibinfo {author} {\bibfnamefont {E}~\bibnamefont {Rowe}}, \bibinfo {author} {\bibfnamefont {M}~\bibnamefont {Fechner}}, \ and\ \bibinfo {author} {\bibfnamefont {A}~\bibnamefont {Cavalleri}},\ }\bibfield  {title} {\enquote {\bibinfo {title} {Nonlocal nonlinear phononics},}\ }\href {https://www.nature.com/articles/s41567-022-01512-3} {\bibfield  {journal} {\bibinfo  {journal} {Nat. Phys.}\ }\textbf {\bibinfo {volume} {18}},\ \bibinfo {pages} {457--461} (\bibinfo {year} {2022})}\BibitemShut {NoStop}%
\bibitem [{\citenamefont {Sous}\ \emph {et~al.}(2021)\citenamefont {Sous}, \citenamefont {Kloss}, \citenamefont {Kennes}, \citenamefont {Reichman},\ and\ \citenamefont {Millis}}]{Sous2021}%
  \BibitemOpen
  \bibfield  {author} {\bibinfo {author} {\bibfnamefont {J.}~\bibnamefont {Sous}}, \bibinfo {author} {\bibfnamefont {B.}~\bibnamefont {Kloss}}, \bibinfo {author} {\bibfnamefont {D.~M.}\ \bibnamefont {Kennes}}, \bibinfo {author} {\bibfnamefont {D.~R.}\ \bibnamefont {Reichman}}, \ and\ \bibinfo {author} {\bibfnamefont {A.~J.}\ \bibnamefont {Millis}},\ }\bibfield  {title} {\enquote {\bibinfo {title} {Phonon-induced disorder in dynamics of optically pumped metals from nonlinear electron-phonon coupling},}\ }\href {https://doi.org/10.1038/s41467-021-26030-3} {\bibfield  {journal} {\bibinfo  {journal} {Nat. Commun.}\ }\textbf {\bibinfo {volume} {12}},\ \bibinfo {pages} {5803} (\bibinfo {year} {2021})}\BibitemShut {NoStop}%
\bibitem [{\citenamefont {Sentef}\ \emph {et~al.}(2013)\citenamefont {Sentef}, \citenamefont {Kemper}, \citenamefont {Moritz}, \citenamefont {Freericks}, \citenamefont {Shen},\ and\ \citenamefont {Devereaux}}]{PhysRevX.3.041033}%
  \BibitemOpen
  \bibfield  {author} {\bibinfo {author} {\bibfnamefont {Michael}\ \bibnamefont {Sentef}}, \bibinfo {author} {\bibfnamefont {Alexander~F.}\ \bibnamefont {Kemper}}, \bibinfo {author} {\bibfnamefont {Brian}\ \bibnamefont {Moritz}}, \bibinfo {author} {\bibfnamefont {James~K.}\ \bibnamefont {Freericks}}, \bibinfo {author} {\bibfnamefont {Zhi-Xun}\ \bibnamefont {Shen}}, \ and\ \bibinfo {author} {\bibfnamefont {Thomas~P.}\ \bibnamefont {Devereaux}},\ }\bibfield  {title} {\enquote {\bibinfo {title} {Examining electron-boson coupling using time-resolved spectroscopy},}\ }\href {https://link.aps.org/doi/10.1103/PhysRevX.3.041033} {\bibfield  {journal} {\bibinfo  {journal} {Phys. Rev. X}\ }\textbf {\bibinfo {volume} {3}},\ \bibinfo {pages} {041033} (\bibinfo {year} {2013})}\BibitemShut {NoStop}%
\bibitem [{\citenamefont {Kemper}\ \emph {et~al.}(2015)\citenamefont {Kemper}, \citenamefont {Sentef}, \citenamefont {Moritz}, \citenamefont {Freericks},\ and\ \citenamefont {Devereaux}}]{PhysRevB.92.224517}%
  \BibitemOpen
  \bibfield  {author} {\bibinfo {author} {\bibfnamefont {A.~F.}\ \bibnamefont {Kemper}}, \bibinfo {author} {\bibfnamefont {M.~A.}\ \bibnamefont {Sentef}}, \bibinfo {author} {\bibfnamefont {B.}~\bibnamefont {Moritz}}, \bibinfo {author} {\bibfnamefont {J.~K.}\ \bibnamefont {Freericks}}, \ and\ \bibinfo {author} {\bibfnamefont {T.~P.}\ \bibnamefont {Devereaux}},\ }\bibfield  {title} {\enquote {\bibinfo {title} {Direct observation of higgs mode oscillations in the pump-probe photoemission spectra of electron-phonon mediated superconductors},}\ }\href {https://link.aps.org/doi/10.1103/PhysRevB.92.224517} {\bibfield  {journal} {\bibinfo  {journal} {Phys. Rev. B}\ }\textbf {\bibinfo {volume} {92}},\ \bibinfo {pages} {224517} (\bibinfo {year} {2015})}\BibitemShut {NoStop}%
\bibitem [{\citenamefont {Sentef}\ \emph {et~al.}(2016)\citenamefont {Sentef}, \citenamefont {Kemper}, \citenamefont {Georges},\ and\ \citenamefont {Kollath}}]{PhysRevB.93.144506}%
  \BibitemOpen
  \bibfield  {author} {\bibinfo {author} {\bibfnamefont {M.~A.}\ \bibnamefont {Sentef}}, \bibinfo {author} {\bibfnamefont {A.~F.}\ \bibnamefont {Kemper}}, \bibinfo {author} {\bibfnamefont {A.}~\bibnamefont {Georges}}, \ and\ \bibinfo {author} {\bibfnamefont {C.}~\bibnamefont {Kollath}},\ }\bibfield  {title} {\enquote {\bibinfo {title} {Theory of light-enhanced phonon-mediated superconductivity},}\ }\href {https://link.aps.org/doi/10.1103/PhysRevB.93.144506} {\bibfield  {journal} {\bibinfo  {journal} {Phys. Rev. B}\ }\textbf {\bibinfo {volume} {93}},\ \bibinfo {pages} {144506} (\bibinfo {year} {2016})}\BibitemShut {NoStop}%
\bibitem [{\citenamefont {Knap}\ \emph {et~al.}(2016)\citenamefont {Knap}, \citenamefont {Babadi}, \citenamefont {Refael}, \citenamefont {Martin},\ and\ \citenamefont {Demler}}]{PhysRevB.94.214504_2016}%
  \BibitemOpen
  \bibfield  {author} {\bibinfo {author} {\bibfnamefont {Michael}\ \bibnamefont {Knap}}, \bibinfo {author} {\bibfnamefont {Mehrtash}\ \bibnamefont {Babadi}}, \bibinfo {author} {\bibfnamefont {Gil}\ \bibnamefont {Refael}}, \bibinfo {author} {\bibfnamefont {Ivar}\ \bibnamefont {Martin}}, \ and\ \bibinfo {author} {\bibfnamefont {Eugene}\ \bibnamefont {Demler}},\ }\bibfield  {title} {\enquote {\bibinfo {title} {Dynamical cooper pairing in nonequilibrium electron-phonon systems},}\ }\href {https://link.aps.org/doi/10.1103/PhysRevB.94.214504} {\bibfield  {journal} {\bibinfo  {journal} {Phys. Rev. B}\ }\textbf {\bibinfo {volume} {94}},\ \bibinfo {pages} {214504} (\bibinfo {year} {2016})}\BibitemShut {NoStop}%
\bibitem [{\citenamefont {Babadi}\ \emph {et~al.}(2017)\citenamefont {Babadi}, \citenamefont {Knap}, \citenamefont {Martin}, \citenamefont {Refael},\ and\ \citenamefont {Demler}}]{PhysRevB.96.014512_2017}%
  \BibitemOpen
  \bibfield  {author} {\bibinfo {author} {\bibfnamefont {Mehrtash}\ \bibnamefont {Babadi}}, \bibinfo {author} {\bibfnamefont {Michael}\ \bibnamefont {Knap}}, \bibinfo {author} {\bibfnamefont {Ivar}\ \bibnamefont {Martin}}, \bibinfo {author} {\bibfnamefont {Gil}\ \bibnamefont {Refael}}, \ and\ \bibinfo {author} {\bibfnamefont {Eugene}\ \bibnamefont {Demler}},\ }\bibfield  {title} {\enquote {\bibinfo {title} {Theory of parametrically amplified electron-phonon superconductivity},}\ }\href {https://link.aps.org/doi/10.1103/PhysRevB.96.014512} {\bibfield  {journal} {\bibinfo  {journal} {Phys. Rev. B}\ }\textbf {\bibinfo {volume} {96}},\ \bibinfo {pages} {014512} (\bibinfo {year} {2017})}\BibitemShut {NoStop}%
\bibitem [{\citenamefont {Kennes}\ \emph {et~al.}(2017{\natexlab{a}})\citenamefont {Kennes}, \citenamefont {Wilner}, \citenamefont {Reichman},\ and\ \citenamefont {Millis}}]{Kennes2017}%
  \BibitemOpen
  \bibfield  {author} {\bibinfo {author} {\bibfnamefont {D.~M.}\ \bibnamefont {Kennes}}, \bibinfo {author} {\bibfnamefont {E.~Y.}\ \bibnamefont {Wilner}}, \bibinfo {author} {\bibfnamefont {D.~R.}\ \bibnamefont {Reichman}}, \ and\ \bibinfo {author} {\bibfnamefont {A.J.}\ \bibnamefont {Millis}},\ }\bibfield  {title} {\enquote {\bibinfo {title} {Transient superconductivity from electronic squeezing of optically pumped phonons},}\ }\href {https://doi.org/10.1038/nphys4024} {\bibfield  {journal} {\bibinfo  {journal} {Nat. Phys.}\ }\textbf {\bibinfo {volume} {13}},\ \bibinfo {pages} {479--483} (\bibinfo {year} {2017}{\natexlab{a}})}\BibitemShut {NoStop}%
\bibitem [{\citenamefont {Kennes}\ \emph {et~al.}(2017{\natexlab{b}})\citenamefont {Kennes}, \citenamefont {Wilner}, \citenamefont {Reichman},\ and\ \citenamefont {Millis}}]{PhysRevB.96.054506}%
  \BibitemOpen
  \bibfield  {author} {\bibinfo {author} {\bibfnamefont {D.~M.}\ \bibnamefont {Kennes}}, \bibinfo {author} {\bibfnamefont {E.~Y.}\ \bibnamefont {Wilner}}, \bibinfo {author} {\bibfnamefont {D.~R.}\ \bibnamefont {Reichman}}, \ and\ \bibinfo {author} {\bibfnamefont {A.~J.}\ \bibnamefont {Millis}},\ }\bibfield  {title} {\enquote {\bibinfo {title} {Nonequilibrium optical conductivity: General theory and application to transient phases},}\ }\href {\doibase 10.1103/PhysRevB.96.054506} {\bibfield  {journal} {\bibinfo  {journal} {Phys. Rev. B}\ }\textbf {\bibinfo {volume} {96}},\ \bibinfo {pages} {054506} (\bibinfo {year} {2017}{\natexlab{b}})}\BibitemShut {NoStop}%
\bibitem [{\citenamefont {Murakami}\ \emph {et~al.}(2017)\citenamefont {Murakami}, \citenamefont {Tsuji}, \citenamefont {Eckstein},\ and\ \citenamefont {Werner}}]{PhysRevB.96.045125}%
  \BibitemOpen
  \bibfield  {author} {\bibinfo {author} {\bibfnamefont {Yuta}\ \bibnamefont {Murakami}}, \bibinfo {author} {\bibfnamefont {Naoto}\ \bibnamefont {Tsuji}}, \bibinfo {author} {\bibfnamefont {Martin}\ \bibnamefont {Eckstein}}, \ and\ \bibinfo {author} {\bibfnamefont {Philipp}\ \bibnamefont {Werner}},\ }\bibfield  {title} {\enquote {\bibinfo {title} {Nonequilibrium steady states and transient dynamics of conventional superconductors under phonon driving},}\ }\href {\doibase 10.1103/PhysRevB.96.045125} {\bibfield  {journal} {\bibinfo  {journal} {Phys. Rev. B}\ }\textbf {\bibinfo {volume} {96}},\ \bibinfo {pages} {045125} (\bibinfo {year} {2017})}\BibitemShut {NoStop}%
\bibitem [{\citenamefont {Wang}\ \emph {et~al.}(2018)\citenamefont {Wang}, \citenamefont {Chen}, \citenamefont {Moritz},\ and\ \citenamefont {Devereaux}}]{PhysRevLett.120.246402}%
  \BibitemOpen
  \bibfield  {author} {\bibinfo {author} {\bibfnamefont {Yao}\ \bibnamefont {Wang}}, \bibinfo {author} {\bibfnamefont {Cheng-Chien}\ \bibnamefont {Chen}}, \bibinfo {author} {\bibfnamefont {B.}~\bibnamefont {Moritz}}, \ and\ \bibinfo {author} {\bibfnamefont {T.~P.}\ \bibnamefont {Devereaux}},\ }\bibfield  {title} {\enquote {\bibinfo {title} {Light-enhanced spin fluctuations and $d$-wave superconductivity at a phase boundary},}\ }\href {\doibase 10.1103/PhysRevLett.120.246402} {\bibfield  {journal} {\bibinfo  {journal} {Phys. Rev. Lett.}\ }\textbf {\bibinfo {volume} {120}},\ \bibinfo {pages} {246402} (\bibinfo {year} {2018})}\BibitemShut {NoStop}%
\bibitem [{\citenamefont {H{\"u}bener}\ \emph {et~al.}(2018)\citenamefont {H{\"u}bener}, \citenamefont {De~Giovannini},\ and\ \citenamefont {Rubio}}]{Hubener2018-cd}%
  \BibitemOpen
  \bibfield  {author} {\bibinfo {author} {\bibfnamefont {Hannes}\ \bibnamefont {H{\"u}bener}}, \bibinfo {author} {\bibfnamefont {Umberto}\ \bibnamefont {De~Giovannini}}, \ and\ \bibinfo {author} {\bibfnamefont {Angel}\ \bibnamefont {Rubio}},\ }\bibfield  {title} {\enquote {\bibinfo {title} {Phonon driven floquet matter},}\ }\href {https://doi.org/10.1021/acs.nanolett.7b05391} {\bibfield  {journal} {\bibinfo  {journal} {Nano Lett.}\ }\textbf {\bibinfo {volume} {18}},\ \bibinfo {pages} {1535--1542} (\bibinfo {year} {2018})}\BibitemShut {NoStop}%
\bibitem [{\citenamefont {Paeckel}\ \emph {et~al.}(2020)\citenamefont {Paeckel}, \citenamefont {Fauseweh}, \citenamefont {Osterkorn}, \citenamefont {K\"ohler}, \citenamefont {Manske},\ and\ \citenamefont {Manmana}}]{Paeckel2020}%
  \BibitemOpen
  \bibfield  {author} {\bibinfo {author} {\bibfnamefont {S.}~\bibnamefont {Paeckel}}, \bibinfo {author} {\bibfnamefont {B.}~\bibnamefont {Fauseweh}}, \bibinfo {author} {\bibfnamefont {A.}~\bibnamefont {Osterkorn}}, \bibinfo {author} {\bibfnamefont {T.}~\bibnamefont {K\"ohler}}, \bibinfo {author} {\bibfnamefont {D.}~\bibnamefont {Manske}}, \ and\ \bibinfo {author} {\bibfnamefont {S.~R.}\ \bibnamefont {Manmana}},\ }\bibfield  {title} {\enquote {\bibinfo {title} {Detecting superconductivity out of equilibrium},}\ }\href {https://link.aps.org/doi/10.1103/PhysRevB.101.180507} {\bibfield  {journal} {\bibinfo  {journal} {Phys. Rev. B}\ }\textbf {\bibinfo {volume} {101}},\ \bibinfo {pages} {180507} (\bibinfo {year} {2020})}\BibitemShut {NoStop}%
\bibitem [{\citenamefont {Wang}\ \emph {et~al.}(2021)\citenamefont {Wang}, \citenamefont {Shi},\ and\ \citenamefont {Chen}}]{PhysRevX.11.041028}%
  \BibitemOpen
  \bibfield  {author} {\bibinfo {author} {\bibfnamefont {Yao}\ \bibnamefont {Wang}}, \bibinfo {author} {\bibfnamefont {Tao}\ \bibnamefont {Shi}}, \ and\ \bibinfo {author} {\bibfnamefont {Cheng-Chien}\ \bibnamefont {Chen}},\ }\bibfield  {title} {\enquote {\bibinfo {title} {Fluctuating nature of light-enhanced $d$-wave superconductivity: A time-dependent variational non-gaussian exact diagonalization study},}\ }\href {https://link.aps.org/doi/10.1103/PhysRevX.11.041028} {\bibfield  {journal} {\bibinfo  {journal} {Phys. Rev. X}\ }\textbf {\bibinfo {volume} {11}},\ \bibinfo {pages} {041028} (\bibinfo {year} {2021})}\BibitemShut {NoStop}%
\bibitem [{\citenamefont {Eckhardt}\ \emph {et~al.}(2024)\citenamefont {Eckhardt}, \citenamefont {Chattopadhyay}, \citenamefont {Kennes}, \citenamefont {Demler}, \citenamefont {Sentef},\ and\ \citenamefont {Michael}}]{eckhardt2023theory}%
  \BibitemOpen
  \bibfield  {author} {\bibinfo {author} {\bibfnamefont {Christian~J.}\ \bibnamefont {Eckhardt}}, \bibinfo {author} {\bibfnamefont {Sambuddha}\ \bibnamefont {Chattopadhyay}}, \bibinfo {author} {\bibfnamefont {Dante~M.}\ \bibnamefont {Kennes}}, \bibinfo {author} {\bibfnamefont {Eugene~A.}\ \bibnamefont {Demler}}, \bibinfo {author} {\bibfnamefont {Michael~A.}\ \bibnamefont {Sentef}}, \ and\ \bibinfo {author} {\bibfnamefont {Marios~H.}\ \bibnamefont {Michael}},\ }\bibfield  {title} {\enquote {\bibinfo {title} {Theory of resonantly enhanced photo-induced superconductivity},}\ }\href {\doibase DO - 10.1038/s41467-024-46632-x} {\bibfield  {journal} {\bibinfo  {journal} {Nat. Commun.}\ }\textbf {\bibinfo {volume} {15}},\ \bibinfo {pages} {2300} (\bibinfo {year} {2024})}\BibitemShut {NoStop}%
\bibitem [{\citenamefont {Trigo}\ \emph {et~al.}(2010)\citenamefont {Trigo}, \citenamefont {Chen}, \citenamefont {Vishwanath}, \citenamefont {Sheu}, \citenamefont {Graber}, \citenamefont {Henning},\ and\ \citenamefont {Reis}}]{Trigo2010imaging}%
  \BibitemOpen
  \bibfield  {author} {\bibinfo {author} {\bibfnamefont {M.}~\bibnamefont {Trigo}}, \bibinfo {author} {\bibfnamefont {J.}~\bibnamefont {Chen}}, \bibinfo {author} {\bibfnamefont {V.~H.}\ \bibnamefont {Vishwanath}}, \bibinfo {author} {\bibfnamefont {Y.~M.}\ \bibnamefont {Sheu}}, \bibinfo {author} {\bibfnamefont {T.}~\bibnamefont {Graber}}, \bibinfo {author} {\bibfnamefont {R.}~\bibnamefont {Henning}}, \ and\ \bibinfo {author} {\bibfnamefont {D.~A.}\ \bibnamefont {Reis}},\ }\bibfield  {title} {\enquote {\bibinfo {title} {Imaging nonequilibrium atomic vibrations with x-ray diffuse scattering},}\ }\href {https://link.aps.org/doi/10.1103/PhysRevB.82.235205} {\bibfield  {journal} {\bibinfo  {journal} {Phys. Rev. B}\ }\textbf {\bibinfo {volume} {82}},\ \bibinfo {pages} {235205} (\bibinfo {year} {2010})}\BibitemShut {NoStop}%
\bibitem [{\citenamefont {Trigo}\ \emph {et~al.}(2013)\citenamefont {Trigo}, \citenamefont {Fuchs}, \citenamefont {Chen}, \citenamefont {Jiang}, \citenamefont {Cammarata}, \citenamefont {Fahy}, \citenamefont {Fritz}, \citenamefont {Gaffney}, \citenamefont {Ghimire}, \citenamefont {Higginbotham}, \citenamefont {Johnson}, \citenamefont {Kozina}, \citenamefont {Larsson}, \citenamefont {Lemke}, \citenamefont {Lindenberg}, \citenamefont {Ndabashimiye}, \citenamefont {Quirin}, \citenamefont {Sokolowski-Tinten}, \citenamefont {Uher}, \citenamefont {Wang}, \citenamefont {Wark}, \citenamefont {Zhu},\ and\ \citenamefont {Reis}}]{Trigo2013fourier}%
  \BibitemOpen
  \bibfield  {author} {\bibinfo {author} {\bibfnamefont {M}~\bibnamefont {Trigo}}, \bibinfo {author} {\bibfnamefont {M}~\bibnamefont {Fuchs}}, \bibinfo {author} {\bibfnamefont {J}~\bibnamefont {Chen}}, \bibinfo {author} {\bibfnamefont {M~P}\ \bibnamefont {Jiang}}, \bibinfo {author} {\bibfnamefont {M}~\bibnamefont {Cammarata}}, \bibinfo {author} {\bibfnamefont {S}~\bibnamefont {Fahy}}, \bibinfo {author} {\bibfnamefont {D~M}\ \bibnamefont {Fritz}}, \bibinfo {author} {\bibfnamefont {K}~\bibnamefont {Gaffney}}, \bibinfo {author} {\bibfnamefont {S}~\bibnamefont {Ghimire}}, \bibinfo {author} {\bibfnamefont {A}~\bibnamefont {Higginbotham}}, \bibinfo {author} {\bibfnamefont {S~L}\ \bibnamefont {Johnson}}, \bibinfo {author} {\bibfnamefont {M~E}\ \bibnamefont {Kozina}}, \bibinfo {author} {\bibfnamefont {J}~\bibnamefont {Larsson}}, \bibinfo {author} {\bibfnamefont {H}~\bibnamefont {Lemke}}, \bibinfo {author} {\bibfnamefont {A~M}\ \bibnamefont {Lindenberg}}, \bibinfo {author} {\bibfnamefont {G}~\bibnamefont
  {Ndabashimiye}}, \bibinfo {author} {\bibfnamefont {F}~\bibnamefont {Quirin}}, \bibinfo {author} {\bibfnamefont {K}~\bibnamefont {Sokolowski-Tinten}}, \bibinfo {author} {\bibfnamefont {C}~\bibnamefont {Uher}}, \bibinfo {author} {\bibfnamefont {G}~\bibnamefont {Wang}}, \bibinfo {author} {\bibfnamefont {J~S}\ \bibnamefont {Wark}}, \bibinfo {author} {\bibfnamefont {D}~\bibnamefont {Zhu}}, \ and\ \bibinfo {author} {\bibfnamefont {D~A}\ \bibnamefont {Reis}},\ }\bibfield  {title} {\enquote {\bibinfo {title} {Fourier-transform inelastic x-ray scattering from time- and momentum-dependent phonon{\textendash}phonon correlations},}\ }\href {https://doi.org/10.1038/nphys2788} {\bibfield  {journal} {\bibinfo  {journal} {Nat. Phys.}\ }\textbf {\bibinfo {volume} {9}},\ \bibinfo {pages} {790--794} (\bibinfo {year} {2013})}\BibitemShut {NoStop}%
\bibitem [{\citenamefont {Konstantinova}\ \emph {et~al.}(2018)\citenamefont {Konstantinova}, \citenamefont {Rameau}, \citenamefont {Reid}, \citenamefont {Abdurazakov}, \citenamefont {Wu}, \citenamefont {Li}, \citenamefont {Shen}, \citenamefont {Gu}, \citenamefont {Huang}, \citenamefont {Rettig}, \citenamefont {Avigo}, \citenamefont {Ligges}, \citenamefont {Freericks}, \citenamefont {Kemper}, \citenamefont {D{\"u}rr}, \citenamefont {Bovensiepen}, \citenamefont {Johnson}, \citenamefont {Wang},\ and\ \citenamefont {Zhu}}]{Konstantinova2018nonequilbrium}%
  \BibitemOpen
  \bibfield  {author} {\bibinfo {author} {\bibfnamefont {Tatiana}\ \bibnamefont {Konstantinova}}, \bibinfo {author} {\bibfnamefont {Jonathan~D}\ \bibnamefont {Rameau}}, \bibinfo {author} {\bibfnamefont {Alexander~H}\ \bibnamefont {Reid}}, \bibinfo {author} {\bibfnamefont {Omadillo}\ \bibnamefont {Abdurazakov}}, \bibinfo {author} {\bibfnamefont {Lijun}\ \bibnamefont {Wu}}, \bibinfo {author} {\bibfnamefont {Renkai}\ \bibnamefont {Li}}, \bibinfo {author} {\bibfnamefont {Xiaozhe}\ \bibnamefont {Shen}}, \bibinfo {author} {\bibfnamefont {Genda}\ \bibnamefont {Gu}}, \bibinfo {author} {\bibfnamefont {Yuan}\ \bibnamefont {Huang}}, \bibinfo {author} {\bibfnamefont {Laurenz}\ \bibnamefont {Rettig}}, \bibinfo {author} {\bibfnamefont {Isabella}\ \bibnamefont {Avigo}}, \bibinfo {author} {\bibfnamefont {Manuel}\ \bibnamefont {Ligges}}, \bibinfo {author} {\bibfnamefont {James~K}\ \bibnamefont {Freericks}}, \bibinfo {author} {\bibfnamefont {Alexander~F}\ \bibnamefont {Kemper}}, \bibinfo {author} {\bibfnamefont {Hermann~A}\
  \bibnamefont {D{\"u}rr}}, \bibinfo {author} {\bibfnamefont {Uwe}\ \bibnamefont {Bovensiepen}}, \bibinfo {author} {\bibfnamefont {Peter~D}\ \bibnamefont {Johnson}}, \bibinfo {author} {\bibfnamefont {Xijie}\ \bibnamefont {Wang}}, \ and\ \bibinfo {author} {\bibfnamefont {Yimei}\ \bibnamefont {Zhu}},\ }\bibfield  {title} {\enquote {\bibinfo {title} {Nonequilibrium electron and lattice dynamics of strongly correlated bi2sr2cacu2o8+$\delta$ single crystals},}\ }\href {https://www.science.org/doi/abs/10.1126/sciadv.aap7427} {\bibfield  {journal} {\bibinfo  {journal} {Sci. Adv.}\ }\textbf {\bibinfo {volume} {4}},\ \bibinfo {pages} {eaap7427} (\bibinfo {year} {2018})}\BibitemShut {NoStop}%
\bibitem [{\citenamefont {K\"ohler}\ \emph {et~al.}(2021)\citenamefont {K\"ohler}, \citenamefont {Stolpp},\ and\ \citenamefont {Paeckel}}]{Koehler2021}%
  \BibitemOpen
  \bibfield  {author} {\bibinfo {author} {\bibfnamefont {T.}~\bibnamefont {K\"ohler}}, \bibinfo {author} {\bibfnamefont {J.}~\bibnamefont {Stolpp}}, \ and\ \bibinfo {author} {\bibfnamefont {S.}~\bibnamefont {Paeckel}},\ }\bibfield  {title} {\enquote {\bibinfo {title} {Efficient and flexible approach to simulate low-dimensional quantum lattice models with large local {H}ilbert spaces},}\ }\href {http://dx.doi.org/10.21468/SciPostPhys.10.3.058} {\bibfield  {journal} {\bibinfo  {journal} {SciPost Phys.}\ }\textbf {\bibinfo {volume} {10}} (\bibinfo {year} {2021})}\BibitemShut {NoStop}%
\bibitem [{\citenamefont {Stolpp}\ \emph {et~al.}(2021)\citenamefont {Stolpp}, \citenamefont {K{\"o}hler}, \citenamefont {Manmana}, \citenamefont {Jeckelmann}, \citenamefont {Heidrich-Meisner},\ and\ \citenamefont {Paeckel}}]{Stolpp2021}%
  \BibitemOpen
  \bibfield  {author} {\bibinfo {author} {\bibfnamefont {J.}~\bibnamefont {Stolpp}}, \bibinfo {author} {\bibfnamefont {T.}~\bibnamefont {K{\"o}hler}}, \bibinfo {author} {\bibfnamefont {S.~R.}\ \bibnamefont {Manmana}}, \bibinfo {author} {\bibfnamefont {E.}~\bibnamefont {Jeckelmann}}, \bibinfo {author} {\bibfnamefont {F.}~\bibnamefont {Heidrich-Meisner}}, \ and\ \bibinfo {author} {\bibfnamefont {S.}~\bibnamefont {Paeckel}},\ }\bibfield  {title} {\enquote {\bibinfo {title} {Comparative study of state-of-the-art matrix-product-state methods for lattice models with large local {H}ilbert spaces without {$\mathrm{U(1)}$} symmetry},}\ }\href {https://www.sciencedirect.com/science/article/pii/S0010465521002186} {\bibfield  {journal} {\bibinfo  {journal} {Comput. Phys. Commun.}\ }\textbf {\bibinfo {volume} {269}},\ \bibinfo {pages} {108106} (\bibinfo {year} {2021})}\BibitemShut {NoStop}%
\bibitem [{\citenamefont {Hu}\ \emph {et~al.}(2014)\citenamefont {Hu}, \citenamefont {Kaiser}, \citenamefont {Nicoletti}, \citenamefont {Hunt}, \citenamefont {Gierz}, \citenamefont {Hoffmann}, \citenamefont {Le~Tacon}, \citenamefont {Loew}, \citenamefont {Keimer},\ and\ \citenamefont {Cavalleri}}]{Hu2014optically}%
  \BibitemOpen
  \bibfield  {author} {\bibinfo {author} {\bibfnamefont {W.}~\bibnamefont {Hu}}, \bibinfo {author} {\bibfnamefont {S.}~\bibnamefont {Kaiser}}, \bibinfo {author} {\bibfnamefont {D.}~\bibnamefont {Nicoletti}}, \bibinfo {author} {\bibfnamefont {C.~R.}\ \bibnamefont {Hunt}}, \bibinfo {author} {\bibfnamefont {I.}~\bibnamefont {Gierz}}, \bibinfo {author} {\bibfnamefont {M.~C.}\ \bibnamefont {Hoffmann}}, \bibinfo {author} {\bibfnamefont {M.}~\bibnamefont {Le~Tacon}}, \bibinfo {author} {\bibfnamefont {T.}~\bibnamefont {Loew}}, \bibinfo {author} {\bibfnamefont {B.}~\bibnamefont {Keimer}}, \ and\ \bibinfo {author} {\bibfnamefont {A.}~\bibnamefont {Cavalleri}},\ }\bibfield  {title} {\enquote {\bibinfo {title} {{Optically enhanced coherent transport in {YBa$_2$Cu$_3$O$_{6.5}$} by ultrafast redistribution of interlayer coupling}},}\ }\href {https://doi.org/10.1038/nmat3963} {\bibfield  {journal} {\bibinfo  {journal} {Nat. Mater.}\ }\textbf {\bibinfo {volume} {13}},\ \bibinfo {pages} {705} (\bibinfo {year}
  {2014})}\BibitemShut {NoStop}%
\bibitem [{\citenamefont {Kaiser}\ \emph {et~al.}(2014)\citenamefont {Kaiser}, \citenamefont {Hunt}, \citenamefont {Nicoletti}, \citenamefont {Hu}, \citenamefont {Gierz}, \citenamefont {Liu}, \citenamefont {Le~Tacon}, \citenamefont {Loew}, \citenamefont {Haug},\ and\ \citenamefont {Keimer}}]{Kaiser2014optically}%
  \BibitemOpen
  \bibfield  {author} {\bibinfo {author} {\bibfnamefont {S}~\bibnamefont {Kaiser}}, \bibinfo {author} {\bibfnamefont {C~R}\ \bibnamefont {Hunt}}, \bibinfo {author} {\bibfnamefont {D}~\bibnamefont {Nicoletti}}, \bibinfo {author} {\bibfnamefont {W}~\bibnamefont {Hu}}, \bibinfo {author} {\bibfnamefont {I}~\bibnamefont {Gierz}}, \bibinfo {author} {\bibfnamefont {H~Y}\ \bibnamefont {Liu}}, \bibinfo {author} {\bibfnamefont {M}~\bibnamefont {Le~Tacon}}, \bibinfo {author} {\bibfnamefont {T}~\bibnamefont {Loew}}, \bibinfo {author} {\bibfnamefont {D}~\bibnamefont {Haug}}, \ and\ \bibinfo {author} {\bibfnamefont {B}~\bibnamefont {Keimer}},\ }\bibfield  {title} {\enquote {\bibinfo {title} {{Optically induced coherent transport far above {$T_c$} in underdoped {YBa$_2$Cu$_3$O$_{6+\delta}$}}},}\ }\href {https://link.aps.org/doi/10.1103/PhysRevB.89.184516} {\bibfield  {journal} {\bibinfo  {journal} {Phys. Rev. B}\ }\textbf {\bibinfo {volume} {89}},\ \bibinfo {pages} {184516} (\bibinfo {year} {2014})}\BibitemShut {NoStop}%
\bibitem [{\citenamefont {Buzzi}\ \emph {et~al.}(2020)\citenamefont {Buzzi}, \citenamefont {Nicoletti}, \citenamefont {Fechner}, \citenamefont {Tancogne-Dejean}, \citenamefont {Sentef}, \citenamefont {Georges}, \citenamefont {Biesner}, \citenamefont {Uykur}, \citenamefont {Dressel}, \citenamefont {Henderson}, \citenamefont {Siegrist}, \citenamefont {Schlueter}, \citenamefont {Miyagawa}, \citenamefont {Kanoda}, \citenamefont {Nam}, \citenamefont {Ardavan}, \citenamefont {Coulthard}, \citenamefont {Tindall}, \citenamefont {Schlawin}, \citenamefont {Jaksch},\ and\ \citenamefont {Cavalleri}}]{Buzzi2020photomolecular}%
  \BibitemOpen
  \bibfield  {author} {\bibinfo {author} {\bibfnamefont {M.}~\bibnamefont {Buzzi}}, \bibinfo {author} {\bibfnamefont {D.}~\bibnamefont {Nicoletti}}, \bibinfo {author} {\bibfnamefont {M.}~\bibnamefont {Fechner}}, \bibinfo {author} {\bibfnamefont {N.}~\bibnamefont {Tancogne-Dejean}}, \bibinfo {author} {\bibfnamefont {M.~A.}\ \bibnamefont {Sentef}}, \bibinfo {author} {\bibfnamefont {A.}~\bibnamefont {Georges}}, \bibinfo {author} {\bibfnamefont {T.}~\bibnamefont {Biesner}}, \bibinfo {author} {\bibfnamefont {E.}~\bibnamefont {Uykur}}, \bibinfo {author} {\bibfnamefont {M.}~\bibnamefont {Dressel}}, \bibinfo {author} {\bibfnamefont {A.}~\bibnamefont {Henderson}}, \bibinfo {author} {\bibfnamefont {T.}~\bibnamefont {Siegrist}}, \bibinfo {author} {\bibfnamefont {J.~A.}\ \bibnamefont {Schlueter}}, \bibinfo {author} {\bibfnamefont {K.}~\bibnamefont {Miyagawa}}, \bibinfo {author} {\bibfnamefont {K.}~\bibnamefont {Kanoda}}, \bibinfo {author} {\bibfnamefont {M.-S.}\ \bibnamefont {Nam}}, \bibinfo {author} {\bibfnamefont
  {A.}~\bibnamefont {Ardavan}}, \bibinfo {author} {\bibfnamefont {J.}~\bibnamefont {Coulthard}}, \bibinfo {author} {\bibfnamefont {J.}~\bibnamefont {Tindall}}, \bibinfo {author} {\bibfnamefont {F.}~\bibnamefont {Schlawin}}, \bibinfo {author} {\bibfnamefont {D.}~\bibnamefont {Jaksch}}, \ and\ \bibinfo {author} {\bibfnamefont {A.}~\bibnamefont {Cavalleri}},\ }\bibfield  {title} {\enquote {\bibinfo {title} {Photomolecular high-temperature superconductivity},}\ }\href {https://link.aps.org/doi/10.1103/PhysRevX.10.031028} {\bibfield  {journal} {\bibinfo  {journal} {Phys. Rev. X}\ }\textbf {\bibinfo {volume} {10}},\ \bibinfo {pages} {031028} (\bibinfo {year} {2020})}\BibitemShut {NoStop}%
\bibitem [{\citenamefont {Fechner}\ \emph {et~al.}(2024)\citenamefont {Fechner}, \citenamefont {F{\"o}rst}, \citenamefont {Orenstein}, \citenamefont {Krapivin}, \citenamefont {Disa}, \citenamefont {Buzzi}, \citenamefont {von Hoegen}, \citenamefont {de~la Pena}, \citenamefont {Nguyen}, \citenamefont {Mankowsky}, \citenamefont {Sander}, \citenamefont {Lemke}, \citenamefont {Deng}, \citenamefont {Trigo},\ and\ \citenamefont {Cavalleri}}]{Fechner2024-gk}%
  \BibitemOpen
  \bibfield  {author} {\bibinfo {author} {\bibfnamefont {M}~\bibnamefont {Fechner}}, \bibinfo {author} {\bibfnamefont {M}~\bibnamefont {F{\"o}rst}}, \bibinfo {author} {\bibfnamefont {G}~\bibnamefont {Orenstein}}, \bibinfo {author} {\bibfnamefont {V}~\bibnamefont {Krapivin}}, \bibinfo {author} {\bibfnamefont {A~S}\ \bibnamefont {Disa}}, \bibinfo {author} {\bibfnamefont {M}~\bibnamefont {Buzzi}}, \bibinfo {author} {\bibfnamefont {A}~\bibnamefont {von Hoegen}}, \bibinfo {author} {\bibfnamefont {G}~\bibnamefont {de~la Pena}}, \bibinfo {author} {\bibfnamefont {Q~L}\ \bibnamefont {Nguyen}}, \bibinfo {author} {\bibfnamefont {R}~\bibnamefont {Mankowsky}}, \bibinfo {author} {\bibfnamefont {M}~\bibnamefont {Sander}}, \bibinfo {author} {\bibfnamefont {H}~\bibnamefont {Lemke}}, \bibinfo {author} {\bibfnamefont {Y}~\bibnamefont {Deng}}, \bibinfo {author} {\bibfnamefont {M}~\bibnamefont {Trigo}}, \ and\ \bibinfo {author} {\bibfnamefont {A}~\bibnamefont {Cavalleri}},\ }\bibfield  {title} {\enquote {\bibinfo {title}
  {Quenched lattice fluctuations in optically driven {SrTiO3}},}\ }\href {https://doi.org/10.1038/s41563-023-01791-y} {\bibfield  {journal} {\bibinfo  {journal} {Nat. Mater.}\ }\textbf {\bibinfo {volume} {23}},\ \bibinfo {pages} {363--368} (\bibinfo {year} {2024})}\BibitemShut {NoStop}%
\bibitem [{\citenamefont {Disa}\ \emph {et~al.}(2023)\citenamefont {Disa}, \citenamefont {Curtis}, \citenamefont {Fechner}, \citenamefont {Liu}, \citenamefont {von Hoegen}, \citenamefont {F{\"o}rst}, \citenamefont {Nova}, \citenamefont {Narang}, \citenamefont {Maljuk}, \citenamefont {Boris}, \citenamefont {Keimer},\ and\ \citenamefont {Cavalleri}}]{Disa2023-wq}%
  \BibitemOpen
  \bibfield  {author} {\bibinfo {author} {\bibfnamefont {A~S}\ \bibnamefont {Disa}}, \bibinfo {author} {\bibfnamefont {J}~\bibnamefont {Curtis}}, \bibinfo {author} {\bibfnamefont {M}~\bibnamefont {Fechner}}, \bibinfo {author} {\bibfnamefont {A}~\bibnamefont {Liu}}, \bibinfo {author} {\bibfnamefont {A}~\bibnamefont {von Hoegen}}, \bibinfo {author} {\bibfnamefont {M}~\bibnamefont {F{\"o}rst}}, \bibinfo {author} {\bibfnamefont {T~F}\ \bibnamefont {Nova}}, \bibinfo {author} {\bibfnamefont {P}~\bibnamefont {Narang}}, \bibinfo {author} {\bibfnamefont {A}~\bibnamefont {Maljuk}}, \bibinfo {author} {\bibfnamefont {A~V}\ \bibnamefont {Boris}}, \bibinfo {author} {\bibfnamefont {B}~\bibnamefont {Keimer}}, \ and\ \bibinfo {author} {\bibfnamefont {A}~\bibnamefont {Cavalleri}},\ }\bibfield  {title} {\enquote {\bibinfo {title} {Photo-induced high-temperature ferromagnetism in {YTiO3}},}\ }\href {https://doi.org/10.1038/s41586-023-05853-8} {\bibfield  {journal} {\bibinfo  {journal} {Nature}\ }\textbf {\bibinfo {volume}
  {617}},\ \bibinfo {pages} {73--78} (\bibinfo {year} {2023})}\BibitemShut {NoStop}%
\bibitem [{\citenamefont {Liu}\ \emph {et~al.}(2020)\citenamefont {Liu}, \citenamefont {F\"orst}, \citenamefont {Fechner}, \citenamefont {Nicoletti}, \citenamefont {Porras}, \citenamefont {Loew}, \citenamefont {Keimer},\ and\ \citenamefont {Cavalleri}}]{Liu2020pump}%
  \BibitemOpen
  \bibfield  {author} {\bibinfo {author} {\bibfnamefont {B.}~\bibnamefont {Liu}}, \bibinfo {author} {\bibfnamefont {M.}~\bibnamefont {F\"orst}}, \bibinfo {author} {\bibfnamefont {M.}~\bibnamefont {Fechner}}, \bibinfo {author} {\bibfnamefont {D.}~\bibnamefont {Nicoletti}}, \bibinfo {author} {\bibfnamefont {J.}~\bibnamefont {Porras}}, \bibinfo {author} {\bibfnamefont {T.}~\bibnamefont {Loew}}, \bibinfo {author} {\bibfnamefont {B.}~\bibnamefont {Keimer}}, \ and\ \bibinfo {author} {\bibfnamefont {A.}~\bibnamefont {Cavalleri}},\ }\bibfield  {title} {\enquote {\bibinfo {title} {Pump frequency resonances for light-induced incipient superconductivity in ${\mathrm{yba}}_{2}{\mathrm{cu}}_{3}{\mathrm{o}}_{6.5}$},}\ }\href {https://link.aps.org/doi/10.1103/PhysRevX.10.011053} {\bibfield  {journal} {\bibinfo  {journal} {Phys. Rev. X}\ }\textbf {\bibinfo {volume} {10}},\ \bibinfo {pages} {011053} (\bibinfo {year} {2020})}\BibitemShut {NoStop}%
\bibitem [{\citenamefont {Rowe}\ \emph {et~al.}(2023)\citenamefont {Rowe}, \citenamefont {Yuan}, \citenamefont {Buzzi}, \citenamefont {Jotzu}, \citenamefont {Zhu}, \citenamefont {Fechner}, \citenamefont {F{\"o}rst}, \citenamefont {Liu}, \citenamefont {Pontiroli}, \citenamefont {Ricc{\`o}},\ and\ \citenamefont {Cavalleri}}]{Rowe2023resonant}%
  \BibitemOpen
  \bibfield  {author} {\bibinfo {author} {\bibfnamefont {E.}~\bibnamefont {Rowe}}, \bibinfo {author} {\bibfnamefont {B.}~\bibnamefont {Yuan}}, \bibinfo {author} {\bibfnamefont {M.}~\bibnamefont {Buzzi}}, \bibinfo {author} {\bibfnamefont {G.}~\bibnamefont {Jotzu}}, \bibinfo {author} {\bibfnamefont {Y.}~\bibnamefont {Zhu}}, \bibinfo {author} {\bibfnamefont {M.}~\bibnamefont {Fechner}}, \bibinfo {author} {\bibfnamefont {M.}~\bibnamefont {F{\"o}rst}}, \bibinfo {author} {\bibfnamefont {B.}~\bibnamefont {Liu}}, \bibinfo {author} {\bibfnamefont {D.}~\bibnamefont {Pontiroli}}, \bibinfo {author} {\bibfnamefont {M.}~\bibnamefont {Ricc{\`o}}}, \ and\ \bibinfo {author} {\bibfnamefont {A.}~\bibnamefont {Cavalleri}},\ }\bibfield  {title} {\enquote {\bibinfo {title} {{Resonant enhancement of photo-induced superconductivity in K$_3$C$_{60}$}},}\ }\href {https://doi.org/10.1038/s41567-023-02235-9} {\bibfield  {journal} {\bibinfo  {journal} {Nat. Phys.}\ }\textbf {\bibinfo {volume} {19}},\ \bibinfo {pages} {1821--1826}
  (\bibinfo {year} {2023})}\BibitemShut {NoStop}%
\bibitem [{\citenamefont {Mankowsky}\ \emph {et~al.}(2014)\citenamefont {Mankowsky}, \citenamefont {Subedi}, \citenamefont {F{\"o}rst}, \citenamefont {Mariager}, \citenamefont {Chollet}, \citenamefont {Lemke}, \citenamefont {Robinson}, \citenamefont {Glownia}, \citenamefont {Minitti}, \citenamefont {Frano}, \citenamefont {Fechner}, \citenamefont {Spaldin}, \citenamefont {Loew}, \citenamefont {Keimer}, \citenamefont {Georges},\ and\ \citenamefont {Cavalleri}}]{Mankowsky2014nonlinear}%
  \BibitemOpen
  \bibfield  {author} {\bibinfo {author} {\bibfnamefont {R.}~\bibnamefont {Mankowsky}}, \bibinfo {author} {\bibfnamefont {A.}~\bibnamefont {Subedi}}, \bibinfo {author} {\bibfnamefont {M.}~\bibnamefont {F{\"o}rst}}, \bibinfo {author} {\bibfnamefont {S.~O.}\ \bibnamefont {Mariager}}, \bibinfo {author} {\bibfnamefont {M.}~\bibnamefont {Chollet}}, \bibinfo {author} {\bibfnamefont {H.~T.}\ \bibnamefont {Lemke}}, \bibinfo {author} {\bibfnamefont {J.~S.}\ \bibnamefont {Robinson}}, \bibinfo {author} {\bibfnamefont {J.~M.}\ \bibnamefont {Glownia}}, \bibinfo {author} {\bibfnamefont {M.~P.}\ \bibnamefont {Minitti}}, \bibinfo {author} {\bibfnamefont {A.}~\bibnamefont {Frano}}, \bibinfo {author} {\bibfnamefont {M.}~\bibnamefont {Fechner}}, \bibinfo {author} {\bibfnamefont {N.~A.}\ \bibnamefont {Spaldin}}, \bibinfo {author} {\bibfnamefont {T.}~\bibnamefont {Loew}}, \bibinfo {author} {\bibfnamefont {B.}~\bibnamefont {Keimer}}, \bibinfo {author} {\bibfnamefont {A.}~\bibnamefont {Georges}}, \ and\ \bibinfo {author}
  {\bibfnamefont {A.}~\bibnamefont {Cavalleri}},\ }\bibfield  {title} {\enquote {\bibinfo {title} {{Nonlinear lattice dynamics as a basis for enhanced superconductivity in YBa$_2$Cu$_3$O$_6.5$}},}\ }\href {https://doi.org/10.1038/nature13875} {\bibfield  {journal} {\bibinfo  {journal} {Nature}\ }\textbf {\bibinfo {volume} {516}},\ \bibinfo {pages} {71--73} (\bibinfo {year} {2014})}\BibitemShut {NoStop}%
\bibitem [{\citenamefont {Mankowsky}\ \emph {et~al.}(2015)\citenamefont {Mankowsky}, \citenamefont {F{\"o}rst}, \citenamefont {Loew}, \citenamefont {Porras}, \citenamefont {Keimer},\ and\ \citenamefont {Cavalleri}}]{Mankowsky2015coherent}%
  \BibitemOpen
  \bibfield  {author} {\bibinfo {author} {\bibfnamefont {R}~\bibnamefont {Mankowsky}}, \bibinfo {author} {\bibfnamefont {M}~\bibnamefont {F{\"o}rst}}, \bibinfo {author} {\bibfnamefont {T}~\bibnamefont {Loew}}, \bibinfo {author} {\bibfnamefont {J}~\bibnamefont {Porras}}, \bibinfo {author} {\bibfnamefont {B}~\bibnamefont {Keimer}}, \ and\ \bibinfo {author} {\bibfnamefont {A}~\bibnamefont {Cavalleri}},\ }\bibfield  {title} {\enquote {\bibinfo {title} {{Coherent modulation of the YBa$_2$Cu$_3$O$_{6+x}$ atomic structure by displacive stimulated ionic Raman scattering}},}\ }\href@noop {} {\bibfield  {journal} {\bibinfo  {journal} {Phys. Rev. B}\ }\textbf {\bibinfo {volume} {91}},\ \bibinfo {pages} {094308} (\bibinfo {year} {2015})}\BibitemShut {NoStop}%
\bibitem [{\citenamefont {Xu}\ and\ \citenamefont {Chiang}(2005)}]{Xu2005determination}%
  \BibitemOpen
  \bibfield  {author} {\bibinfo {author} {\bibfnamefont {Ruqing}\ \bibnamefont {Xu}}\ and\ \bibinfo {author} {\bibfnamefont {Tai~C.}\ \bibnamefont {Chiang}},\ }\bibfield  {title} {\enquote {\bibinfo {title} {Determination of phonon dispersion relations by x-ray thermal diffuse scattering},}\ }\href {https://doi.org/10.1524/zkri.2005.220.12.1009} {\bibfield  {journal} {\bibinfo  {journal} {Z. Kristallogr. - Cryst. Mater.}\ }\textbf {\bibinfo {volume} {220}},\ \bibinfo {pages} {1009--1016} (\bibinfo {year} {2005})}\BibitemShut {NoStop}%
\bibitem [{\citenamefont {White}(1992)}]{white_1992}%
  \BibitemOpen
  \bibfield  {author} {\bibinfo {author} {\bibfnamefont {S.~R.}\ \bibnamefont {White}},\ }\bibfield  {title} {\enquote {\bibinfo {title} {Density matrix formulation for quantum renormalization groups},}\ }\href {https://link.aps.org/doi/10.1103/PhysRevLett.69.2863} {\bibfield  {journal} {\bibinfo  {journal} {Phys. Rev. Lett.}\ }\textbf {\bibinfo {volume} {69}},\ \bibinfo {pages} {2863} (\bibinfo {year} {1992})}\BibitemShut {NoStop}%
\bibitem [{\citenamefont {White}(1993)}]{white_1993}%
  \BibitemOpen
  \bibfield  {author} {\bibinfo {author} {\bibfnamefont {S.~R.}\ \bibnamefont {White}},\ }\bibfield  {title} {\enquote {\bibinfo {title} {Density-matrix algorithms for quantum renormalization groups},}\ }\href {https://link.aps.org/doi/10.1103/PhysRevB.48.10345} {\bibfield  {journal} {\bibinfo  {journal} {Phys. Rev. B}\ }\textbf {\bibinfo {volume} {48}},\ \bibinfo {pages} {10345} (\bibinfo {year} {1993})}\BibitemShut {NoStop}%
\bibitem [{\citenamefont {Schollw\"ock}(2005)}]{schollwoeck_2005}%
  \BibitemOpen
  \bibfield  {author} {\bibinfo {author} {\bibfnamefont {U.}~\bibnamefont {Schollw\"ock}},\ }\bibfield  {title} {\enquote {\bibinfo {title} {The density-matrix renormalization group},}\ }\href {https://link.aps.org/doi/10.1103/RevModPhys.77.259} {\bibfield  {journal} {\bibinfo  {journal} {Rev. Mod. Phys.}\ }\textbf {\bibinfo {volume} {77}},\ \bibinfo {pages} {259} (\bibinfo {year} {2005})}\BibitemShut {NoStop}%
\bibitem [{\citenamefont {Schollwöck}(2011)}]{schollwoeck_2011}%
  \BibitemOpen
  \bibfield  {author} {\bibinfo {author} {\bibfnamefont {U.}~\bibnamefont {Schollwöck}},\ }\bibfield  {title} {\enquote {\bibinfo {title} {The density-matrix renormalization group in the age of matrix product states},}\ }\href {https://www.sciencedirect.com/science/article/pii/S0003491610001752} {\bibfield  {journal} {\bibinfo  {journal} {Ann. Phys.}\ }\textbf {\bibinfo {volume} {326}},\ \bibinfo {pages} {96} (\bibinfo {year} {2011})}\BibitemShut {NoStop}%
\bibitem [{\citenamefont {Ferris}\ and\ \citenamefont {Vidal}(2012)}]{Ferris2012}%
  \BibitemOpen
  \bibfield  {author} {\bibinfo {author} {\bibfnamefont {A.~J.}\ \bibnamefont {Ferris}}\ and\ \bibinfo {author} {\bibfnamefont {G.}~\bibnamefont {Vidal}},\ }\bibfield  {title} {\enquote {\bibinfo {title} {Perfect sampling with unitary tensor networks},}\ }\href {https://doi.org/10.1103%2Fphysrevb.85.165146} {\bibfield  {journal} {\bibinfo  {journal} {Phys. Rev. B}\ }\textbf {\bibinfo {volume} {85}} (\bibinfo {year} {2012})}\BibitemShut {NoStop}%
\bibitem [{\citenamefont {Bohrdt}\ \emph {et~al.}(2019)\citenamefont {Bohrdt}, \citenamefont {Chiu}, \citenamefont {Ji}, \citenamefont {Xu}, \citenamefont {Greif}, \citenamefont {Greiner}, \citenamefont {Demler}, \citenamefont {Grusdt},\ and\ \citenamefont {Knap}}]{Bohrdt2019}%
  \BibitemOpen
  \bibfield  {author} {\bibinfo {author} {\bibfnamefont {A.}~\bibnamefont {Bohrdt}}, \bibinfo {author} {\bibfnamefont {C.~S.}\ \bibnamefont {Chiu}}, \bibinfo {author} {\bibfnamefont {G.}~\bibnamefont {Ji}}, \bibinfo {author} {\bibfnamefont {M.}~\bibnamefont {Xu}}, \bibinfo {author} {\bibfnamefont {D.}~\bibnamefont {Greif}}, \bibinfo {author} {\bibfnamefont {M.}~\bibnamefont {Greiner}}, \bibinfo {author} {\bibfnamefont {E.}~\bibnamefont {Demler}}, \bibinfo {author} {\bibfnamefont {F.}~\bibnamefont {Grusdt}}, \ and\ \bibinfo {author} {\bibfnamefont {M.}~\bibnamefont {Knap}},\ }\bibfield  {title} {\enquote {\bibinfo {title} {Classifying snapshots of the doped {H}ubbard model with machine learning},}\ }\href {https://doi.org/10.1038%2Fs41567-019-0565-x} {\bibfield  {journal} {\bibinfo  {journal} {Nat. Phys.}\ }\textbf {\bibinfo {volume} {15}},\ \bibinfo {pages} {921} (\bibinfo {year} {2019})}\BibitemShut {NoStop}%
\bibitem [{\citenamefont {Bohrdt}\ \emph {et~al.}(2021{\natexlab{a}})\citenamefont {Bohrdt}, \citenamefont {Kim}, \citenamefont {Lukin}, \citenamefont {Rispoli}, \citenamefont {Schittko}, \citenamefont {Knap}, \citenamefont {Greiner},\ and\ \citenamefont {L\'eonard}}]{Bohrdt2021}%
  \BibitemOpen
  \bibfield  {author} {\bibinfo {author} {\bibfnamefont {A.}~\bibnamefont {Bohrdt}}, \bibinfo {author} {\bibfnamefont {S.}~\bibnamefont {Kim}}, \bibinfo {author} {\bibfnamefont {A.}~\bibnamefont {Lukin}}, \bibinfo {author} {\bibfnamefont {M.}~\bibnamefont {Rispoli}}, \bibinfo {author} {\bibfnamefont {R.}~\bibnamefont {Schittko}}, \bibinfo {author} {\bibfnamefont {M.}~\bibnamefont {Knap}}, \bibinfo {author} {\bibfnamefont {M.}~\bibnamefont {Greiner}}, \ and\ \bibinfo {author} {\bibfnamefont {J.}~\bibnamefont {L\'eonard}},\ }\bibfield  {title} {\enquote {\bibinfo {title} {Analyzing nonequilibrium quantum states through snapshots with artificial neural networks},}\ }\href {https://link.aps.org/doi/10.1103/PhysRevLett.127.150504} {\bibfield  {journal} {\bibinfo  {journal} {Phys. Rev. Lett.}\ }\textbf {\bibinfo {volume} {127}},\ \bibinfo {pages} {150504} (\bibinfo {year} {2021}{\natexlab{a}})}\BibitemShut {NoStop}%
\bibitem [{\citenamefont {Bohrdt}\ \emph {et~al.}(2021{\natexlab{b}})\citenamefont {Bohrdt}, \citenamefont {Homeier}, \citenamefont {Reinmoser}, \citenamefont {Demler},\ and\ \citenamefont {Grusdt}}]{Bohrdt2021b}%
  \BibitemOpen
  \bibfield  {author} {\bibinfo {author} {\bibfnamefont {Annabelle}\ \bibnamefont {Bohrdt}}, \bibinfo {author} {\bibfnamefont {Lukas}\ \bibnamefont {Homeier}}, \bibinfo {author} {\bibfnamefont {Christian}\ \bibnamefont {Reinmoser}}, \bibinfo {author} {\bibfnamefont {Eugene}\ \bibnamefont {Demler}}, \ and\ \bibinfo {author} {\bibfnamefont {Fabian}\ \bibnamefont {Grusdt}},\ }\bibfield  {title} {\enquote {\bibinfo {title} {Exploration of doped quantum magnets with ultracold atoms},}\ }\href {https://www.sciencedirect.com/science/article/pii/S0003491621002578} {\bibfield  {journal} {\bibinfo  {journal} {Ann. Phys.}\ }\textbf {\bibinfo {volume} {435}},\ \bibinfo {pages} {168651} (\bibinfo {year} {2021}{\natexlab{b}})}\BibitemShut {NoStop}%
\bibitem [{\citenamefont {Buser}\ \emph {et~al.}(2022)\citenamefont {Buser}, \citenamefont {Schollwöck},\ and\ \citenamefont {Grusdt}}]{Buser2022}%
  \BibitemOpen
  \bibfield  {author} {\bibinfo {author} {\bibfnamefont {Maximilian}\ \bibnamefont {Buser}}, \bibinfo {author} {\bibfnamefont {Ulrich}\ \bibnamefont {Schollwöck}}, \ and\ \bibinfo {author} {\bibfnamefont {Fabian}\ \bibnamefont {Grusdt}},\ }\bibfield  {title} {\enquote {\bibinfo {title} {Snapshot-based characterization of particle currents and the hall response in synthetic flux lattices},}\ }\href {https://doi.org/10.1103%2Fphysreva.105.033303} {\bibfield  {journal} {\bibinfo  {journal} {Phys. Rev. A}\ }\textbf {\bibinfo {volume} {105}} (\bibinfo {year} {2022})}\BibitemShut {NoStop}%
\bibitem [{\citenamefont {Palm}\ \emph {et~al.}(2022)\citenamefont {Palm}, \citenamefont {Mardazad}, \citenamefont {Bohrdt}, \citenamefont {Schollwöck},\ and\ \citenamefont {Grusdt}}]{Palm2022}%
  \BibitemOpen
  \bibfield  {author} {\bibinfo {author} {\bibfnamefont {F.~A.}\ \bibnamefont {Palm}}, \bibinfo {author} {\bibfnamefont {S.}~\bibnamefont {Mardazad}}, \bibinfo {author} {\bibfnamefont {A.}~\bibnamefont {Bohrdt}}, \bibinfo {author} {\bibfnamefont {U.}~\bibnamefont {Schollwöck}}, \ and\ \bibinfo {author} {\bibfnamefont {F.}~\bibnamefont {Grusdt}},\ }\bibfield  {title} {\enquote {\bibinfo {title} {Snapshot-based detection of $\ensuremath{\nu}=\frac{1}{2}$ {L}aughlin states: {C}oupled chains and central charge},}\ }\href {https://link.aps.org/doi/10.1103/PhysRevB.106.L081108} {\bibfield  {journal} {\bibinfo  {journal} {Phys. Rev. B}\ }\textbf {\bibinfo {volume} {106}},\ \bibinfo {pages} {L081108} (\bibinfo {year} {2022})}\BibitemShut {NoStop}%
\bibitem [{\citenamefont {Pauw}\ \emph {et~al.}(2024)\citenamefont {Pauw}, \citenamefont {Palm}, \citenamefont {Schollw\"ock}, \citenamefont {Bohrdt}, \citenamefont {Paeckel},\ and\ \citenamefont {Grusdt}}]{pauw2023}%
  \BibitemOpen
  \bibfield  {author} {\bibinfo {author} {\bibfnamefont {F.}~\bibnamefont {Pauw}}, \bibinfo {author} {\bibfnamefont {F.~A.}\ \bibnamefont {Palm}}, \bibinfo {author} {\bibfnamefont {U.}~\bibnamefont {Schollw\"ock}}, \bibinfo {author} {\bibfnamefont {A.}~\bibnamefont {Bohrdt}}, \bibinfo {author} {\bibfnamefont {S.}~\bibnamefont {Paeckel}}, \ and\ \bibinfo {author} {\bibfnamefont {F.}~\bibnamefont {Grusdt}},\ }\bibfield  {title} {\enquote {\bibinfo {title} {Detecting hidden order in fractional chern insulators},}\ }\href {\doibase 10.1103/PhysRevResearch.6.023180} {\bibfield  {journal} {\bibinfo  {journal} {Phys. Rev. Res.}\ }\textbf {\bibinfo {volume} {6}},\ \bibinfo {pages} {023180} (\bibinfo {year} {2024})}\BibitemShut {NoStop}%
\bibitem [{\citenamefont {Hirthe}\ \emph {et~al.}(2023)\citenamefont {Hirthe}, \citenamefont {Chalopin}, \citenamefont {Bourgund}, \citenamefont {Bojovi{\'c}}, \citenamefont {Bohrdt}, \citenamefont {Demler}, \citenamefont {Grusdt}, \citenamefont {Bloch},\ and\ \citenamefont {Hilker}}]{Hirthe2023}%
  \BibitemOpen
  \bibfield  {author} {\bibinfo {author} {\bibfnamefont {Sarah}\ \bibnamefont {Hirthe}}, \bibinfo {author} {\bibfnamefont {Thomas}\ \bibnamefont {Chalopin}}, \bibinfo {author} {\bibfnamefont {Dominik}\ \bibnamefont {Bourgund}}, \bibinfo {author} {\bibfnamefont {Petar}\ \bibnamefont {Bojovi{\'c}}}, \bibinfo {author} {\bibfnamefont {Annabelle}\ \bibnamefont {Bohrdt}}, \bibinfo {author} {\bibfnamefont {Eugene}\ \bibnamefont {Demler}}, \bibinfo {author} {\bibfnamefont {Fabian}\ \bibnamefont {Grusdt}}, \bibinfo {author} {\bibfnamefont {Immanuel}\ \bibnamefont {Bloch}}, \ and\ \bibinfo {author} {\bibfnamefont {Timon~A.}\ \bibnamefont {Hilker}},\ }\bibfield  {title} {\enquote {\bibinfo {title} {Magnetically mediated hole pairing in fermionic ladders of ultracold atoms},}\ }\href {https://doi.org/10.1038/s41586-022-05437-y} {\bibfield  {journal} {\bibinfo  {journal} {Nature}\ }\textbf {\bibinfo {volume} {613}},\ \bibinfo {pages} {463--467} (\bibinfo {year} {2023})}\BibitemShut {NoStop}%
\bibitem [{\citenamefont {Mankowsky}\ \emph {et~al.}(2016)\citenamefont {Mankowsky}, \citenamefont {F{\"o}rst},\ and\ \citenamefont {Cavalleri}}]{Mankowsky2016}%
  \BibitemOpen
  \bibfield  {author} {\bibinfo {author} {\bibfnamefont {Roman}\ \bibnamefont {Mankowsky}}, \bibinfo {author} {\bibfnamefont {Michael}\ \bibnamefont {F{\"o}rst}}, \ and\ \bibinfo {author} {\bibfnamefont {Andrea}\ \bibnamefont {Cavalleri}},\ }\bibfield  {title} {\enquote {\bibinfo {title} {Non-equilibrium control of complex solids by nonlinear phononics},}\ }\href {https://dx.doi.org/10.1088/0034-4885/79/6/064503} {\bibfield  {journal} {\bibinfo  {journal} {Rep. Prog. Phys.}\ }\textbf {\bibinfo {volume} {79}},\ \bibinfo {pages} {064503} (\bibinfo {year} {2016})}\BibitemShut {NoStop}%
\bibitem [{\citenamefont {Sentef}(2017)}]{PhysRevB.95.205111}%
  \BibitemOpen
  \bibfield  {author} {\bibinfo {author} {\bibfnamefont {M.~A.}\ \bibnamefont {Sentef}},\ }\bibfield  {title} {\enquote {\bibinfo {title} {Light-enhanced electron-phonon coupling from nonlinear electron-phonon coupling},}\ }\href {https://link.aps.org/doi/10.1103/PhysRevB.95.205111} {\bibfield  {journal} {\bibinfo  {journal} {Phys. Rev. B}\ }\textbf {\bibinfo {volume} {95}},\ \bibinfo {pages} {205111} (\bibinfo {year} {2017})}\BibitemShut {NoStop}%
\bibitem [{\citenamefont {Haegeman}\ \emph {et~al.}(2011)\citenamefont {Haegeman}, \citenamefont {Cirac}, \citenamefont {Osborne}, \citenamefont {Pi\ifmmode~\check{z}\else \v{z}\fi{}orn}, \citenamefont {Verschelde},\ and\ \citenamefont {Verstraete}}]{Haegeman2011}%
  \BibitemOpen
  \bibfield  {author} {\bibinfo {author} {\bibfnamefont {Jutho}\ \bibnamefont {Haegeman}}, \bibinfo {author} {\bibfnamefont {J.~Ignacio}\ \bibnamefont {Cirac}}, \bibinfo {author} {\bibfnamefont {Tobias~J.}\ \bibnamefont {Osborne}}, \bibinfo {author} {\bibfnamefont {Iztok}\ \bibnamefont {Pi\ifmmode~\check{z}\else \v{z}\fi{}orn}}, \bibinfo {author} {\bibfnamefont {Henri}\ \bibnamefont {Verschelde}}, \ and\ \bibinfo {author} {\bibfnamefont {Frank}\ \bibnamefont {Verstraete}},\ }\bibfield  {title} {\enquote {\bibinfo {title} {Time-dependent variational principle for quantum lattices},}\ }\href {https://link.aps.org/doi/10.1103/PhysRevLett.107.070601} {\bibfield  {journal} {\bibinfo  {journal} {Phys. Rev. Lett.}\ }\textbf {\bibinfo {volume} {107}},\ \bibinfo {pages} {070601} (\bibinfo {year} {2011})}\BibitemShut {NoStop}%
\bibitem [{\citenamefont {Haegeman}\ \emph {et~al.}(2016)\citenamefont {Haegeman}, \citenamefont {Lubich}, \citenamefont {Oseledets}, \citenamefont {Vandereycken},\ and\ \citenamefont {Verstraete}}]{Haegeman2016}%
  \BibitemOpen
  \bibfield  {author} {\bibinfo {author} {\bibfnamefont {Jutho}\ \bibnamefont {Haegeman}}, \bibinfo {author} {\bibfnamefont {Christian}\ \bibnamefont {Lubich}}, \bibinfo {author} {\bibfnamefont {Ivan}\ \bibnamefont {Oseledets}}, \bibinfo {author} {\bibfnamefont {Bart}\ \bibnamefont {Vandereycken}}, \ and\ \bibinfo {author} {\bibfnamefont {Frank}\ \bibnamefont {Verstraete}},\ }\bibfield  {title} {\enquote {\bibinfo {title} {Unifying time evolution and optimization with matrix product states},}\ }\href {https://link.aps.org/doi/10.1103/PhysRevB.94.165116} {\bibfield  {journal} {\bibinfo  {journal} {Phys. Rev. B}\ }\textbf {\bibinfo {volume} {94}},\ \bibinfo {pages} {165116} (\bibinfo {year} {2016})}\BibitemShut {NoStop}%
\bibitem [{\citenamefont {Paeckel}\ \emph {et~al.}(2019)\citenamefont {Paeckel}, \citenamefont {Köhler}, \citenamefont {Swoboda}, \citenamefont {Manmana}, \citenamefont {Schollwöck},\ and\ \citenamefont {Hubig}}]{mps_time_ev_methods_paeckel}%
  \BibitemOpen
  \bibfield  {author} {\bibinfo {author} {\bibfnamefont {Sebastian}\ \bibnamefont {Paeckel}}, \bibinfo {author} {\bibfnamefont {Thomas}\ \bibnamefont {Köhler}}, \bibinfo {author} {\bibfnamefont {Andreas}\ \bibnamefont {Swoboda}}, \bibinfo {author} {\bibfnamefont {Salvatore~R.}\ \bibnamefont {Manmana}}, \bibinfo {author} {\bibfnamefont {Ulrich}\ \bibnamefont {Schollwöck}}, \ and\ \bibinfo {author} {\bibfnamefont {Claudius}\ \bibnamefont {Hubig}},\ }\bibfield  {title} {\enquote {\bibinfo {title} {Time-evolution methods for matrix-product states},}\ }\href {http://dx.doi.org/10.1016/j.aop.2019.167998} {\bibfield  {journal} {\bibinfo  {journal} {Ann. Phys.}\ }\textbf {\bibinfo {volume} {411}},\ \bibinfo {pages} {167998} (\bibinfo {year} {2019})}\BibitemShut {NoStop}%
\bibitem [{\citenamefont {Yang}\ and\ \citenamefont {White}(2020)}]{Yang_2020prb}%
  \BibitemOpen
  \bibfield  {author} {\bibinfo {author} {\bibfnamefont {M.}~\bibnamefont {Yang}}\ and\ \bibinfo {author} {\bibfnamefont {S.~R.}\ \bibnamefont {White}},\ }\bibfield  {title} {\enquote {\bibinfo {title} {Time-dependent variational principle with ancillary krylov subspace},}\ }\href {https://doi.org/10.1103%2Fphysrevb.102.094315} {\bibfield  {journal} {\bibinfo  {journal} {Phys. Rev. B}\ }\textbf {\bibinfo {volume} {102}} (\bibinfo {year} {2020})}\BibitemShut {NoStop}%
\end{thebibliography}%


\begin{thebibliography}{18}%
\makeatletter
\providecommand \@ifxundefined [1]{%
 \@ifx{#1\undefined}
}%
\providecommand \@ifnum [1]{%
 \ifnum #1\expandafter \@firstoftwo
 \else \expandafter \@secondoftwo
 \fi
}%
\providecommand \@ifx [1]{%
 \ifx #1\expandafter \@firstoftwo
 \else \expandafter \@secondoftwo
 \fi
}%
\providecommand \natexlab [1]{#1}%
\providecommand \enquote  [1]{``#1''}%
\providecommand \bibnamefont  [1]{#1}%
\providecommand \bibfnamefont [1]{#1}%
\providecommand \citenamefont [1]{#1}%
\providecommand \href@noop [0]{\@secondoftwo}%
\providecommand \href [0]{\begingroup \@sanitize@url \@href}%
\providecommand \@href[1]{\@@startlink{#1}\@@href}%
\providecommand \@@href[1]{\endgroup#1\@@endlink}%
\providecommand \@sanitize@url [0]{\catcode `\\12\catcode `\$12\catcode `\&12\catcode `\#12\catcode `\^12\catcode `\_12\catcode `\%12\relax}%
\providecommand \@@startlink[1]{}%
\providecommand \@@endlink[0]{}%
\providecommand \url  [0]{\begingroup\@sanitize@url \@url }%
\providecommand \@url [1]{\endgroup\@href {#1}{\urlprefix }}%
\providecommand \urlprefix  [0]{URL }%
\providecommand \Eprint [0]{\href }%
\providecommand \doibase [0]{http://dx.doi.org/}%
\providecommand \selectlanguage [0]{\@gobble}%
\providecommand \bibinfo  [0]{\@secondoftwo}%
\providecommand \bibfield  [0]{\@secondoftwo}%
\providecommand \translation [1]{[#1]}%
\providecommand \BibitemOpen [0]{}%
\providecommand \bibitemStop [0]{}%
\providecommand \bibitemNoStop [0]{.\EOS\space}%
\providecommand \EOS [0]{\spacefactor3000\relax}%
\providecommand \BibitemShut  [1]{\csname bibitem#1\endcsname}%
\let\auto@bib@innerbib\@empty
\bibitem [{\citenamefont {Holstein}(1959)}]{HOLSTEIN1959325}%
  \BibitemOpen
  \bibfield  {author} {\bibinfo {author} {\bibfnamefont {T.}~\bibnamefont {Holstein}},\ }\href {\doibase https://doi.org/10.1016/0003-4916(59)90002-8} {\bibfield  {journal} {\bibinfo  {journal} {Ann. Phys.}\ }\textbf {\bibinfo {volume} {8}},\ \bibinfo {pages} {325} (\bibinfo {year} {1959})}\BibitemShut {NoStop}%
\bibitem [{\citenamefont {Marchand}\ \emph {et~al.}(2010)\citenamefont {Marchand}, \citenamefont {De~Filippis}, \citenamefont {Cataudella}, \citenamefont {Berciu}, \citenamefont {Nagaosa}, \citenamefont {Prokof'ev}, \citenamefont {Mishchenko},\ and\ \citenamefont {Stamp}}]{Marchand2010}%
  \BibitemOpen
  \bibfield  {author} {\bibinfo {author} {\bibfnamefont {D.~J.~J.}\ \bibnamefont {Marchand}}, \bibinfo {author} {\bibfnamefont {G.}~\bibnamefont {De~Filippis}}, \bibinfo {author} {\bibfnamefont {V.}~\bibnamefont {Cataudella}}, \bibinfo {author} {\bibfnamefont {M.}~\bibnamefont {Berciu}}, \bibinfo {author} {\bibfnamefont {N.}~\bibnamefont {Nagaosa}}, \bibinfo {author} {\bibfnamefont {N.~V.}\ \bibnamefont {Prokof'ev}}, \bibinfo {author} {\bibfnamefont {A.~S.}\ \bibnamefont {Mishchenko}}, \ and\ \bibinfo {author} {\bibfnamefont {P.~C.~E.}\ \bibnamefont {Stamp}},\ }\href {\doibase 10.1103/PhysRevLett.105.266605} {\bibfield  {journal} {\bibinfo  {journal} {Phys. Rev. Lett.}\ }\textbf {\bibinfo {volume} {105}},\ \bibinfo {pages} {266605} (\bibinfo {year} {2010})}\BibitemShut {NoStop}%
\bibitem [{\citenamefont {Sous}\ \emph {et~al.}(2018)\citenamefont {Sous}, \citenamefont {Chakraborty}, \citenamefont {Krems},\ and\ \citenamefont {Berciu}}]{Sous2018}%
  \BibitemOpen
  \bibfield  {author} {\bibinfo {author} {\bibfnamefont {J.}~\bibnamefont {Sous}}, \bibinfo {author} {\bibfnamefont {M.}~\bibnamefont {Chakraborty}}, \bibinfo {author} {\bibfnamefont {R.~V.}\ \bibnamefont {Krems}}, \ and\ \bibinfo {author} {\bibfnamefont {M.}~\bibnamefont {Berciu}},\ }\href {\doibase 10.1103/PhysRevLett.121.247001} {\bibfield  {journal} {\bibinfo  {journal} {Phys. Rev. Lett.}\ }\textbf {\bibinfo {volume} {121}},\ \bibinfo {pages} {247001} (\bibinfo {year} {2018})}\BibitemShut {NoStop}%
\bibitem [{\citenamefont {Ferris}\ and\ \citenamefont {Vidal}(2012)}]{Ferris2012}%
  \BibitemOpen
  \bibfield  {author} {\bibinfo {author} {\bibfnamefont {A.~J.}\ \bibnamefont {Ferris}}\ and\ \bibinfo {author} {\bibfnamefont {G.}~\bibnamefont {Vidal}},\ }\href {https://doi.org/10.1103%2Fphysrevb.85.165146} {\bibfield  {journal} {\bibinfo  {journal} {Phys. Rev. B}\ }\textbf {\bibinfo {volume} {85}} (\bibinfo {year} {2012})}\BibitemShut {NoStop}%
\bibitem [{\citenamefont {Bohrdt}\ \emph {et~al.}(2019)\citenamefont {Bohrdt}, \citenamefont {Chiu}, \citenamefont {Ji}, \citenamefont {Xu}, \citenamefont {Greif}, \citenamefont {Greiner}, \citenamefont {Demler}, \citenamefont {Grusdt},\ and\ \citenamefont {Knap}}]{Bohrdt2019}%
  \BibitemOpen
  \bibfield  {author} {\bibinfo {author} {\bibfnamefont {A.}~\bibnamefont {Bohrdt}}, \bibinfo {author} {\bibfnamefont {C.~S.}\ \bibnamefont {Chiu}}, \bibinfo {author} {\bibfnamefont {G.}~\bibnamefont {Ji}}, \bibinfo {author} {\bibfnamefont {M.}~\bibnamefont {Xu}}, \bibinfo {author} {\bibfnamefont {D.}~\bibnamefont {Greif}}, \bibinfo {author} {\bibfnamefont {M.}~\bibnamefont {Greiner}}, \bibinfo {author} {\bibfnamefont {E.}~\bibnamefont {Demler}}, \bibinfo {author} {\bibfnamefont {F.}~\bibnamefont {Grusdt}}, \ and\ \bibinfo {author} {\bibfnamefont {M.}~\bibnamefont {Knap}},\ }\href {https://doi.org/10.1038%2Fs41567-019-0565-x} {\bibfield  {journal} {\bibinfo  {journal} {Nat. Phys.}\ }\textbf {\bibinfo {volume} {15}},\ \bibinfo {pages} {921} (\bibinfo {year} {2019})}\BibitemShut {NoStop}%
\bibitem [{\citenamefont {White}(1992)}]{white_1992}%
  \BibitemOpen
  \bibfield  {author} {\bibinfo {author} {\bibfnamefont {S.~R.}\ \bibnamefont {White}},\ }\href {https://link.aps.org/doi/10.1103/PhysRevLett.69.2863} {\bibfield  {journal} {\bibinfo  {journal} {Phys. Rev. Lett.}\ }\textbf {\bibinfo {volume} {69}},\ \bibinfo {pages} {2863} (\bibinfo {year} {1992})}\BibitemShut {NoStop}%
\bibitem [{\citenamefont {White}(1993)}]{white_1993}%
  \BibitemOpen
  \bibfield  {author} {\bibinfo {author} {\bibfnamefont {S.~R.}\ \bibnamefont {White}},\ }\href {https://link.aps.org/doi/10.1103/PhysRevB.48.10345} {\bibfield  {journal} {\bibinfo  {journal} {Phys. Rev. B}\ }\textbf {\bibinfo {volume} {48}},\ \bibinfo {pages} {10345} (\bibinfo {year} {1993})}\BibitemShut {NoStop}%
\bibitem [{\citenamefont {Schollw\"ock}(2005)}]{schollwoeck_2005}%
  \BibitemOpen
  \bibfield  {author} {\bibinfo {author} {\bibfnamefont {U.}~\bibnamefont {Schollw\"ock}},\ }\href {https://link.aps.org/doi/10.1103/RevModPhys.77.259} {\bibfield  {journal} {\bibinfo  {journal} {Rev. Mod. Phys.}\ }\textbf {\bibinfo {volume} {77}},\ \bibinfo {pages} {259} (\bibinfo {year} {2005})}\BibitemShut {NoStop}%
\bibitem [{\citenamefont {Schollwöck}(2011)}]{schollwoeck_2011}%
  \BibitemOpen
  \bibfield  {author} {\bibinfo {author} {\bibfnamefont {U.}~\bibnamefont {Schollwöck}},\ }\href {https://www.sciencedirect.com/science/article/pii/S0003491610001752} {\bibfield  {journal} {\bibinfo  {journal} {Ann. Phys.}\ }\textbf {\bibinfo {volume} {326}},\ \bibinfo {pages} {96} (\bibinfo {year} {2011})}\BibitemShut {NoStop}%
\bibitem [{\citenamefont {K\"ohler}\ \emph {et~al.}(2021)\citenamefont {K\"ohler}, \citenamefont {Stolpp},\ and\ \citenamefont {Paeckel}}]{Koehler2021}%
  \BibitemOpen
  \bibfield  {author} {\bibinfo {author} {\bibfnamefont {T.}~\bibnamefont {K\"ohler}}, \bibinfo {author} {\bibfnamefont {J.}~\bibnamefont {Stolpp}}, \ and\ \bibinfo {author} {\bibfnamefont {S.}~\bibnamefont {Paeckel}},\ }\href {http://dx.doi.org/10.21468/SciPostPhys.10.3.058} {\bibfield  {journal} {\bibinfo  {journal} {SciPost Phys.}\ }\textbf {\bibinfo {volume} {10}} (\bibinfo {year} {2021})}\BibitemShut {NoStop}%
\bibitem [{\citenamefont {Stolpp}\ \emph {et~al.}(2021)\citenamefont {Stolpp}, \citenamefont {K{\"o}hler}, \citenamefont {Manmana}, \citenamefont {Jeckelmann}, \citenamefont {Heidrich-Meisner},\ and\ \citenamefont {Paeckel}}]{Stolpp2021}%
  \BibitemOpen
  \bibfield  {author} {\bibinfo {author} {\bibfnamefont {J.}~\bibnamefont {Stolpp}}, \bibinfo {author} {\bibfnamefont {T.}~\bibnamefont {K{\"o}hler}}, \bibinfo {author} {\bibfnamefont {S.~R.}\ \bibnamefont {Manmana}}, \bibinfo {author} {\bibfnamefont {E.}~\bibnamefont {Jeckelmann}}, \bibinfo {author} {\bibfnamefont {F.}~\bibnamefont {Heidrich-Meisner}}, \ and\ \bibinfo {author} {\bibfnamefont {S.}~\bibnamefont {Paeckel}},\ }\href {https://www.sciencedirect.com/science/article/pii/S0010465521002186} {\bibfield  {journal} {\bibinfo  {journal} {Comput. Phys. Commun.}\ }\textbf {\bibinfo {volume} {269}},\ \bibinfo {pages} {108106} (\bibinfo {year} {2021})}\BibitemShut {NoStop}%
\bibitem [{\citenamefont {Moroder}\ \emph {et~al.}(2023)\citenamefont {Moroder}, \citenamefont {Grundner}, \citenamefont {Damanet}, \citenamefont {Schollw\"ock}, \citenamefont {Mardazad}, \citenamefont {Flannigan}, \citenamefont {K\"ohler},\ and\ \citenamefont {Paeckel}}]{PhysRevB.107.214310}%
  \BibitemOpen
  \bibfield  {author} {\bibinfo {author} {\bibfnamefont {M.}~\bibnamefont {Moroder}}, \bibinfo {author} {\bibfnamefont {M.}~\bibnamefont {Grundner}}, \bibinfo {author} {\bibfnamefont {F.}~\bibnamefont {Damanet}}, \bibinfo {author} {\bibfnamefont {U.}~\bibnamefont {Schollw\"ock}}, \bibinfo {author} {\bibfnamefont {S.}~\bibnamefont {Mardazad}}, \bibinfo {author} {\bibfnamefont {S.}~\bibnamefont {Flannigan}}, \bibinfo {author} {\bibfnamefont {T.}~\bibnamefont {K\"ohler}}, \ and\ \bibinfo {author} {\bibfnamefont {S.}~\bibnamefont {Paeckel}},\ }\href {https://link.aps.org/doi/10.1103/PhysRevB.107.214310} {\bibfield  {journal} {\bibinfo  {journal} {Phys. Rev. B}\ }\textbf {\bibinfo {volume} {107}},\ \bibinfo {pages} {214310} (\bibinfo {year} {2023})}\BibitemShut {NoStop}%
\bibitem [{\citenamefont {Singh}\ \emph {et~al.}(2011)\citenamefont {Singh}, \citenamefont {Pfeifer},\ and\ \citenamefont {Vidal}}]{Singh2011}%
  \BibitemOpen
  \bibfield  {author} {\bibinfo {author} {\bibfnamefont {S.}~\bibnamefont {Singh}}, \bibinfo {author} {\bibfnamefont {R.~N.~C.}\ \bibnamefont {Pfeifer}}, \ and\ \bibinfo {author} {\bibfnamefont {G.}~\bibnamefont {Vidal}},\ }\href {\doibase 10.1103/PhysRevB.83.115125} {\bibfield  {journal} {\bibinfo  {journal} {Phys. Rev. B}\ }\textbf {\bibinfo {volume} {83}},\ \bibinfo {pages} {115125} (\bibinfo {year} {2011})}\BibitemShut {NoStop}%
\bibitem [{\citenamefont {Yang}\ and\ \citenamefont {White}()}]{GSE_with_LSE_arxiv}%
  \BibitemOpen
  \bibfield  {author} {\bibinfo {author} {\bibfnamefont {M.}~\bibnamefont {Yang}}\ and\ \bibinfo {author} {\bibfnamefont {S.~R.}\ \bibnamefont {White}},\ }\href@noop {} {\enquote {\bibinfo {title} {Time dependent variational principle with ancillary krylov subspace},}\ }\bibinfo {howpublished} {arXiv (cond-mat)},\ \bibinfo {note} {submitted 2020-05-13. \url{https://arxiv.org/abs/2005.06104} (accessed 2024-10-25)}\BibitemShut {NoStop}%
\bibitem [{\citenamefont {Grundner}\ \emph {et~al.}()\citenamefont {Grundner}, \citenamefont {Blatz}, \citenamefont {Sous}, \citenamefont {Schollwöck},\ and\ \citenamefont {Paeckel}}]{grundner2023cooperpaired}%
  \BibitemOpen
  \bibfield  {author} {\bibinfo {author} {\bibfnamefont {M.}~\bibnamefont {Grundner}}, \bibinfo {author} {\bibfnamefont {T.}~\bibnamefont {Blatz}}, \bibinfo {author} {\bibfnamefont {J.}~\bibnamefont {Sous}}, \bibinfo {author} {\bibfnamefont {U.}~\bibnamefont {Schollwöck}}, \ and\ \bibinfo {author} {\bibfnamefont {S.}~\bibnamefont {Paeckel}},\ }\href@noop {} {\enquote {\bibinfo {title} {Cooper-paired bipolaronic superconductors},}\ }\bibinfo {howpublished} {arXiv (cond-mat)},\ \bibinfo {note} {submitted 2023-08-25. \url{https://arxiv.org/abs/2308.13427} (accessed 2024-10-25)}\BibitemShut {NoStop}%
\bibitem [{\citenamefont {Hubig}\ \emph {et~al.}(2015)\citenamefont {Hubig}, \citenamefont {McCulloch}, \citenamefont {Schollw\"ock},\ and\ \citenamefont {Wolf}}]{PhysRevB.91.155115}%
  \BibitemOpen
  \bibfield  {author} {\bibinfo {author} {\bibfnamefont {C.}~\bibnamefont {Hubig}}, \bibinfo {author} {\bibfnamefont {I.~P.}\ \bibnamefont {McCulloch}}, \bibinfo {author} {\bibfnamefont {U.}~\bibnamefont {Schollw\"ock}}, \ and\ \bibinfo {author} {\bibfnamefont {F.~A.}\ \bibnamefont {Wolf}},\ }\href {\doibase 10.1103/PhysRevB.91.155115} {\bibfield  {journal} {\bibinfo  {journal} {Phys. Rev. B}\ }\textbf {\bibinfo {volume} {91}},\ \bibinfo {pages} {155115} (\bibinfo {year} {2015})}\BibitemShut {NoStop}%
\bibitem [{\citenamefont {Yang}\ and\ \citenamefont {White}(2020)}]{Yang_2020prb}%
  \BibitemOpen
  \bibfield  {author} {\bibinfo {author} {\bibfnamefont {M.}~\bibnamefont {Yang}}\ and\ \bibinfo {author} {\bibfnamefont {S.~R.}\ \bibnamefont {White}},\ }\href {https://doi.org/10.1103%2Fphysrevb.102.094315} {\bibfield  {journal} {\bibinfo  {journal} {Phys. Rev. B}\ }\textbf {\bibinfo {volume} {102}} (\bibinfo {year} {2020})}\BibitemShut {NoStop}%
\bibitem [{\citenamefont {Sous}\ \emph {et~al.}(2021)\citenamefont {Sous}, \citenamefont {Kloss}, \citenamefont {Kennes}, \citenamefont {Reichman},\ and\ \citenamefont {Millis}}]{Sous2021}%
  \BibitemOpen
  \bibfield  {author} {\bibinfo {author} {\bibfnamefont {J.}~\bibnamefont {Sous}}, \bibinfo {author} {\bibfnamefont {B.}~\bibnamefont {Kloss}}, \bibinfo {author} {\bibfnamefont {D.~M.}\ \bibnamefont {Kennes}}, \bibinfo {author} {\bibfnamefont {D.~R.}\ \bibnamefont {Reichman}}, \ and\ \bibinfo {author} {\bibfnamefont {A.~J.}\ \bibnamefont {Millis}},\ }\href {https://doi.org/10.1038/s41467-021-26030-3} {\bibfield  {journal} {\bibinfo  {journal} {Nat. Commun.}\ }\textbf {\bibinfo {volume} {12}},\ \bibinfo {pages} {5803} (\bibinfo {year} {2021})}\BibitemShut {NoStop}%
\end{thebibliography}%
